\documentclass[twocolumn,  trackchanges]{aastex631}

\usepackage{amsmath}
\usepackage{float}

\graphicspath{{./}{figures/}}

\begin{document}

\title{Time-Series Photometric Detection and Physical Characterization of Variable Stars in Four Intermediate- to Old-Age Galactic Open Clusters}

\author{K. Belwal}\thanks{E-mail: kuldeepbelwal1997@gmail.com}
\affiliation{Indian Centre for Space Physics, 466 Barakhola, Singabari Road, Netai Nagar, Kolkata 700099, India}

\author[0000-0002-8988-8434]{D. Bisht}\thanks{E-mail: devendrabisht297@gmail.com}
\affiliation{Indian Centre for Space Physics, 466 Barakhola, Singabari Road, Netai Nagar, Kolkata 700099, India}

\author[0000-0000-0000-0000]{Ing-Guey Jiang}\thanks{E-mail: jiang@phys.nthu.edu.tw}
\affiliation{Department of Physics and Institute of Astronomy, National Tsing-Hua University, Hsinchu 30013, Taiwan}

\author{Mohit Singh Bisht}
\affiliation{Indian Centre for Space Physics, 466 Barakhola, Singabari Road, Netai Nagar, Kolkata 700099, India}

\author{A. Raj}
\affiliation{Indian Centre for Space Physics, 466 Barakhola, Singabari Road, Netai Nagar, Kolkata 700099, India}

\author{S. K. Chakrabarti}
\affiliation{Indian Centre for Space Physics, 466 Barakhola, Singabari Road, Netai Nagar, Kolkata 700099, India}

\author{D. Bhowmick}
\affiliation{Indian Centre for Space Physics, 466 Barakhola, Singabari Road, Netai Nagar, Kolkata 700099, India}

\begin{abstract}

We present a ground-based time-series photometric study of stellar variability in four intermediate- to old-age open clusters—NGC~2192, NGC~2266, NGC~2509, and IC~1369—based on high-cadence Cousins $R$-band observations obtained with the 0.6\ m VASISTHA telescope at the IERCOO observatory. The monitoring campaign comprises more than $\sim$34\ h of time-series data, providing sensitivity to short-period variability on timescales of $\sim$0.02--2 d. We identified between 190 and 290 probable members in each cluster using a Gaussian Mixture Model. Structural parameters were derived from radial density profiles fitted with King models. Fundamental parameters were further constrained using color--magnitude diagram analysis with PARSEC isochrones, yielding ages of $\sim$0.3--1.6\ Gyr and distances of $\sim$2.5--3.9\ kpc. From the time-series photometry, we identify four new variable stars and seven previously uncharacterized periodic variables, including $\delta$~Scuti and $\gamma$~Doradus pulsators, as well as rotational variables. The detected variables exhibit periods between $\sim$0.12--0.90\ d,  with R-band amplitudes ranging from 0.01 to 0.20 mag. Periods were determined using Lomb-Scargle analysis of calibrated light curves. For a subset of variables, spectral energy distribution fitting was performed to derive effective temperatures ($\sim$4300--10\,000\ K), radii ($\sim$1.3--46\ $R_\odot$), and luminosities ($\sim$2--100\ $L_\odot$), enabling reliable placement on the Hertzsprung-Russell diagram. We present PHOEBE light-curve modelling of the W UMa-type eclipsing binary Gaia DR3 2164531610149292288 in IC 1369, deriving its physical parameters and providing the first detailed characterization beyond its previously reported variability. These results demonstrate that combining dense-cadence ground-based observations with Gaia astrometry provides a reliable approach for identifying and characterizing variable stars in OCs.

\end{abstract}



\section{Introduction}

Open clusters (OCs) serve as excellent testbeds for investigating stellar variability, as their member stars share common distances, ages, and chemical compositions \citep{mowlavi2013stellar, lata2023photometric, elsanhoury2025deeply}. Such homogeneity enables direct comparison of stars at different evolutionary stages and provides valuable constraints on theoretical models of stellar structure and evolution. Photometric monitoring of cluster members reveals variability arising from both intrinsic processes, such as stellar pulsations and rotational modulation, and extrinsic effects, including eclipses in binary systems \citep{xue2019ngc, joshi2020photometric, sandquist2020variability}.

Photometric monitoring programs of OCs have contributed significantly to studies of stellar evolution, star–disk interactions, exoplanet searches, and systematic surveys for variable stars \citep{zhang2009tx, gillen2020ngts, cody2014csi, burke2006survey, pepper2017low, sandquist2011variable, handler2011new, lata2019short, soares2020catalog}. Because cluster parameters such as distance and age are well constrained, variable stars in OCs provide valuable constraints on stellar evolution and enable calibration of fundamental relations such as the period–luminosity relation \citep{iorio2019shape, skowron2019three, minniti2020using, ccinar2025tracing}.

 Intermediate-age OCs are particularly suitable for studying short-period pulsators such as $\delta$ Scuti and $\gamma$ Doradus stars, as their main-sequence turnoff regions lie close to the instability strip. Such clusters commonly host $\delta$ Scuti and $\gamma$ Doradus stars. $\delta$ Scuti stars exhibit short-period pulsations driven by the $\kappa$-mechanism, while $\gamma$ Doradus stars show gravity-mode pulsations that provide important constraints on stellar interiors and internal mixing processes \citep{arentoft2005dozen}.

The clusters NGC~2509, NGC~2266, NGC~2192, and IC~1369 have been 
characterised in several photometric and Gaia-based studies 
\citep{Yontan2023, Maciejewski2008, Park1999, Straizys2020}. Previous studies indicate that these four clusters have ages of approximately 0.3--1.6\,Gyr, distances of about 2.5--3.5\,kpc, and extinctions of $A_V \sim 0.2$--2.6\,mag. These differences make them suitable targets for investigating stellar variability under varying cluster conditions.

 This work presents a ground-based time-series photometric study of stellar variability in the four OCs studied here. These intermediate- to old-age systems possess well-defined main sequences and turnoff regions, and have reliable Gaia~DR3 astrometric memberships, yet their short-period variable-star populations have not been investigated through dedicated, high-cadence observations.

The targets were observed as part of an ongoing time-domain survey using the 0.6 m VASISTHA telescope at the IERCOO observatory, with  a similar instrumental configuration and observing cadence. The study combines dense R-band monitoring with Gaia-based membership probabilities derived via a Gaussian Mixture Model, enabling a consistent search for pulsators and eclipsing systems across all clusters. The properties of selected variables are further constrained through multi-wavelength SED fitting and detailed binary light-curve modelling. Unlike previous works based mainly on archival or sparsely sampled data, the present observations are optimized for periods below $\sim2$~d and provide uniform sensitivity to low-amplitude variability.

The structure of this paper is as follows. Section~\ref{sec: data} describes the observations, data reduction procedures, and photometric analysis. Section~\ref{sec: membership} presents the identification of cluster members using Gaia~DR3 astrometry and the Gaussian Mixture Model. Section~\ref{sec: rdp} discusses the structural properties of the clusters, including radial density profile analysis and King-profile fitting, along with the determination of fundamental parameters through colour-magnitude diagrams in Section~\ref{sec: cmd}. Section~\ref{sec: variable} presents the identification and classification of variable stars based on time-series photometry and period analysis. Section~\ref{sec: variable} also provides the physical characterization of the variable stars, including SED-based parameter estimation, placement on the Hertzsprung-Russell diagram, and detailed light-curve modeling of the contact binary system. Finally, Section~\ref{sec: conclusions} summarises the main conclusions of this study and outlines prospects for future work.

\begin{figure*}[ht]
    \centering

    \includegraphics[width=0.45\textwidth,height=7cm]{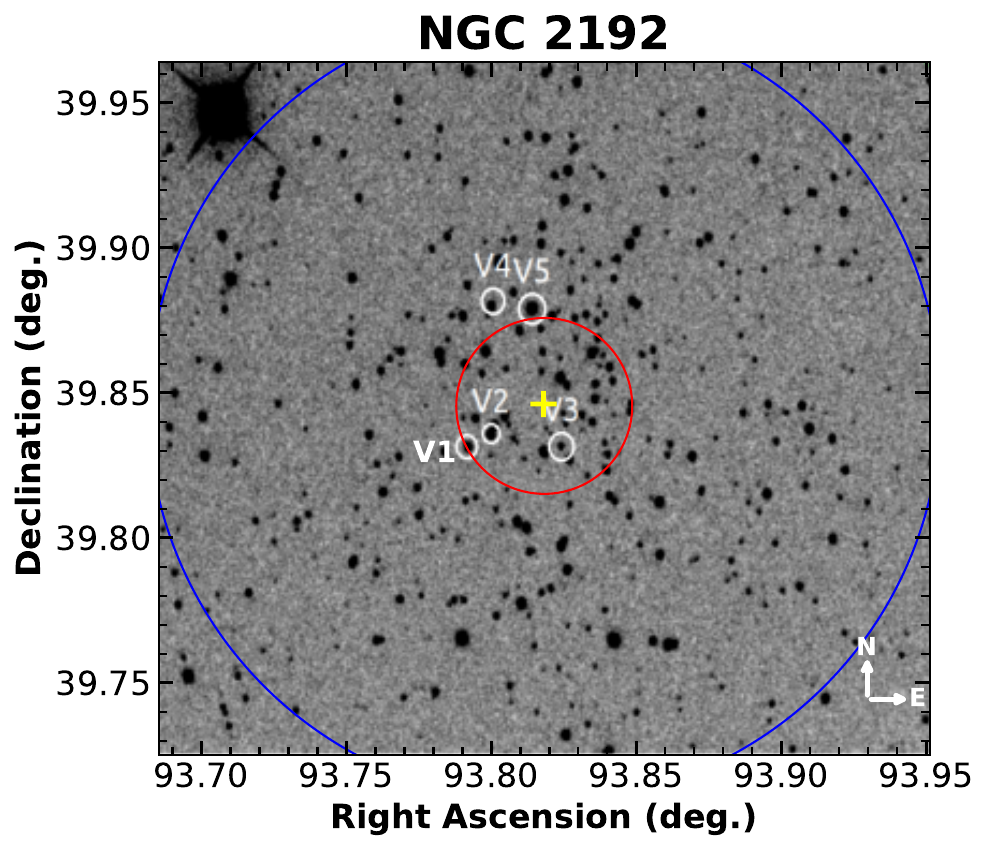}
    \includegraphics[width=0.45\textwidth,height=7cm]{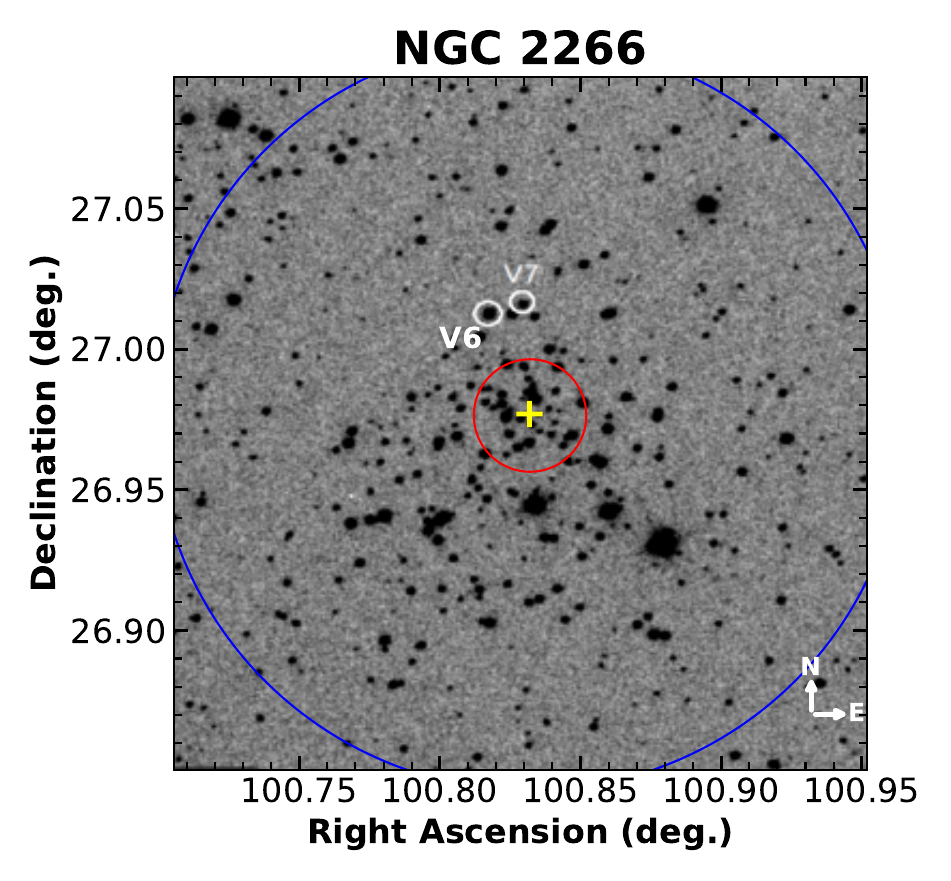}

    \includegraphics[width=0.45\textwidth,height=7cm]{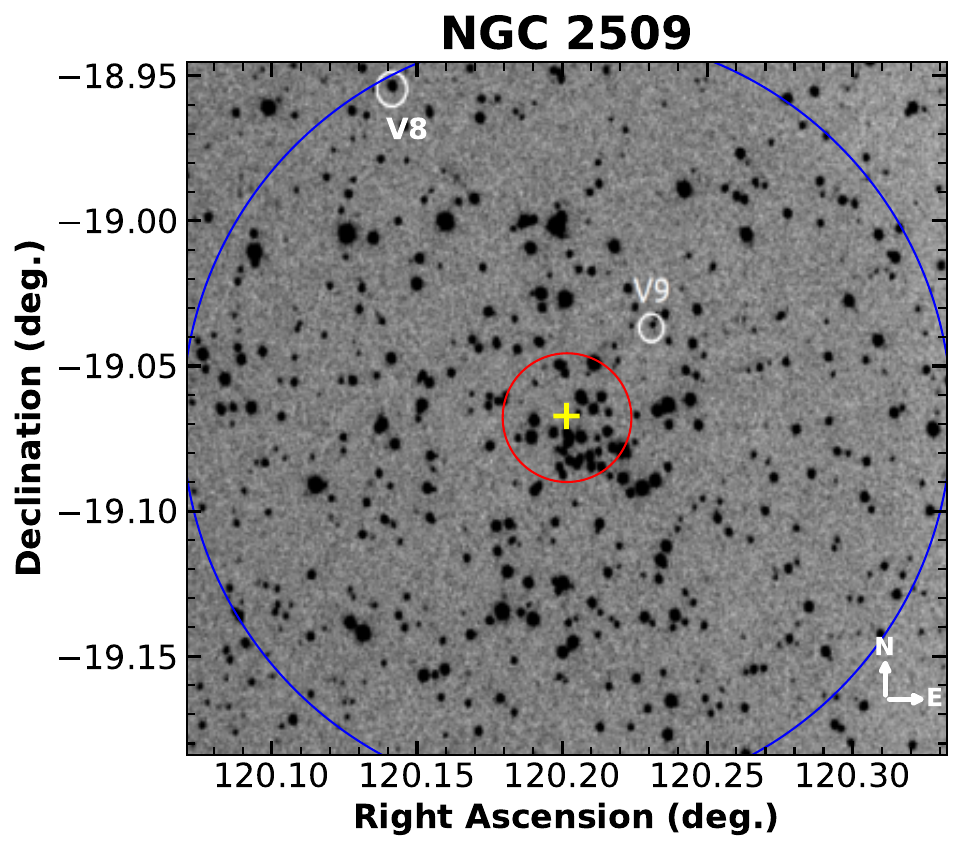}
    \includegraphics[width=0.45\textwidth,height=7cm]{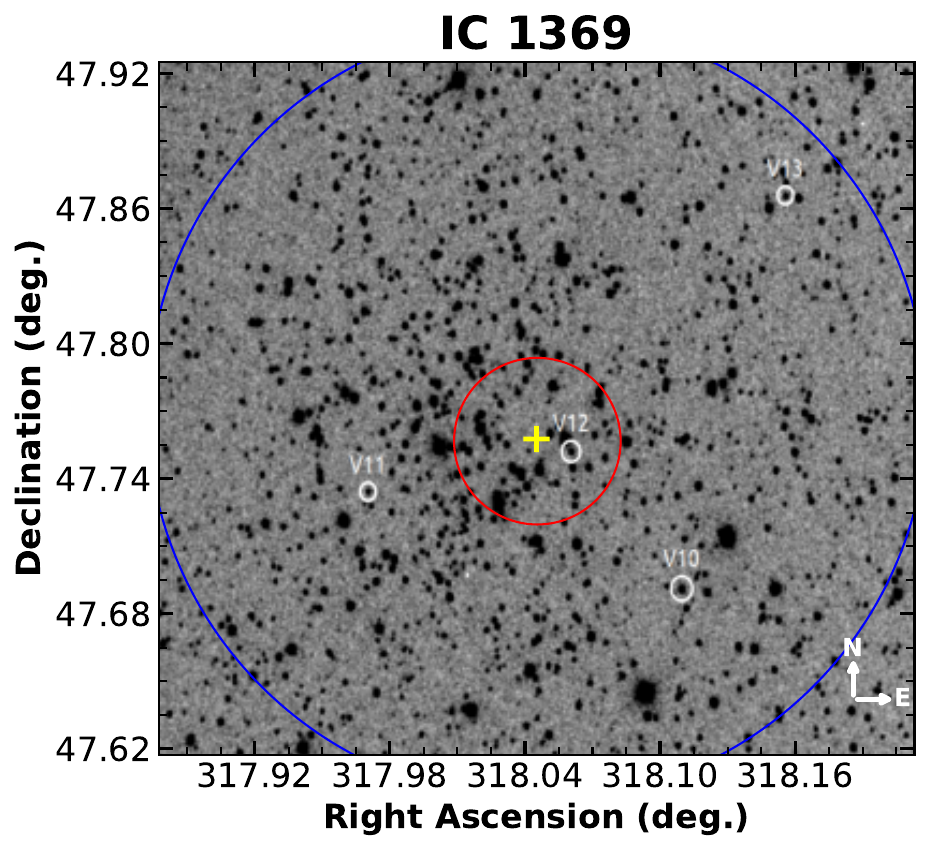}

    \caption{$R$-band finding charts of the observed fields for the OCs NGC~2192, NGC~2266, NGC~2509, and IC~1369.
The images are representative frames from the time-series photometric observations used in this study. Newly identified variable stars are marked with white circles and labelled by their IDs. Yellow plus symbols indicate the cluster centres. The inner red circle denotes the cluster core radius, while the outer blue circle indicates the cluster (limiting) radius, both estimated in the present analysis. The orientation is north up and east to the left, with right ascension and declination shown in degrees.}
\label{fig: id_chart}
\end{figure*}

\section{Data Used} \label{sec: data}
\subsection{Ground-Based Time-Series Photometry
}
The photometric data for the present study were obtained using the 0.6 m VASISTHA reflecting telescope located at the Ionospheric and Earthquake Research Centre and Optical Observatory (IERCOO), established by the Indian Centre for Space Physics (ICSP) at Sitapur, Paschim Medinipur, West Bengal. This telescope, the largest in eastern India, is mounted on an Ascension 200 German equatorial system equipped with high-resolution encoders, allowing precise tracking and pointing. The telescope employs a 0.6 m primary mirror with a focal ratio of $f/6.5$ and is equipped with an Atik 460EX Mono CCD camera. The detector consists of $2749 \times 2199$ pixels, each of size $4.54~\mu\mathrm{m}$, providing an image scale of $0.235~\mathrm{arcsec~pixel}^{-1}$. During all observing nights, the airmass remained within the range 1.0–1.7, ensuring stable atmospheric transmission and minimal differential extinction across the frames. All observations were performed in the Cousins $R$ band, which offers good sensitivity for stellar photometry and minimizes atmospheric extinction effects. OCs NGC 2192, NGC 2266, NGC 2509, and IC 1369 were monitored between January and October 2025 as part of an ongoing time-domain survey.  The ID charts are shown in Fig. \ref{fig: id_chart} and the observation log is provided in Table~\ref{tab:log_table}.  The four clusters are  located at different Galactic latitudes of approximately $|b|\approx2^\circ$ to $10^\circ$, corresponding to vertical heights of $|Z|\approx90$–$400$~pc for the Gaia-based distances. This range samples regions with varying stellar densities and field-star contamination, from low-latitude disk fields affected by crowding and extinction to higher-latitude environments with lower background levels. Such differences in Galactic height, rather than age alone, are relevant for assessing the detectability of low-amplitude variables in our dataset.

Exposure times were set to 30--40\,s, depending on seeing and sky background levels, to ensure adequate signal-to-noise while preserving temporal sampling for short-period variability. A total of 1043 science frames were acquired for NGC 2192, 1160 for NGC 2266, 639 for NGC 2509, and 814 for IC 1369, corresponding to net integration times of approximately 10.38 h, 10.04 h, 5.24 h, and 9.04 h, respectively. The reduced number of frames in NGC 2509 lowers the effective signal-to-noise ratio at faint magnitudes, limiting the reliable detection of faint sources in this field. The exposure time of individual frames primarily determines the photometric signal-to-noise ratio and, therefore, the minimum detectable variability amplitude. In contrast, temporal sampling controls sensitivity to short-period variability, whereas the total observation time determines the longest recoverable period and the quality of phase coverage. The present dataset spans a combined observing baseline of $\sim$34.7 h distributed over multiple nights, which provides adequate sampling for periodicities in the range
$\sim$0.02--2 d targeted in this study.

The data were later processed using standard IRAF routines to produce science-ready images suitable for photometric analysis. Aperture and point-spread function (PSF) photometry were performed using the \texttt{DAOPHOT II} \citep{stetson1987daophot} package in IRAF\footnote{ The Image Reduction and Analysis Facility (IRAF) is distributed by the National Optical Astronomy Observatories (NOAO)} to extract instrumental magnitudes of the detected sources. The instrumental magnitudes were then transformed to the standard photometric system using secondary standard stars within the observed field. Three stars in the observed CCD field with well-established standard magnitudes from the literature were identified and used as calibration standards. The difference between their instrumental and literature magnitudes was used to compute the photometric zero-point offset, which was then applied to standardize the instrumental magnitudes of all detected sources.  The photometric calibration follows a standard secondary--standard zero-point transformation procedure using non-variable field stars within the CCD frame, and the resulting photometric zero-point uncertainty is estimated to be $\le$ 0.03 mag. This implementation is similar to the widely adopted approach described in \citet{subramaniam2005ngc}. The photometric calibration follows a standard zero-point transformation using secondary standard stars within the observed field.

\begin{figure}
    \centering
    \includegraphics[width=1\linewidth,height=12cm]{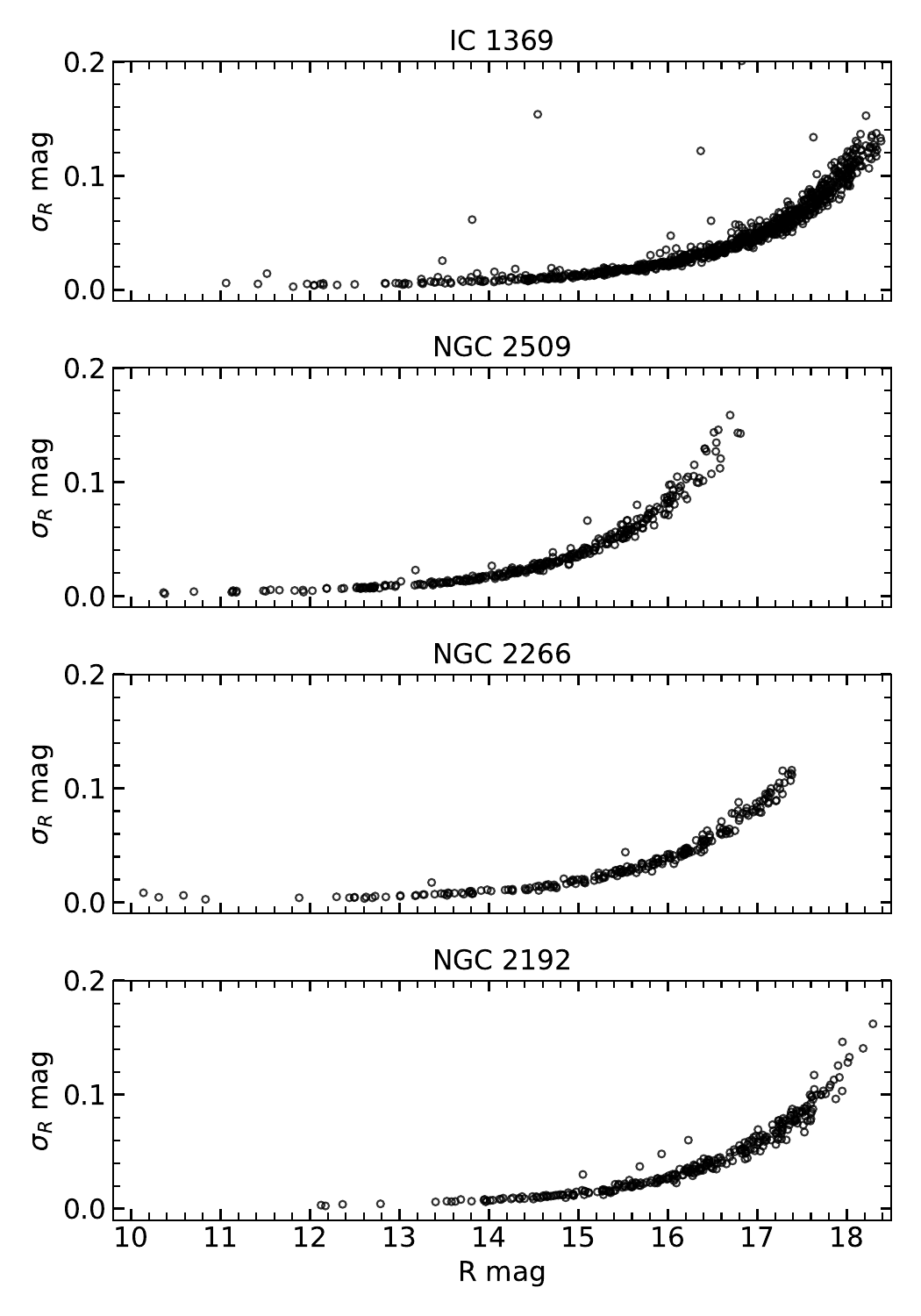}
    \caption{Mean photometric uncertainties as a function of stellar brightness in the $R$ band for the clusters under study. Each panel shows the variation of photometric error with $R$-band magnitude. The plots indicate that the uncertainty remains below $\sim$0.01 mag for the brighter stars and gradually increases up to $\sim$0.15–0.20 mag toward the fainter end. The magnitude distribution for NGC 2509 appears truncated at the faint end compared to the other clusters. This effect arises from differences in observing conditions and the shorter total integration time for this cluster, which limits the detection of faint sources.}
    \label{fig: error_plot}
\end{figure}

The internal photometric errors derived from DAOPHOT are plotted as a function of the R-band magnitude in Fig. \ref{fig: error_plot}.  The photometric precision is strongly magnitude dependent: uncertainties remain $\leq0.02$~mag for $R<16$, increase to $\sim0.05$~mag at $R\approx17.5$, and reach  $\sim0.15$--$0.20$~mag only at the faint limit ($R\gtrsim18$). Adopting a $3\sigma$ detection criterion, the photometric scatter of $\sim0.20$~mag at the faint end corresponds to a practical variability detection threshold of $\approx0.06$~mag. Consequently, the survey is effectively complete for amplitudes $\gtrsim0.05$~mag down to $R\approx17.5$, while the detection efficiency decreases for fainter stars. This limitation primarily affects low-amplitude $\delta$~Scuti and $\gamma$~Doradus candidates, but does not affect eclipsing binaries or rotational variables with larger amplitudes.


Light curves were constructed using the calibrated magnitudes of all candidate stars. Outliers exceeding $3\sigma$ in nightly statistics were removed, and timestamps were converted to Barycentric Julian Date. Period searching was performed using the Lomb--Scargle periodogram \citep{lomb1976least, scargle1982studies}, which is well suited for unevenly sampled observational datasets. The periodograms were computed using the standard Lomb--Scargle normalization, and the false alarm probabilities were estimated analytically assuming Gaussian noise. The dominant period peaks were refined using nonlinear sinusoidal fitting.  Stars showing coherent periodicity, statistically significant Lomb–Scargle peaks, consistent phase-folded light curves, and amplitudes exceeding the photometric detection threshold were classified as bona fide variables.

\subsection{Gaia DR3 Astrometric and Photometric Data}

The \textit{Gaia} mission provides high-precision astrometric and photometric data for more than 1.46 billion sources in its third data release (DR3), including positions ($\alpha$, $\delta$), parallaxes, proper motions ($\mu_{\alpha}\cos\delta$, $\mu_{\delta}$), and photometry in the $G$, $G_{BP}$, and $G_{RP}$ bands \citep{collaboration2023gaia, collaboration2021gaia}. For this study, we retrieved \textit{Gaia} DR3 sources in the direction of each cluster and applied standard quality criteria to ensure reliable astrometry, including Renormalized Unit Weight Error (RUWE) $\leq 1.4$ \citep{lindegren2021gaia}, the availability of five-parameter astrometric solutions, and valid photometry in all three \textit{Gaia} bands. We selected stars with $G \leq 20.1$, ensuring uniform completeness across the observed fields. The variable stars identified from our ground-based monitoring were cross-matched with the \textit{Gaia} DR3 variability catalogue, which independently confirms their photometric variability. However, the sparse and irregular \textit{Gaia} cadence limits reliable period determination for short-period signals ($P\lesssim1$~d), whereas the dense ground-based time series presented here provides more accurate periods and amplitudes.

\begin{table}[]
    \centering
    \begin{tabular}{|c c c|}
    \hline
      Date of  & No. of frames &  Cluster  \\
     Observation &  $\times$ Exposure (sec) & Name \\
    \hline
    2025/Feb/03 &  190 $\times$ 40  & NGC 2192 \\
    2025/Feb/04 & 81 $\times$ 40  & NGC 2192 \\
    2025/Feb/05 & 156 $\times$ 40  & NGC 2192 \\
    2025/Feb/15 & 150 $\times$ 30  & NGC 2192 \\
    2025/Feb/16 & 166 $\times$ 30  & NGC 2192 \\
    2025/Feb/17 & 110 $\times$ 30  & NGC 2192 \\
    2025/Feb/21 & 250 $\times$ 30  & NGC 2192 \\
    
    2025/Jan/17 & 106 $\times$ 40  & NGC 2266 \\
    2025/Jan/18 & 190 $\times$ 30  & NGC 2266 \\
    2025/Jan/20 & 90  $\times$ 30  & NGC 2266 \\
    2025/Jan/21 & 299 $\times$ 30  & NGC 2266 \\
    2025/Jan/22 & 270 $\times$ 30  & NGC 2266 \\
    2025/Feb/06 & 108 $\times$ 30  & NGC 2266 \\
    2025/Feb/06 & 107 $\times$ 30  & NGC 2266 \\

    2025/Feb/06 & 68 $\times$ 30  & NGC 2509 \\
    2025/Feb/07 & 208 $\times$ 30  & NGC 2509 \\
    2025/Mar/06 & 194 $\times$ 30  & NGC 2509 \\
    2025/April/30 & 100 $\times$ 30  & NGC 2509 \\
    2025/May/02 & 59 $\times$ 30  & NGC 2509 \\

    2025/Oct/11 & 108 $\times$ 30  & IC 1369 \\
    2025/Oct/12 & 128 $\times$ 40  & IC 1369 \\
    2025/Oct/13 & 171 $\times$ 40  & IC 1369 \\
    2025/Oct/14 & 160 $\times$ 40  & IC 1369 \\
    2025/Oct/15 & 167 $\times$ 40  & IC 1369 \\
    2025/Oct/16 & 80 $\times$ 40  & IC 1369 \\
 
    \hline
    \end{tabular}
    \caption{Log of CCD observations for the clusters NGC 2192, NGC 2266, NGC 2509, and IC 1369 obtained with the 0.6 m VASISTHA telescope. The table consists of three columns: observation dates (column 1), the total number of frames and exposure times in the R band (column 2), and cluster names (column 3). These data were acquired as part of a time-series photometric campaign to identify and characterize variable stars in the observed clusters.}
    \label{tab:log_table}
\end{table}

\subsection{Multi-wavelength Photometric Data}

For the spectral energy distribution (SED) analysis, we used multi-wavelength photometric data from GALEX (ultraviolet; \citet{martin2005galaxy}) (where available), APASS (optical; \citet{henden2016vizier}), 2MASS (near-infrared; \citet{skrutskie2006two}), and WISE (mid-infrared; \citet{wright2010wise}), providing broad wavelength coverage from the ultraviolet to the infrared. These data were used exclusively for SED fitting to derive stellar physical parameters and were not employed in the variability analysis (see Section \ref{subsec: sed_hr_diagram}).

\section{Membership Probability}\label{sec: membership}

To reliably identify cluster members, we applied a Gaussian Mixture Model (GMM) in the Gaia astrometric parameter space, following the methodology outlined by \citet{agarwal2021ml}. These clusters span a wide range of Galactic locations, with varying stellar densities and field-star contamination levels. Particularly, NGC 2509 and IC 1369 lie toward the outer Galactic disk, where disk stars dominate the background population, while NGC 2266 and NGC 2192 are located at relatively higher Galactic latitudes. Under such conditions, traditional selection methods based on simple parallax or proper-motion thresholds often fail to effectively distinguish between cluster and field populations due to kinematic overlap.

The GMM approach provides a robust statistical framework by modeling the observed astrometric distribution as a combination of multiple Gaussian components—typically one representing the cluster population and the others modeling the field stars. This unsupervised, data-driven method assigns membership probabilities to individual stars based solely on their astrometric parameters ($\mu_{\alpha} \cos \delta$, $\mu_{\delta}$, and $\varpi$)  without requiring explicit a priori constraints on the cluster’s mean motion or dispersion. Consequently, GMM is particularly suitable for intermediate- and old-age OCs, where the contrast in proper motion with the field population is modest. The method has been effectively applied for membership determination in several recent works (e.g., \citet{gao2018machine, qiu2024deeper}).

For each cluster, the input dataset was refined following the procedure described in our earlier work \citep{belwal2025unveiling}. We retrieved Gaia DR3 sources within a circular region centered on the cluster coordinates. We retained only stars with complete five-parameter astrometric solutions and valid photometry in the $G$, $G_{BP}$, and $G_{RP}$ bands. Quality criteria included positive parallaxes, proper motion uncertainties $\leq \epsilon_{\mu}$, and RUWE $\leq$ 1.4 \citep{lindegren2021gaia}. We used the $k$-Nearest Neighbours algorithm  \citep{cover1967nearest} to estimate the cluster centroid in three-dimensional astrometric space ({$\mu_{\alpha} \cos \delta$}, $\mu_{\delta}$, $\varpi$) based on an initial kinematic pre-selection in the proper motion–parallax plane. Here, $\mu_{\alpha} \cos \delta$ and $\mu_{\delta}$ represent the components of stellar proper motion along the right ascension and declination directions, respectively, while $\varpi$ denotes the stellar parallax. This step was used only to provide a robust initial estimate for the GMM initialization and does not affect the final probabilistic classification.

The refined datasets were normalized and analyzed using a two-component GMM implemented via the Expectation–maximization algorithm \citep{dempster1977maximum, mclachlan2000finite}. This configuration adequately separates the dominant cluster and field populations, and adding additional components did not produce a statistically significant improvement. Each star was assigned a membership probability ($p$), and only sources with $p \geq p_\mathrm{min}$ were retained as probable cluster members for subsequent photometric and variability analysis. Because the model includes only two Gaussian components, the membership determination primarily reflects the central cluster population and may not capture extended non-Gaussian structures such as tidal features. The mean astrometric parameters derived from the high-probability members agree well with recent catalog values \citep{cantat2020painting, hunt2023improving}, supporting the reliability of our membership determination.

\begin{figure*}
   \centering
   \includegraphics[width=18cm,height=6.0cm]{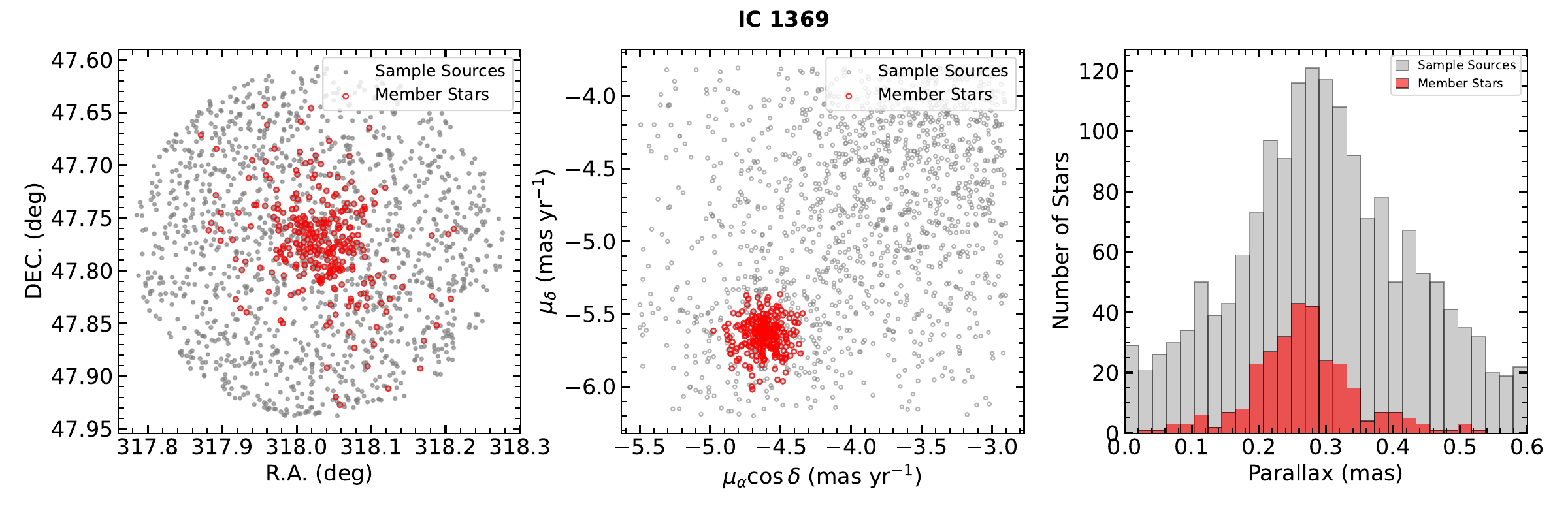}
   \caption{Spatial, kinematic, and parallax distributions of the stars in the field of IC 1369. The grey points represent all sources within the adopted search radius, while the red points denote the stars identified as cluster members. The panels illustrate (left) the spatial concentration of members, (middle) their tight grouping in proper-motion space, and (right) the parallax histogram, which shows a distinct peak corresponding to the cluster. This figure shows IC 1369 as a representative example of the procedure applied to all four clusters analyzed in this study.}
   \label{fig:membership}
\end{figure*}

We adopted a membership probability threshold of 60$\%$, yielding 196 probable members for NGC~2192, 285 for NGC~2266, 273 for NGC~2509, and 293 for IC~1369. Similar probability thresholds have been adopted in several recent Gaia-based cluster studies (e.g., \citep{cantat2020painting, hunt2023improving}), where values between 50$\%$ and 70$\%$ are commonly used to
balance completeness and contamination. Our adopted threshold of
$p \geq 0.60$, therefore, lies within the range typically used in
astrometric membership analyses and provides a statistically stable
sample for subsequent photometric and variability studies. In Fig. \ref{fig:membership} and \ref{fig:membership_appendix}, stars with membership probabilities $\ge$ 60$\%$ are shown in red open circles, while the grey open circles represent the full sample of observed stars. We then cross-matched our member sample with the catalog of \cite{hunt2023improving},  obtaining 192 common stars for NGC~2192, 274 for NGC~2266, 
268 for NGC~2509, and 287 for IC~1369. The high level of agreement ($\ge$ 95$\%$ in all cases) further validates the robustness of our membership determination.

\section{Structural Parameters} \label{sec: rdp}

 To investigate the clusters' structural parameters, we constructed their radial density profiles. The cluster centers were first determined iteratively by fitting a two-dimensional Gaussian function to the stellar surface density distribution. The stellar surface density, $\rho(r)$, was then computed in concentric annuli around the cluster center. The cluster radius ($r_{\rm cl}$) is defined as the distance from the cluster center at which the stellar surface density becomes statistically consistent with the background field level, i.e., where $\rho(r)$ agrees with $\rho_{\rm bg}$ within the $1\sigma$ Poisson uncertainty of the outer annuli. To estimate the cluster radius and assess the level of field-star contamination, we constructed the radial stellar surface density profile following the approach described by \citet{belwal2024exploring} and \citet{bisht2022deep, bisht2025dynamical}. This was accomplished by dividing the region around the cluster center into a series of concentric circular annuli. Within each annulus, we counted the number of stars and normalized this count by the area of the annulus to derive the stellar surface density, $\rho_i = N_i/A_i$, where $N_i$ and $A_i$ represent the number of stars and the area of the $i$th annulus, respectively.
The observed density distributions (red points) were fitted with the empirical three-parameter \cite{king1962structure}, expressed as:

\begin{equation}
\rho(r) = \rho_{bg} + \rho_0 \left[ \frac{1}{\sqrt{1 + (r/r_c)^2}} - \frac{1}{\sqrt{1 + (r_t/r_c)^2}} \right]^2,
\end{equation}

Where $\rho(r)$ is the stellar surface density at a radial distance $r$, $\rho_{bg}$ is the background field density, $\rho_0$ is the central stellar density, $r_c$ is the core radius, and $r_t$ denotes the tidal radius beyond which the cluster is no longer gravitationally bound.

\begin{figure}
    \centering
   \hbox{
    \includegraphics[width=4.2cm,height=4.2cm]{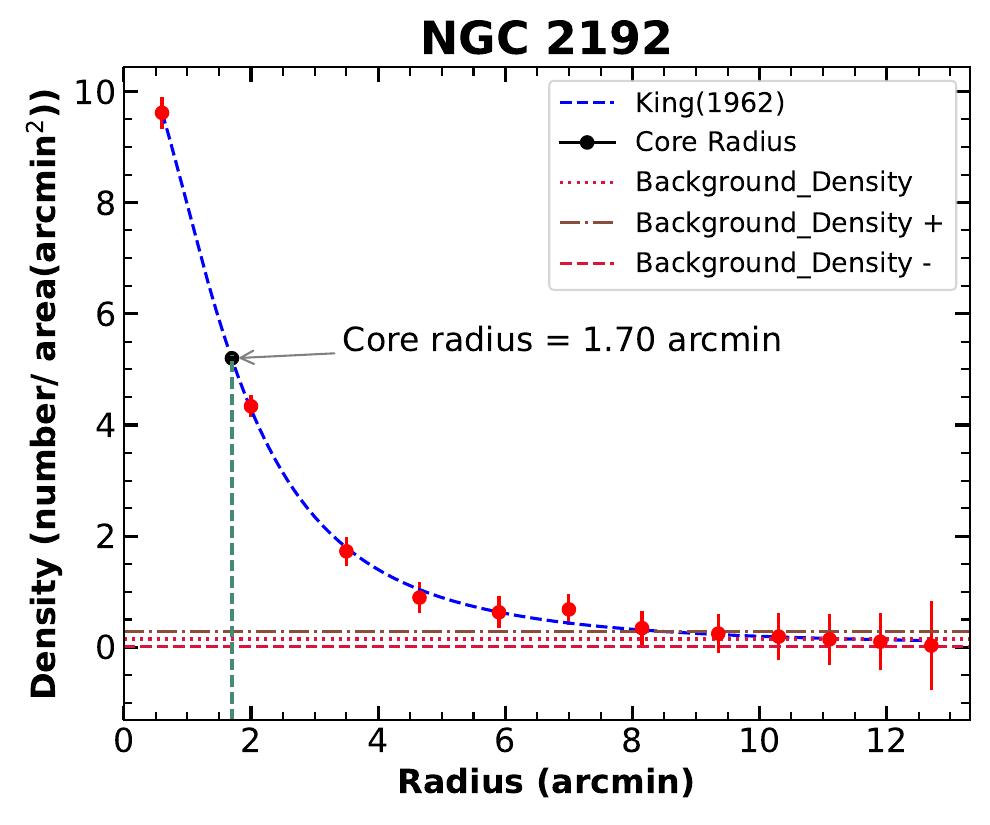}
    \includegraphics[width=4.2cm,height=4.2cm]{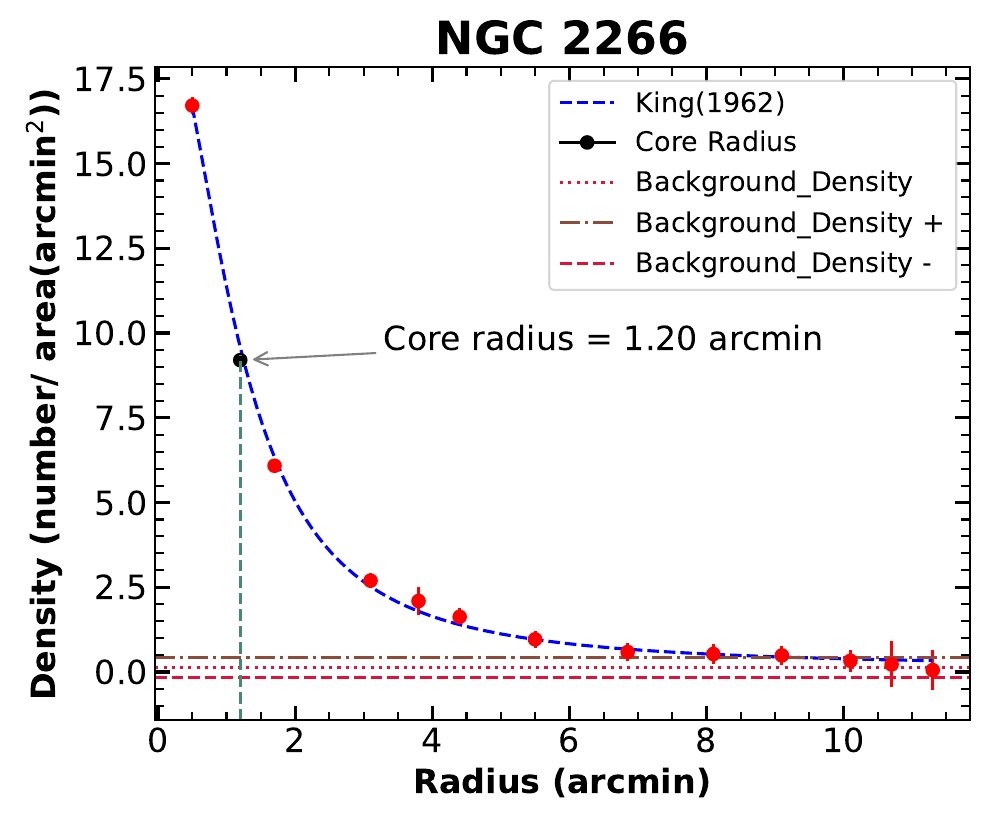}
    }
    \hbox{
    \includegraphics[width=4.2cm,height=4.2cm]{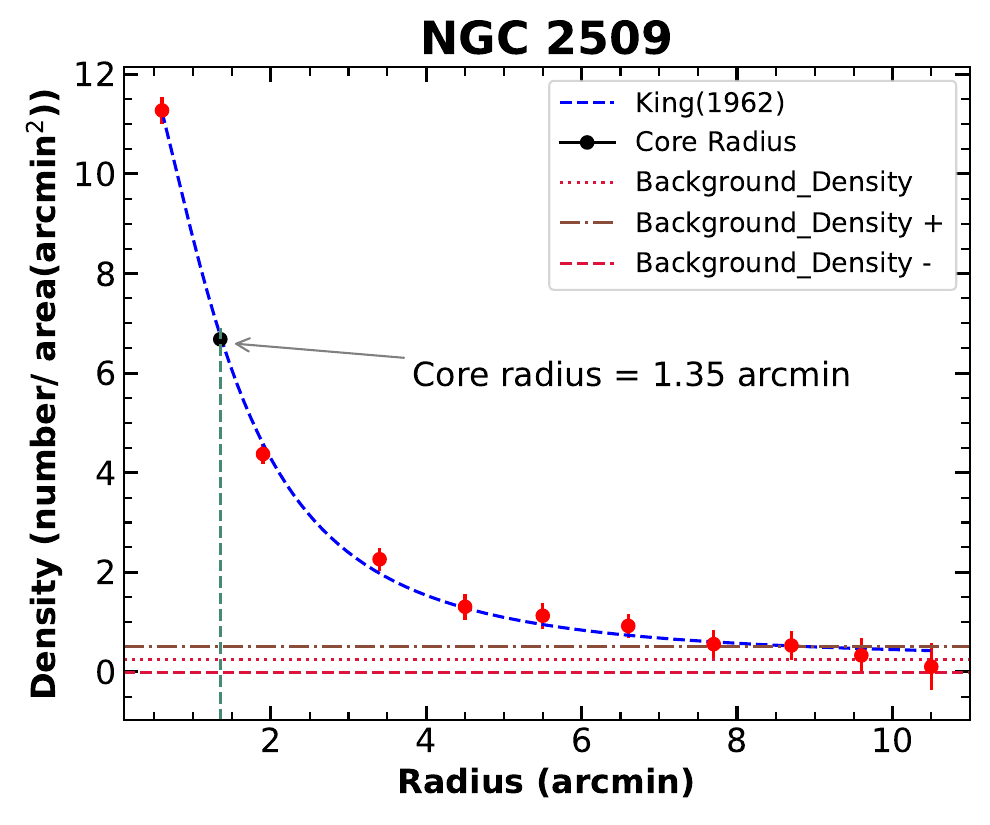}
    \includegraphics[width=4.2cm,height=4.2cm]{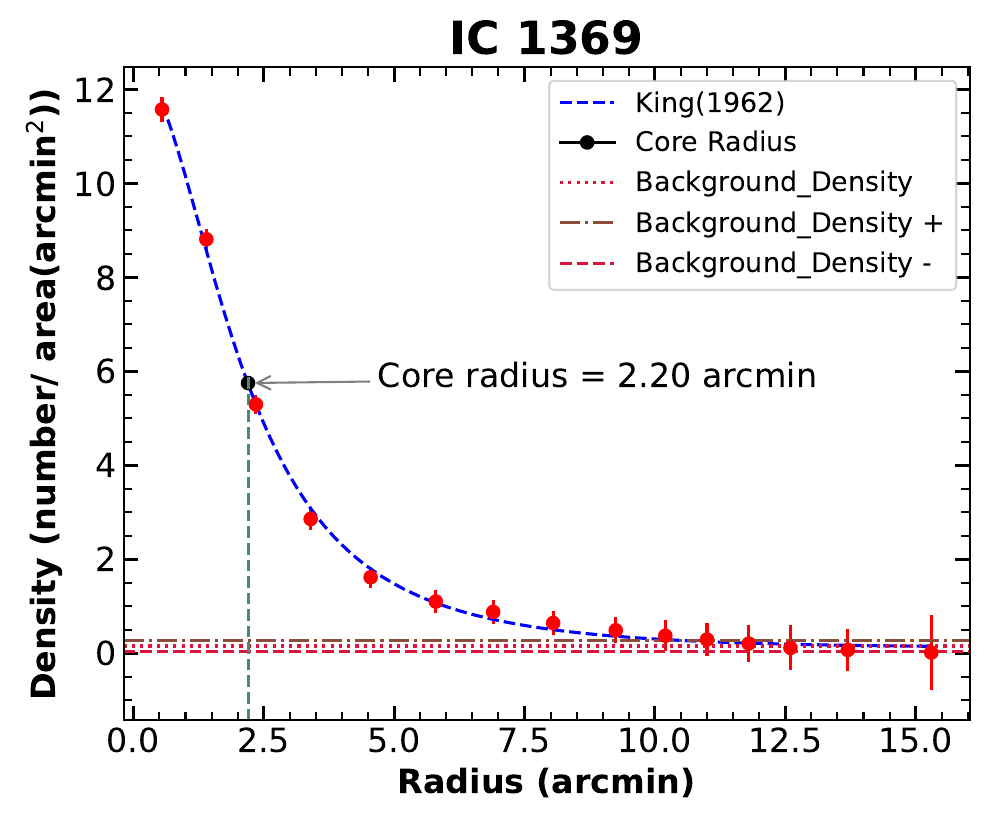}
    }
    \caption{Radial density profiles for the OCs NGC~2192, NGC~2266, NGC~2509 , and IC 1369. The red points with error bars represent the observed stellar density distribution, while the blue dashed curves show the best-fitting three-parameter King (1962) model. The horizontal dashed lines denote the background field stellar density, and the vertical dashed lines mark the core radii. The derived core radii are $1.70$~arcmin for NGC~2192, $1.20$~arcmin for NGC~2266,$1.35$~arcmin for NGC~2509, and $2.22$ arcmin for IC 1369.}
    \label{fig: rdp}
\end{figure}

The best-fitting King profiles (blue dashed curves) reproduce the observed radial density distributions of all four clusters (Fig.~\ref{fig: rdp}). The derived structural parameters (Table~\ref{tab: structural_para}) indicate compact cores with $r_c = 1.70$, $1.20$, $1.35$, and $2.22$ arcmin for NGC~2192, NGC~2266, NGC~2509, and IC~1369, respectively. The corresponding tidal radii range from $r_t \sim 11.6$ to $16.6$ arcmin. The ratios of the core to tidal radii are $r_c/r_t \approx 0.11$, $0.09$, $0.11$, and $0.13$, respectively, consistent with centrally concentrated and dynamically evolved OCs. These structural parameters confirm that the clusters are well defined against the surrounding stellar field and provide the spatial framework for interpreting the detected variable stars.

\begin{table*}[th]
    \centering
     \caption{Structural parameters of the OCs NGC~2192, NGC~2266, NGC~2509, and IC~1369 derived from \cite{king1962structure} model fitting to their stellar Radial density profiles. The table lists the central stellar density ($\rho_0$), background field density ($\rho_{\rm bg}$), core radius ($r_c$), cluster radius ($r_{\rm cl}$), and tidal radius ($r_t$). The quoted uncertainties represent 1$\sigma$ errors obtained from the fitting procedure.}
    \begin{tabular}{|c c c c c|}
    \hline
     Parameters & NGC 2192  &  NGC 2266  &  NGC 2509  &  IC 1369 \\
     \hline
    Central Density($\rho_{0}$) number arcmin$^{-2}$  &  12.30 $\pm$ 1.25  & 19.66 $\pm$ 2.30 &  13.28 $\pm$ 3.80 &   15.70 $\pm$ 1.43 \\
    Background Density($\rho_{bg}$) number arcmin$^{-2}$ & 0.20 $\pm$ 0.15    & 0.18 $\pm$ 0.20   & 0.26 $\pm$ 0.26 &   0.18 $\pm$ 0.15 \\
    Core Radius($r_{c}$) arcmin & 1.70 $\pm$ 0.11  & 1.20  $\pm$ 0.20 & 1.35 $\pm$ 0.30 &  2.22 $\pm$ 0.40 \\
    Core Radius($r_{c}$) parsec & 1.90 $\pm$ 0.30  &  1.21 $\pm$ 0.25 &  0.97 $\pm$ 0.23  & 2.25 $\pm$ 0.50 \\
    Cluster Radius($r_{cl}$) arcmin & 8.10 $\pm$ 1.50 & 8.0 $\pm$ 1.15 & 7.95 $\pm$ 2.0 &  10.40  $\pm$ 1.85 \\
    Cluster Radius($r_{cl}$) parsec & 9.07 $\pm$ 2.14  & 8.08 $\pm$ 1.52  &  5.74 $\pm$ 1.53  &  10.53 $\pm$ 2.32 \\
    Tidal Radius($r_{t}$) arcmin & 15.90 $\pm$ 2.10  & 13.83 $\pm$ 1.85    &  12.60  $\pm$ 1.55  &    16.60  $\pm$ 2.20    \\
    Tidal Radius($r_{t}$) parsec & 17.81 $\pm$ 3.50  & 13.96 $\pm$ 2.52  &  9.09 $\pm$ 1.38  &   16.80 $\pm$ 3.11 \\
    Ratio (r$_c$/r$_t$) &  0.11    & 0.09   & 0.11    &  0.13      \\
    \hline
    
    \end{tabular}
    
    \label{tab: structural_para}
\end{table*}

\section{Fundamental Parameters of the Clusters} \label{sec: cmd}
\subsection{Distances of clusters from trigonometric parallaxes}

Cluster distances were estimated by first selecting stars with membership probabilities greater than 60$\%$ and fitting a Gaussian profile to their parallax distributions to determine the mean trigonometric parallax of each system. The resulting mean parallaxes are 0.232 mas for NGC~2192, 0.258 mas for NGC~2266, 0.377 mas for NGC~2509, and 0.270 mas for IC~1369. Although the distance to a star can formally be obtained by inverting its parallax ($d = 1/\varpi$), this approach is reliable only when the parallax uncertainties are negligible. In practice, Gaia parallaxes contain measurement errors, and directly inverting noisy parallaxes can lead to biased or unrealistic distance estimates, particularly for distant objects. To avoid this problem, we adopt the probabilistic distance estimates of \citet{bailer2018estimating}, which employ a Bayesian framework that incorporates both the measured trigonometric parallax and its associated uncertainty. This method assumes an exponentially decreasing space-density prior and has been shown to provide more robust distance estimates when parallax uncertainties are significant \citep{bailer2015estimating,bailer2018estimating}. The resulting distances are $3.85^{+0.56}_{-0.43}$~kpc for NGC~2192, $3.47^{+0.42}_{-0.30}$~kpc for NGC~2266, $2.48^{+0.22}_{-0.19}$~kpc for NGC~2509, and $3.48^{+0.45}_{-0.36}$~kpc for IC~1369. These values are in good agreement with the distances reported by \citet{cantat2020painting} and \citet{hunt2023improving}. The derived distances were subsequently adopted for all analyses of the cluster parameters presented in this work.

\subsection{Color Magnitude Diagram}

\begin{figure*}
    \includegraphics[width=4cm,height=7cm]{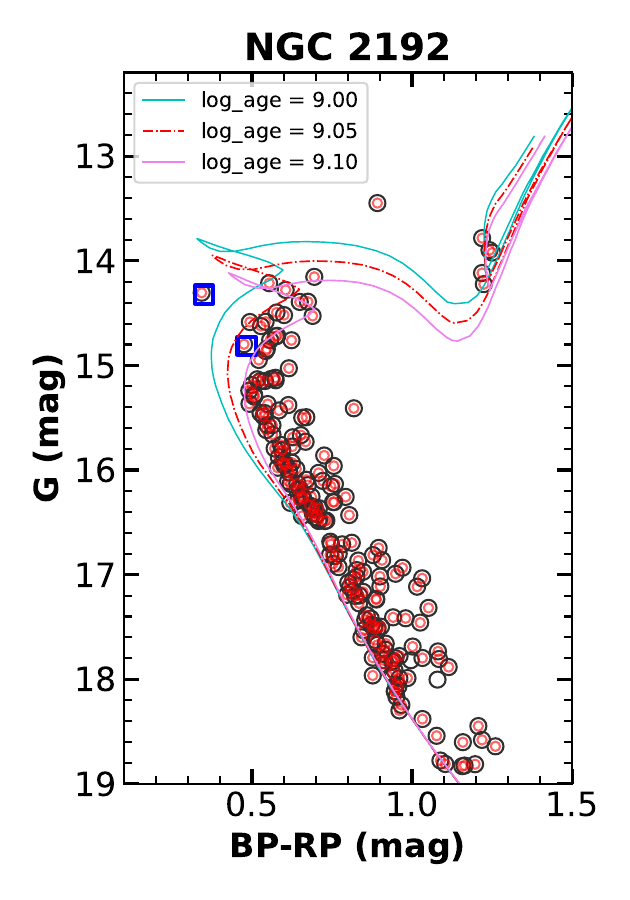}
    \includegraphics[width=4cm,height=7cm]{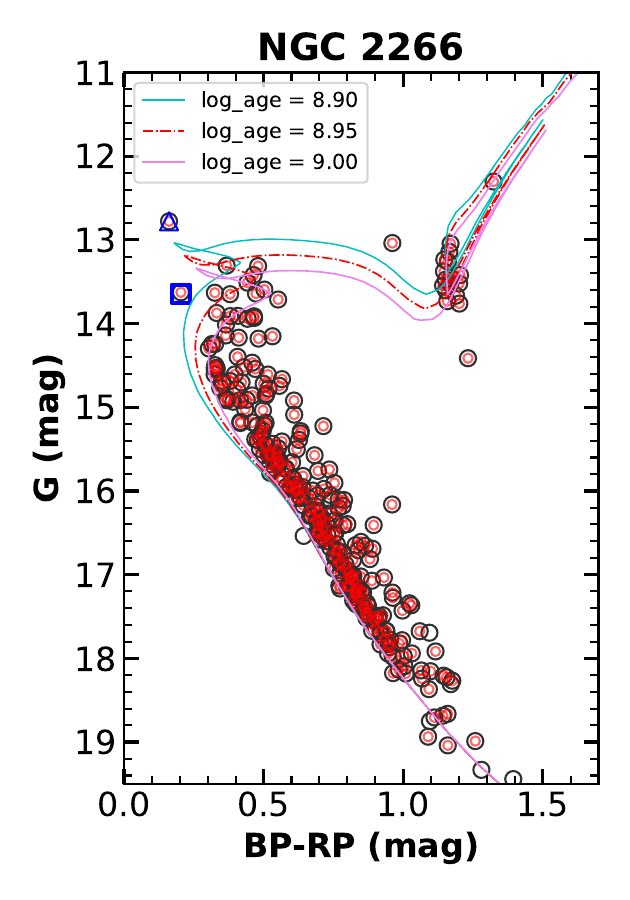}
    \includegraphics[width=4cm,height=7cm]{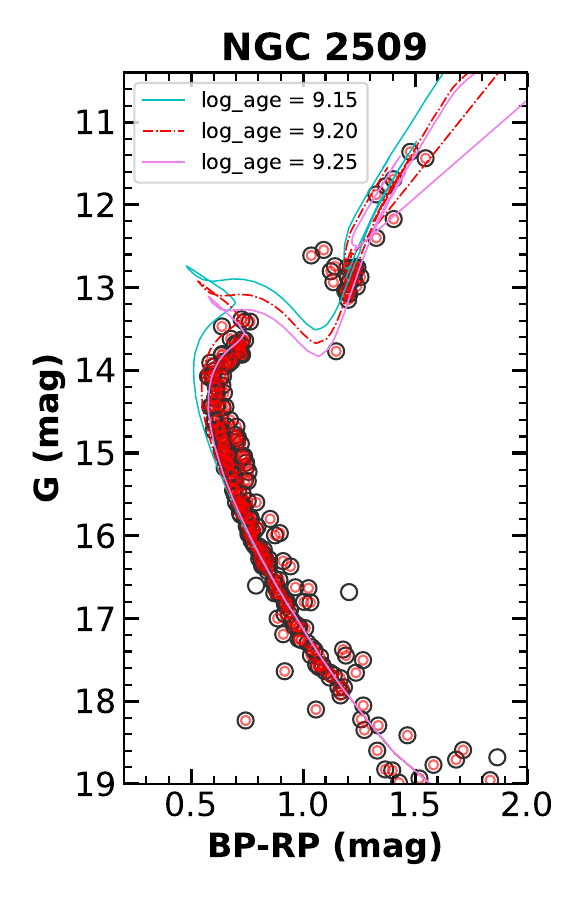}
    \includegraphics[width=4cm,height=7cm]{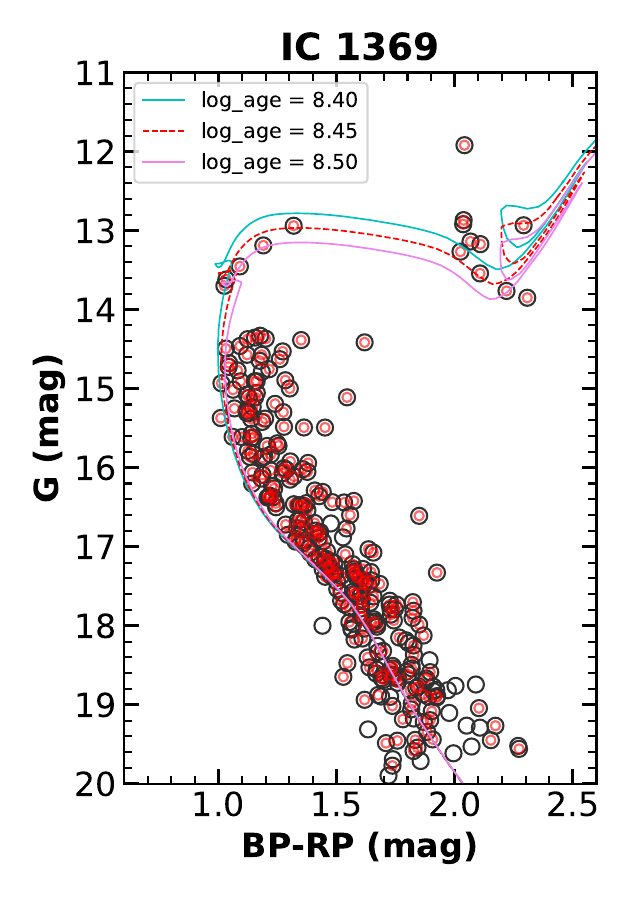}
    \caption{Color--magnitude diagrams (CMDs) of the OCs NGC~2192, NGC~2266, NGC~2509, and IC~1369, shown in the $G$ versus $(\mathrm{BP}-\mathrm{RP})$ plane. Grey points represent stars within the estimated cluster radius, while red filled circles denote the probable cluster members selected using astrometric and photometric criteria. The PARSEC isochrones plotted correspond to three ages for each cluster, with the central value providing the best fit and the younger and older values illustrating the age uncertainty. The adopted ages (in $\log[\mathrm{age/yr}]$) are 9.00, 9.05, and 9.10 for NGC~2192; 8.85, 8.95, and 9.05 for NGC~2266; 9.05, 9.15, and 9.25 for NGC~2509; and 8.30, 8.40, and 8.50 for IC~1369. These CMD-based fits provide the fundamental cluster parameters used throughout this work. In addition, the stars marked by blue boxes and triangles are found to be consistent with the BSSs identified by \citet{jadhav2021high, rain2021new}, with two BSSs located in NGC~2192 and two in NGC~2266.}
    \label{fig: cmd_all}
\end{figure*}

The colour–magnitude diagrams (CMDs) of NGC~2192, NGC~2266, NGC~2509, and IC~1369 are shown in Fig.~\ref{fig: cmd_all}. The CMDs were constructed using \textit{Gaia} DR3 photometry in the $G$ versus $(BP-RP)$ plane, including only high-probability cluster members. All four CMDs show well-defined main sequences, clear turn-off regions, and evolved populations along the red giant branch and red clump, consistent with intermediate- to old-age open clusters. Cluster parameters were estimated by fitting PARSEC stellar isochrones \citep{marigo2017new} to the observed CMDs. The fitting explored a grid of ages, distance moduli, reddening values, and metallicities, with the best solutions constrained mainly by the morphology of the main-sequence and the position of the turn-off region while remaining consistent with the evolved stellar populations.

Among the clusters, NGC~2509 appears the oldest, with a relatively faint and red turn-off, whereas NGC~2192 and NGC~2266 show CMD morphologies typical of intermediate-age systems. IC~1369 displays a broadened main-sequence and prominent red clump, likely reflecting its higher reddening. The CMDs also provide the evolutionary context for the detected variable stars, enabling identification of main-sequence pulsators (e.g., $\delta$~Scuti and $\gamma$~Doradus stars) and evolved variables associated with the red giant branch or red clump.

To estimate the extinction in the Gaia G band, we applied the empirical relation
$A_{G} = 1.86 \times E(G_{BP} -G_{RP})$, as given by \citet{riello2021gaia}, resulting in $A_{G}$ values between 0.20 and 2.10 mag. We use the transformation relation $A_{V} = A_{G} / 0.789$ \citep{casagrande2018use} to derive the visual extinction for each cluster. Using this relation, we obtain $A_{V} = 0.40$ for NGC~2192, $0.30$ for NGC~2266, $0.23$ for NGC~2509, and $2.68$ for IC~1369. The relatively high extinction toward IC 1369 is consistent with its low Galactic latitude and line-of-sight dust concentration, as also noted in previous Gaia-based studies. All the estimated values are in good agreement with those reported by \citet{hunt2023improving, Straizys2020}.

 The cluster distances derived using the probabilistic parallax method of \citet{bailer2018estimating} were adopted as the primary distance estimates. These distances were used to compute the corresponding distance moduli applied in the CMD isochrone fitting. Consequently, the CMD analysis was performed using Gaia-based distance constraints to ensure consistency between the astrometric and photometric determinations of the cluster parameters. The derived distance moduli are $13.30 \pm 0.20$ for NGC 2192, $12.98 \pm 0.21$ for NGC 2266, $12.16 \pm 0.17$ for NGC 2509, and $14.75 \pm 0.30$ for IC 1369. The corresponding colour excess values, $E(BP - RP)$, are $0.20 \pm 0.10$, $0.15 \pm 0.02$, $0.10 \pm 0.05$, and $1.10 \pm 0.20$, respectively. To quantitatively assess the reliability of the isochrone fitting, we employed a reduced chi-square ($\chi^2_{\mathrm{r}}$) analysis following the methodology described by \cite{valle2021goodness}. The reduced chi-square statistic assesses the goodness of fit by evaluating how well the model isochrones reproduce the observed CMDs, accounting for observational uncertainties. The best-fitting distance modulus and colour excess for each cluster were determined by minimising the reduced $\chi^2_{\mathrm{r}}$ value. The minimum reduced chi-square values obtained are 0.95 for NGC~2192, 0.90 for NGC~2266, 0.90 for NGC~2509, and 0.80 for IC~1369, indicating that the adopted isochrone solutions provide a statistically robust and reliable representation of the observed CMDs. The derived cluster ages are $1.12 \pm 0.129$~Gyr for NGC~2192, $0.891 \pm 0.103$~Gyr for NGC~2266, $1.59 \pm 0.183$~Gyr for NGC~2509, and $0.282 \pm 0.033$~Gyr for IC~1369. The corresponding metallicities are $Z = 0.008 \pm 0.002$ for NGC~2192 \cite{tapia2009ubv}, $0.010 \pm 0.004$ for NGC~2266 \cite{angelo2025exploring}, $0.0152 \pm 0.003$ for NGC~2509 \cite{ahmed2025photometry}, and $0.0117 \pm 0.001$ for IC~1369. For NGC 2192, NGC 2266, and NGC 2509, metallicity values are adopted from recent literature, whereas for IC 1369, the metallicity is taken from APOGEE DR16 spectroscopy of confirmed cluster members. Our calculated ages for all clusters are consistent with the estimated values reported by \citet{hunt2023improving, dias2021updated, cantat2020painting}. 

In this work, we identify four Blue Straggler Stars (BSSs), two of which are located in NGC~2192 and two in NGC~2266. For NGC~2192, we cross-match our BSS candidates against the catalogues of \citet{rain2021new} and \citet{jadhav2021high} and find that both BSSs are present in these studies. In the case of NGC~2266, one BSS is matched with the catalogue of \citet{rain2021new}, while the other is matched with \citet{jadhav2021high}. As shown in Fig.~\ref{fig: cmd_all}, the matched BSSs are represented by blue triangles and blue squares, respectively. This indicates that the two catalogues identify different BSSs in NGC~2266.

 The cluster parameters derived above are primarily used to interpret the subset of variable stars that are probable cluster members. Membership probabilities allow us to distinguish cluster variables from field stars, ensuring that physical interpretation is applied only to bona fide members. The derived distances and reddening values are used to estimate reliable luminosities through spectral energy distribution (SED) fitting, while cluster ages provide the evolutionary context for classifying pulsators and binary systems. These parameters, therefore, provide the necessary physical framework for interpreting the detected variable stars.

\section{Identification and Characterization of Variable Stars}
\label{sec: variable}

\begin{figure*}
    
    \centering
    \includegraphics[width=4.9cm,height=3.0cm]{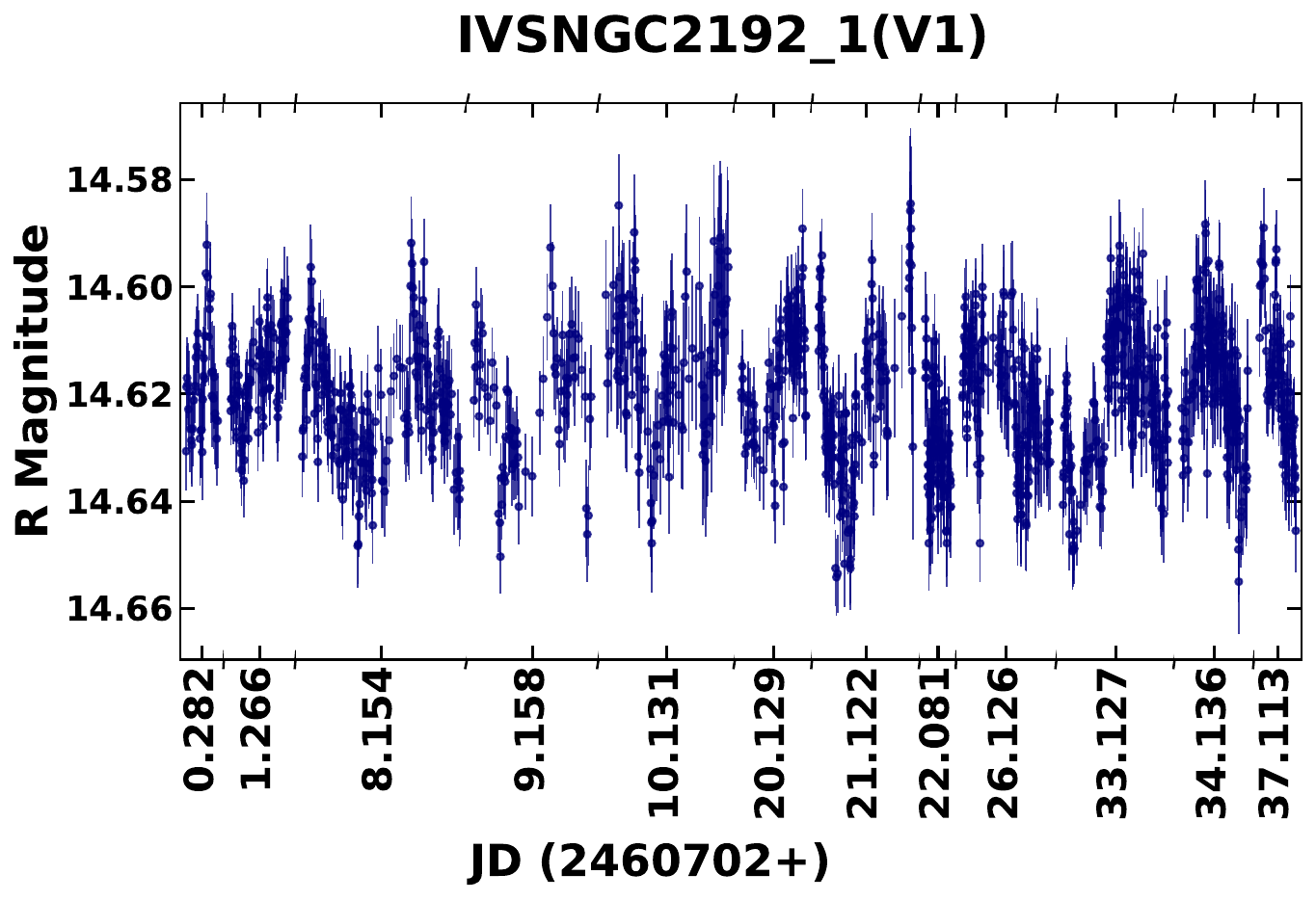}
    \includegraphics[width=4.9cm,height=2.8cm]{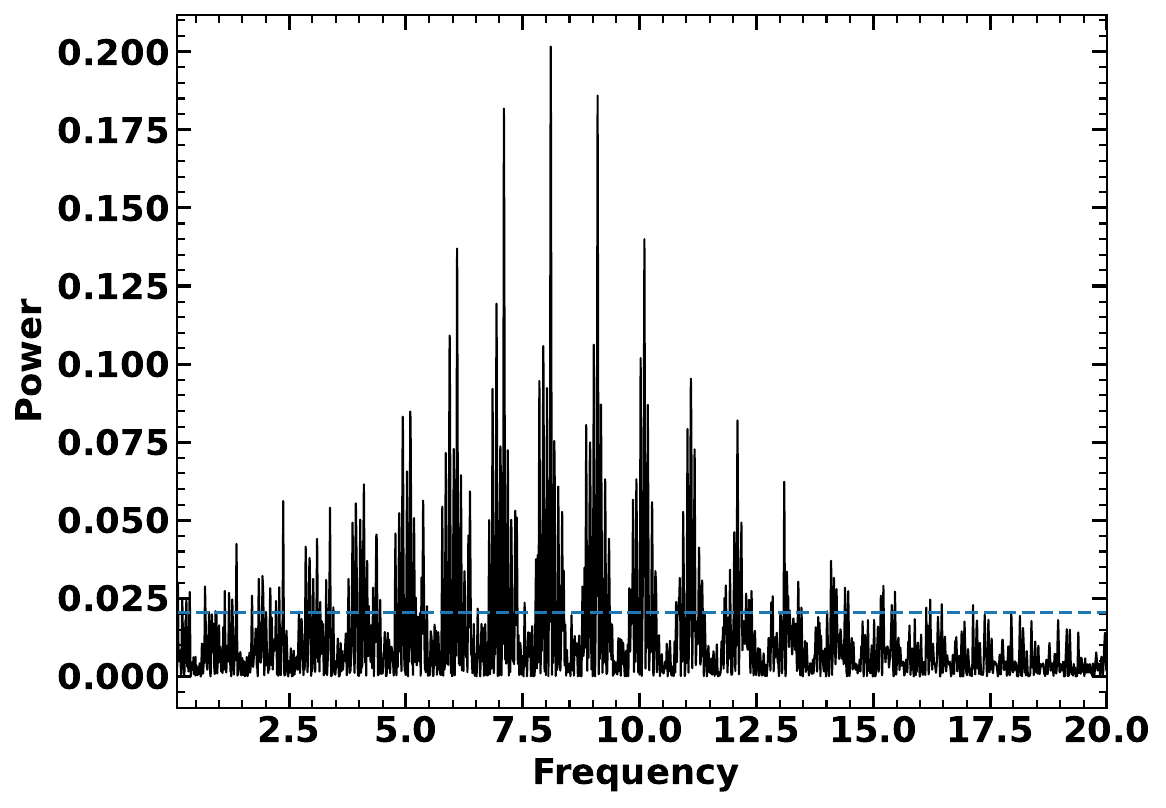}
    \includegraphics[width=4.9cm,height=3.0cm]{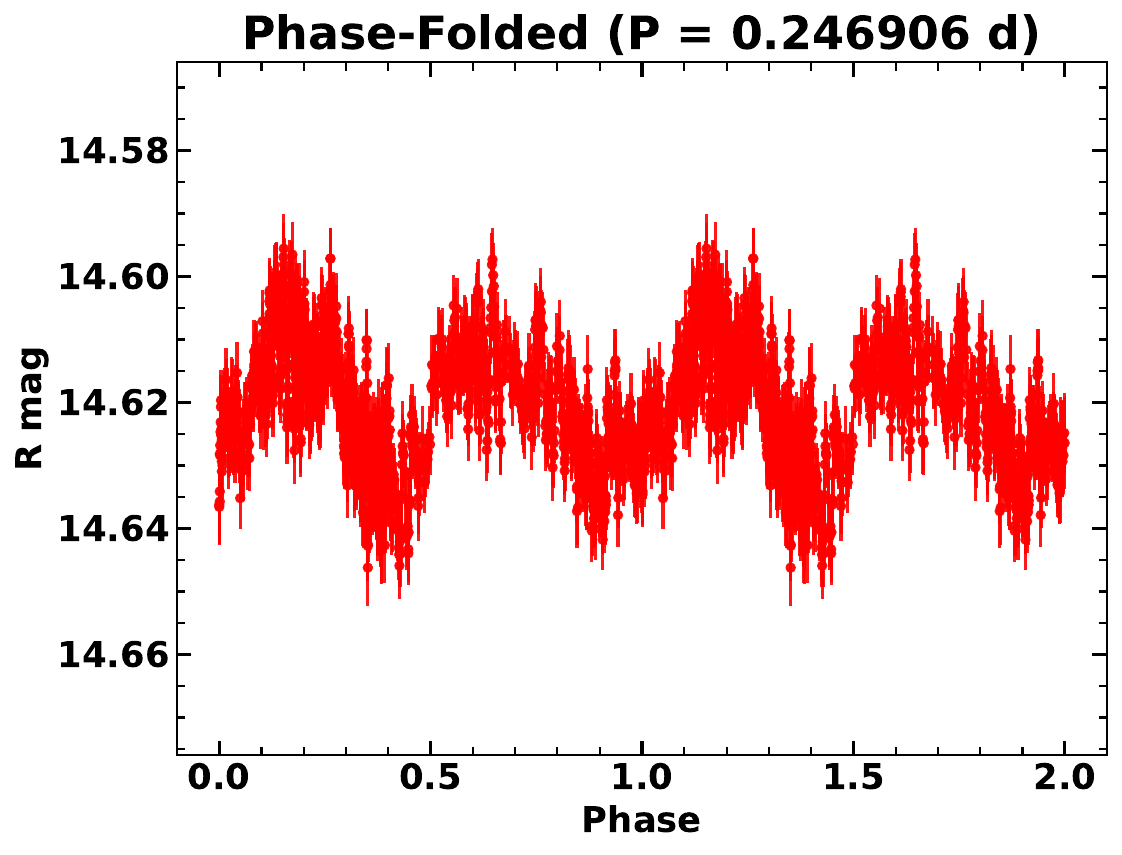}
    
    \includegraphics[width=4.9cm,height=3.0cm]{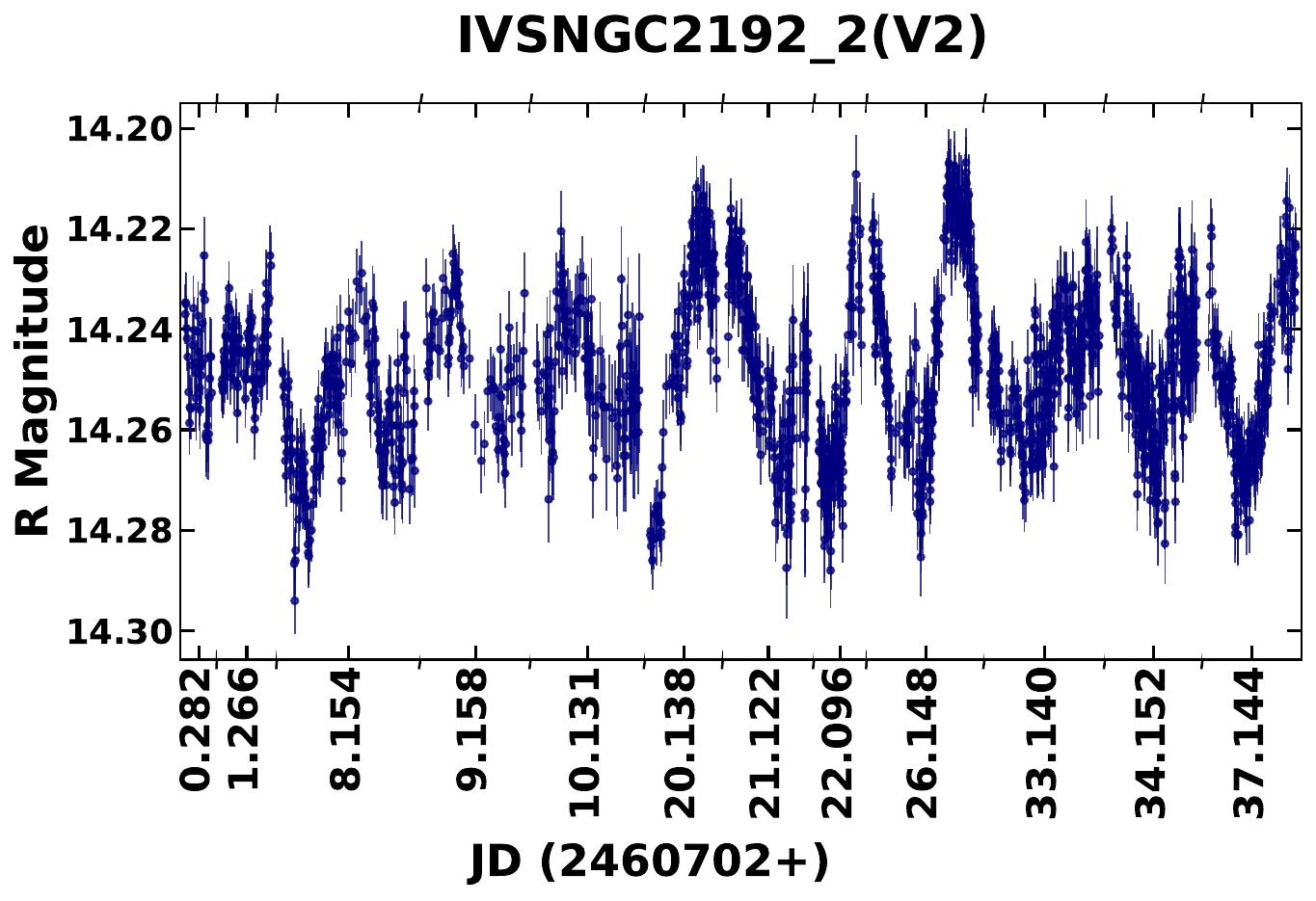}
    \includegraphics[width=4.9cm,height=2.8cm]{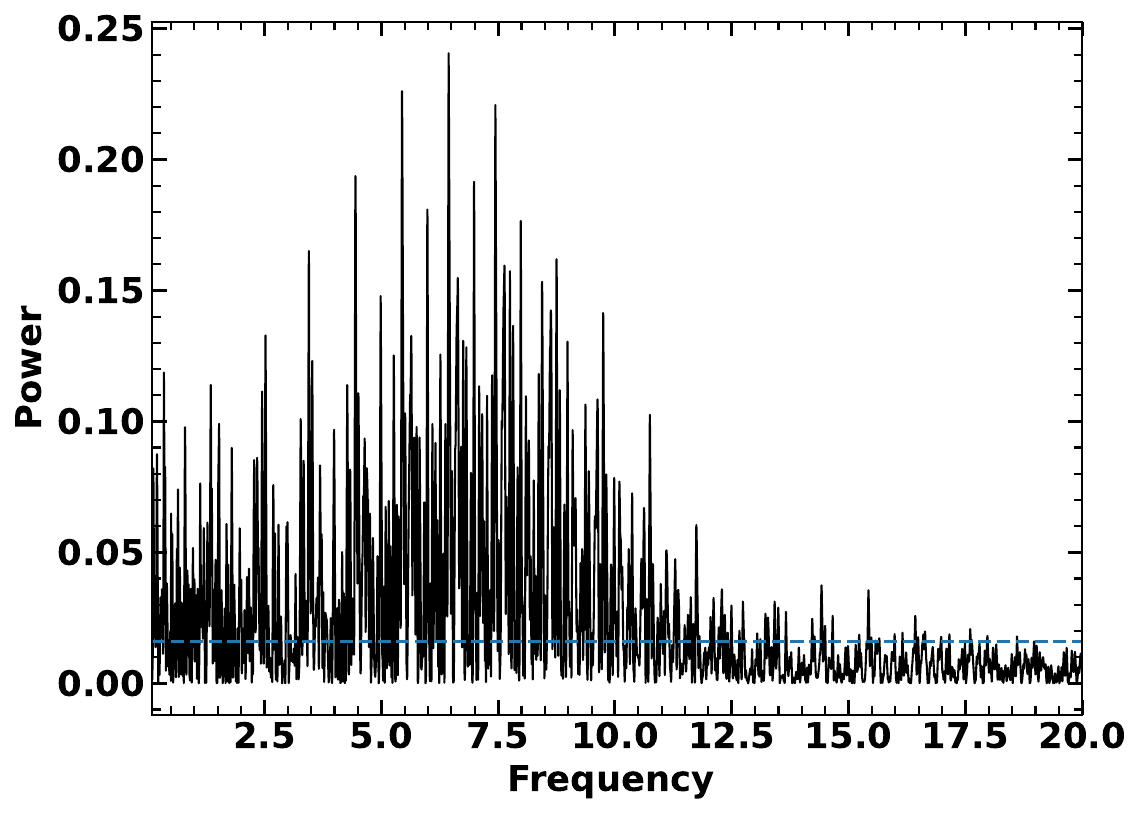}
    \includegraphics[width=4.9cm,height=3.0cm]{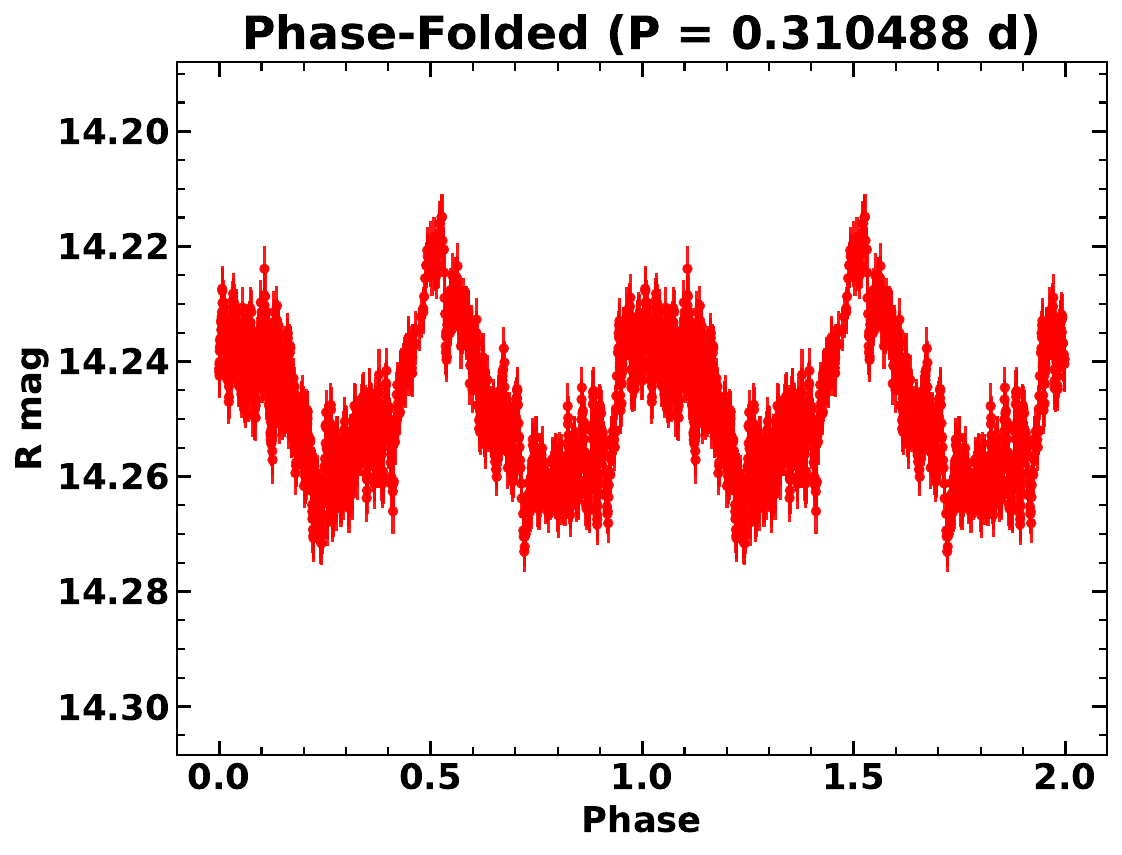}

    \includegraphics[width=4.9cm,height=3.0cm]{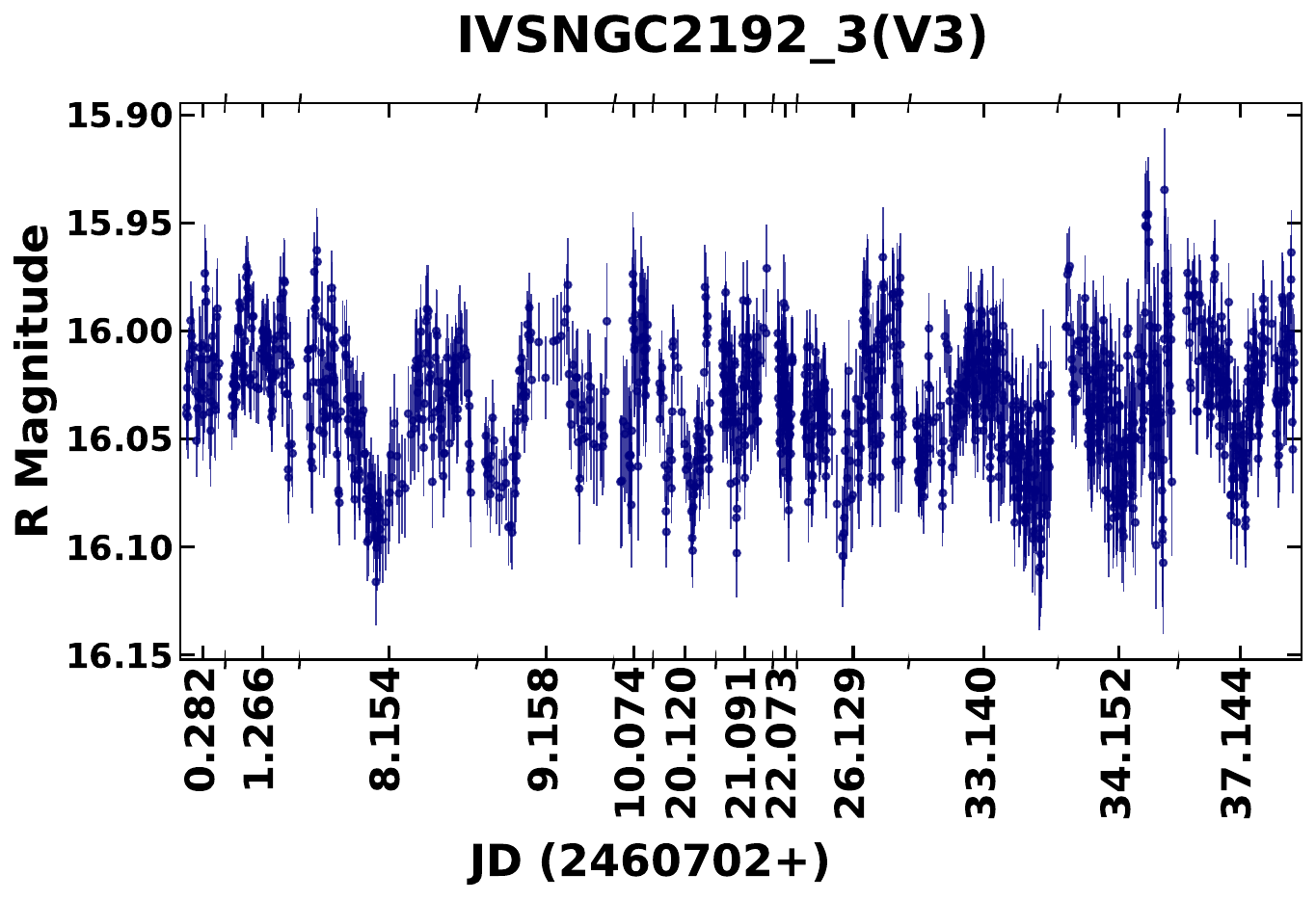}
    \includegraphics[width=4.9cm,height=2.8cm]{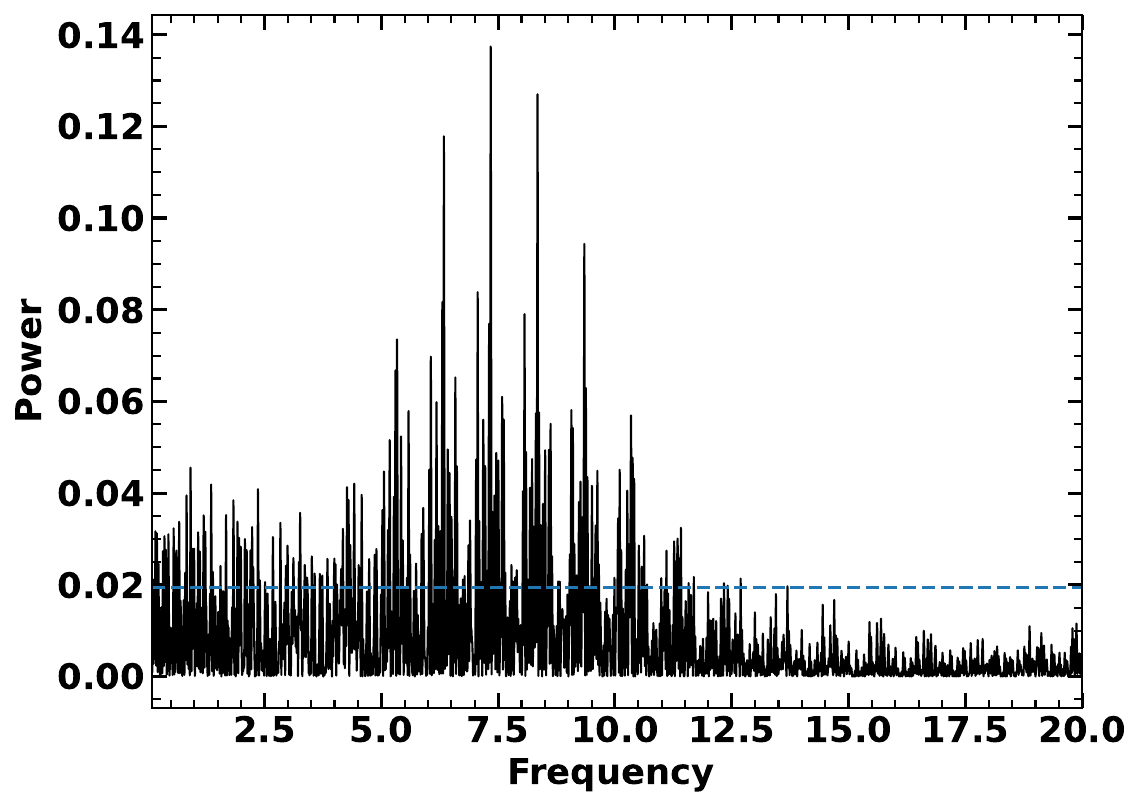}
    \includegraphics[width=4.9cm,height=3.0cm]{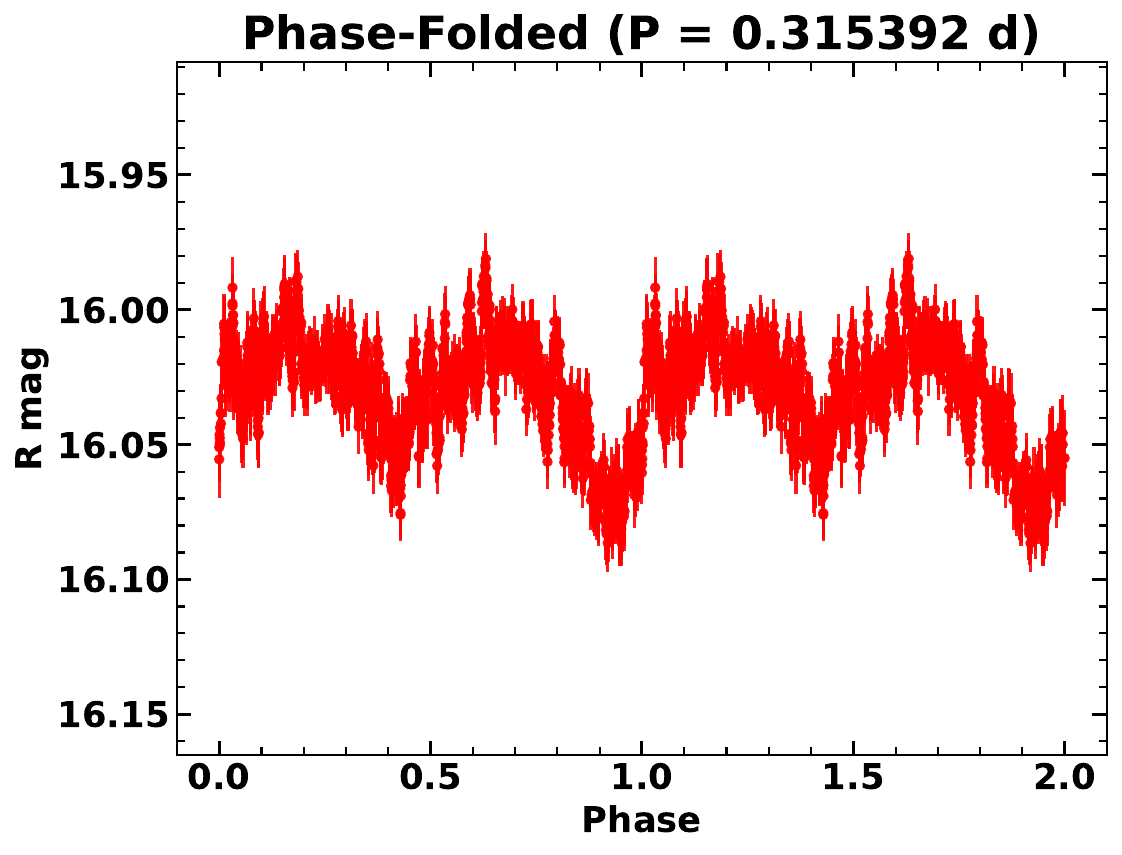}
    
    \includegraphics[width=4.9cm,height=3.0cm]{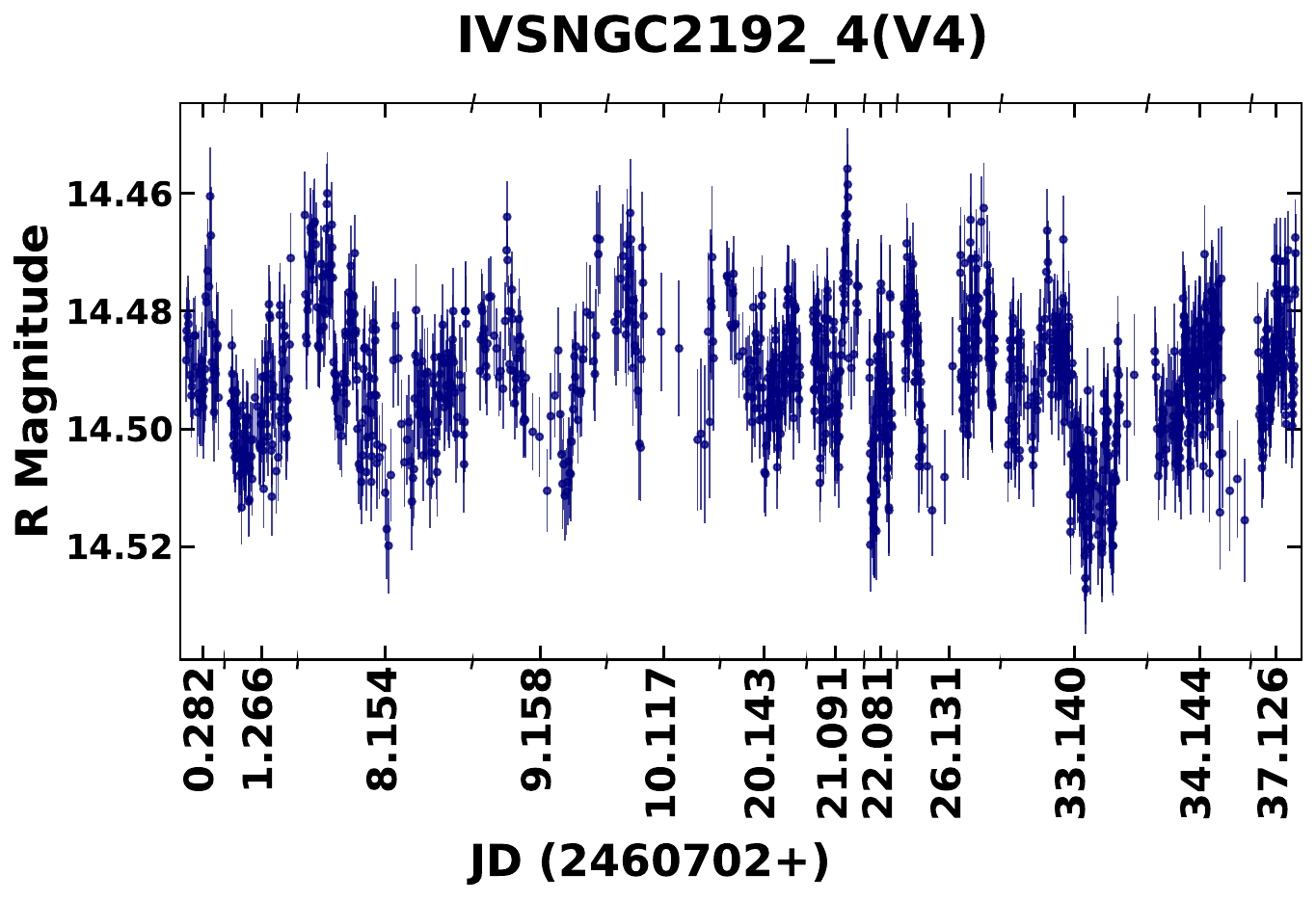}
    \includegraphics[width=4.9cm,height=2.8cm]{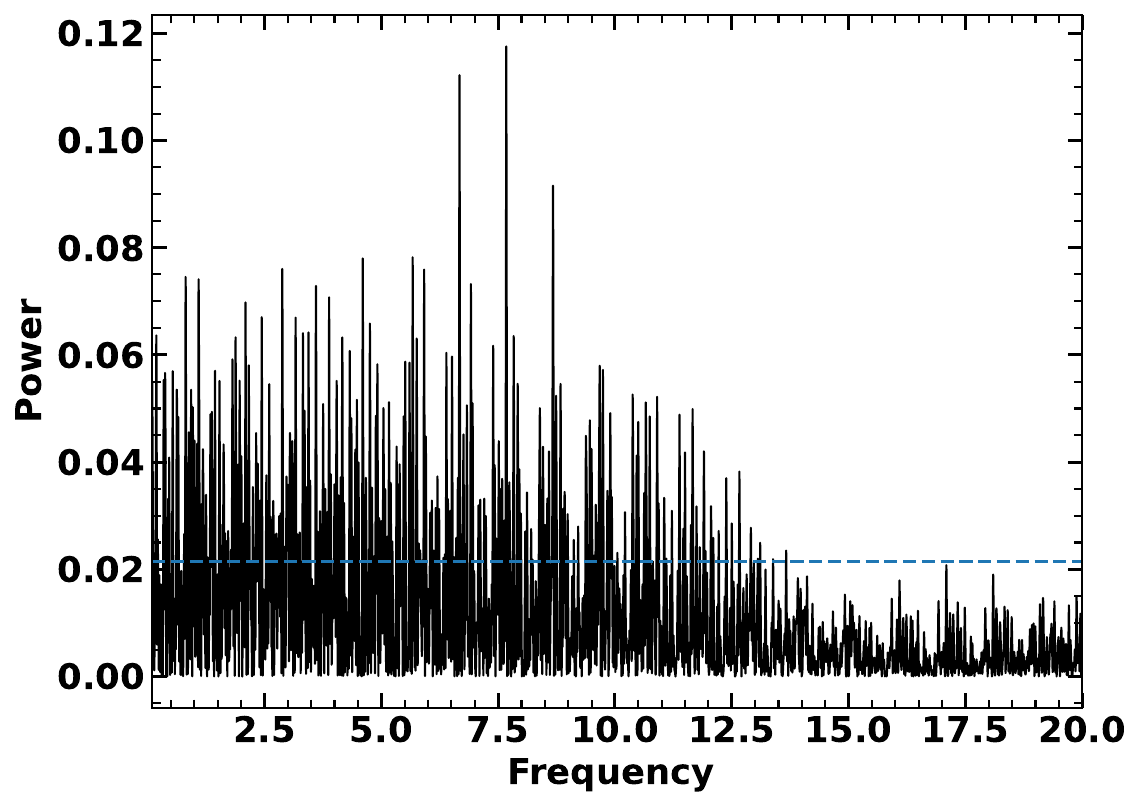}
    \includegraphics[width=4.9cm,height=3.0cm]{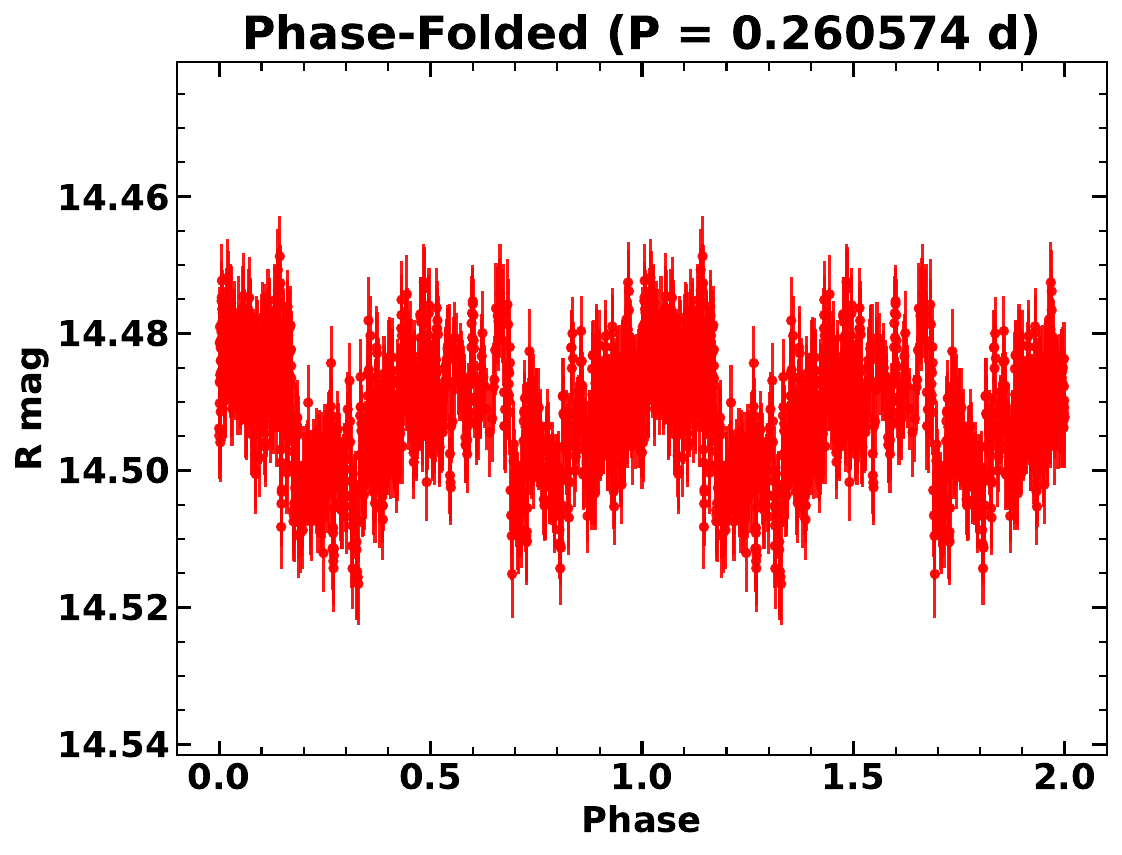}

    \includegraphics[width=4.9cm,height=3.0cm]{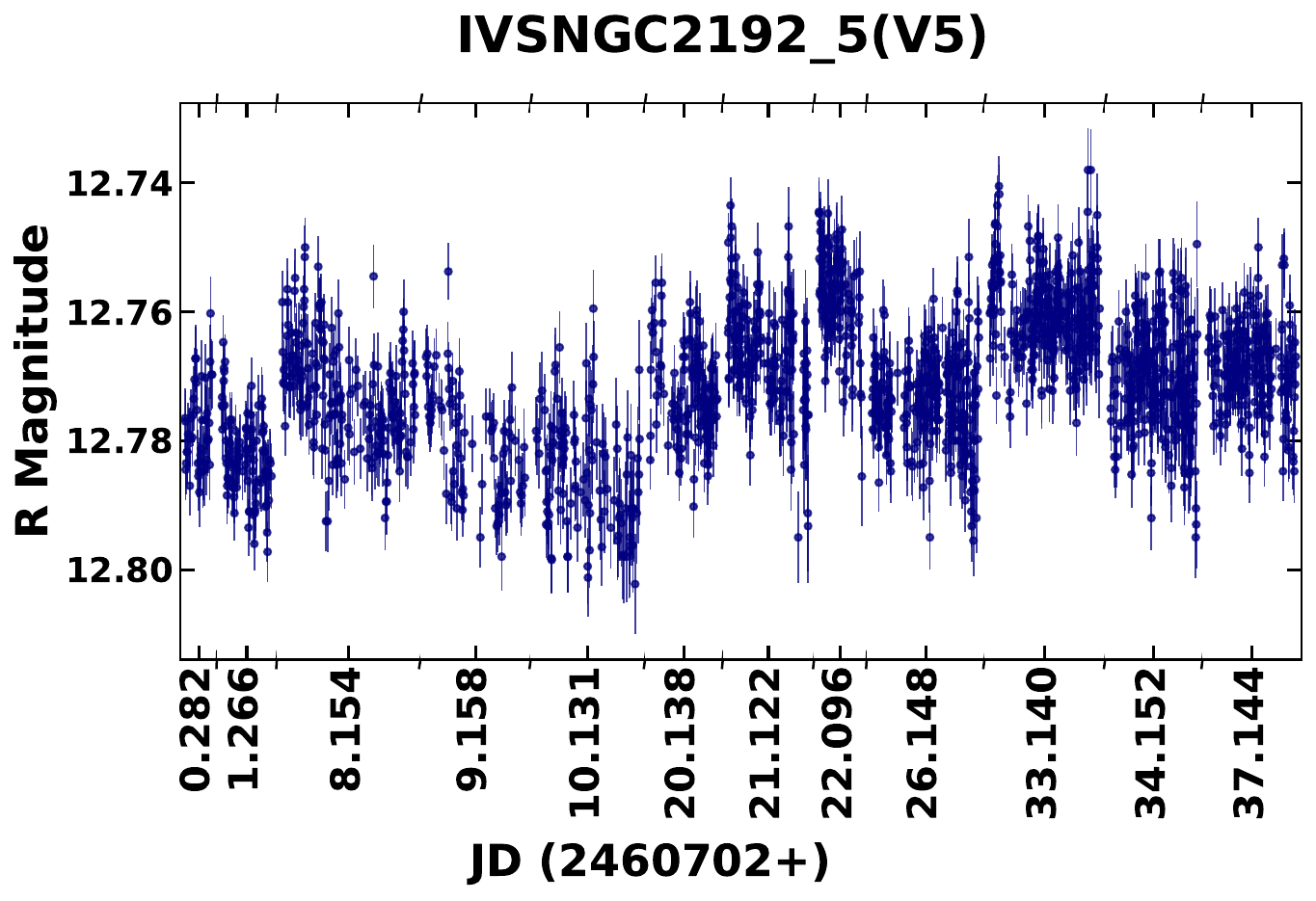}
    \includegraphics[width=4.9cm,height=2.8cm]{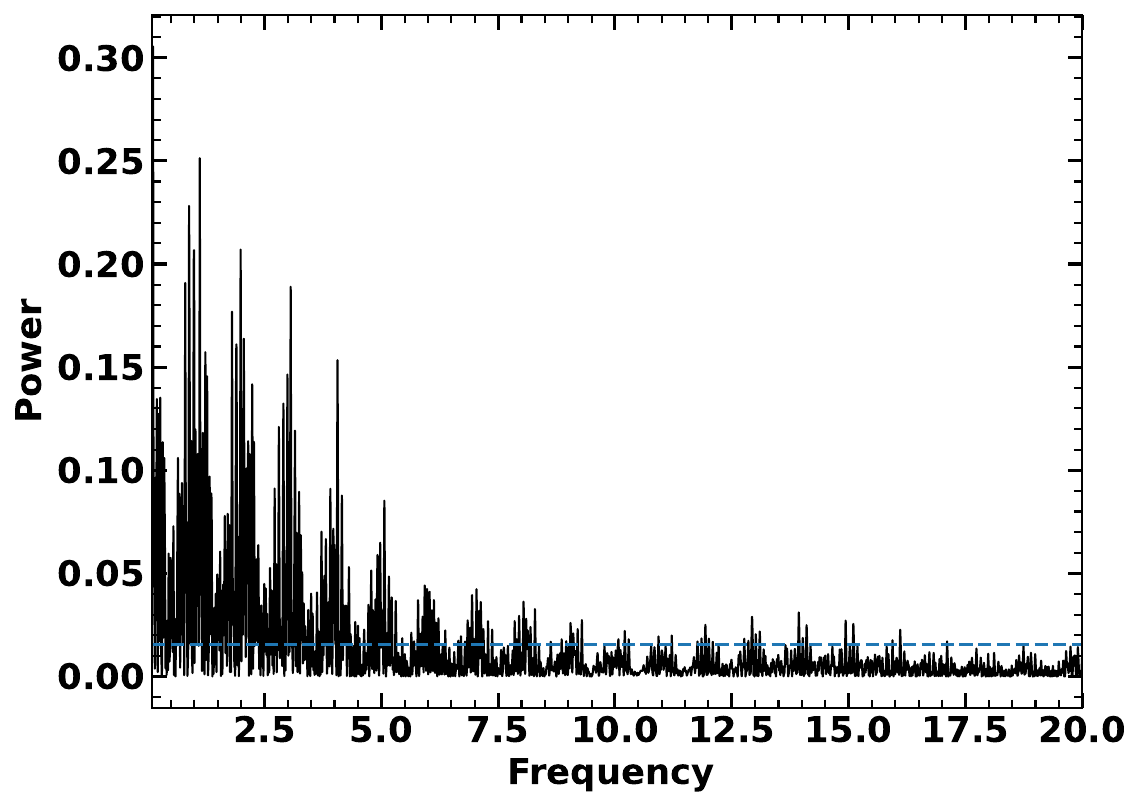}
    \includegraphics[width=4.9cm,height=3.0cm]{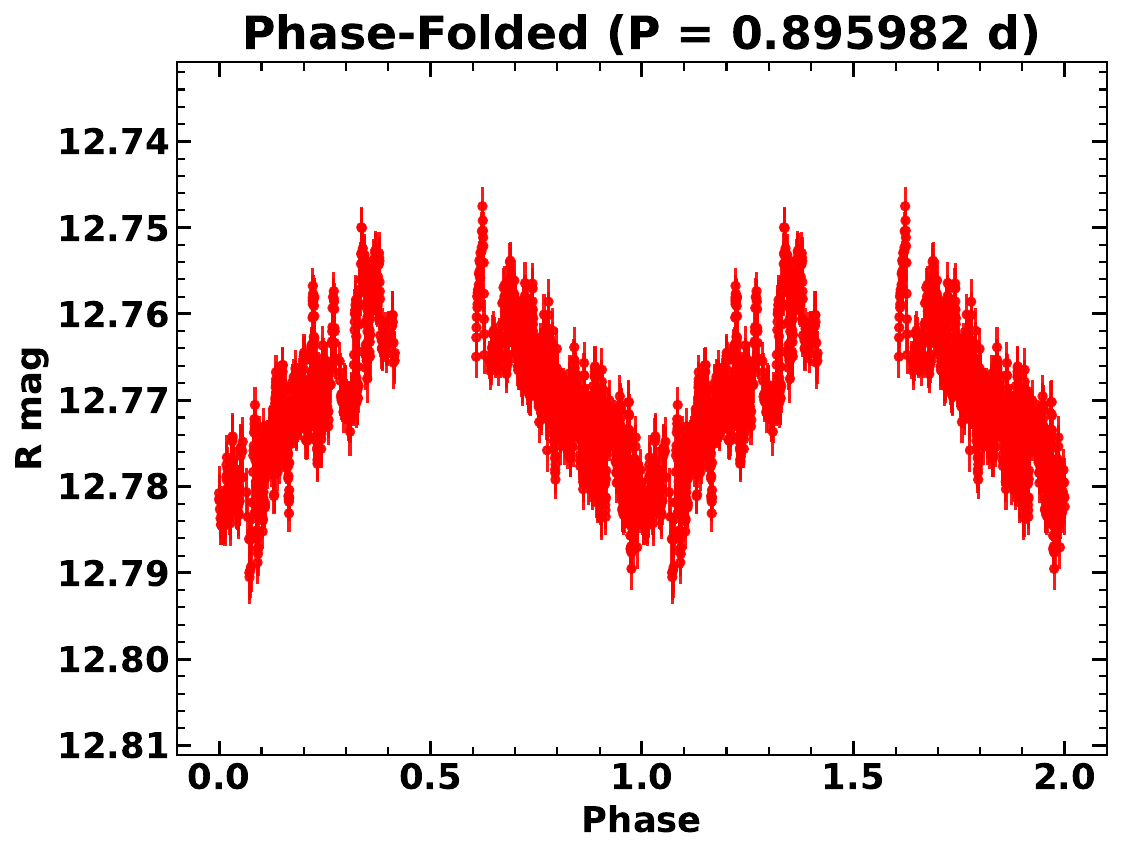}

    \includegraphics[width=4.9cm,height=3.0cm]{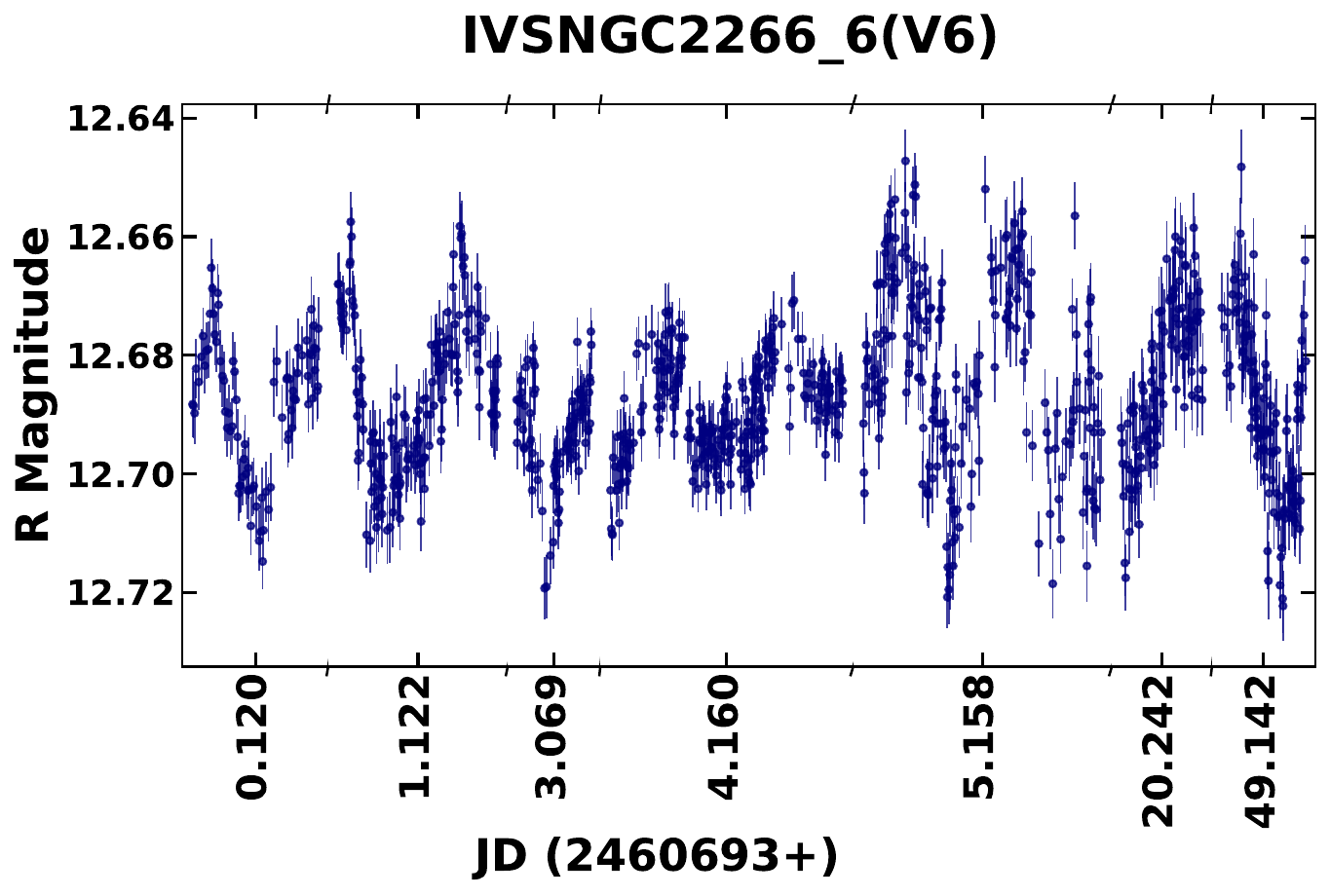}
    \includegraphics[width=4.9cm,height=2.8cm]{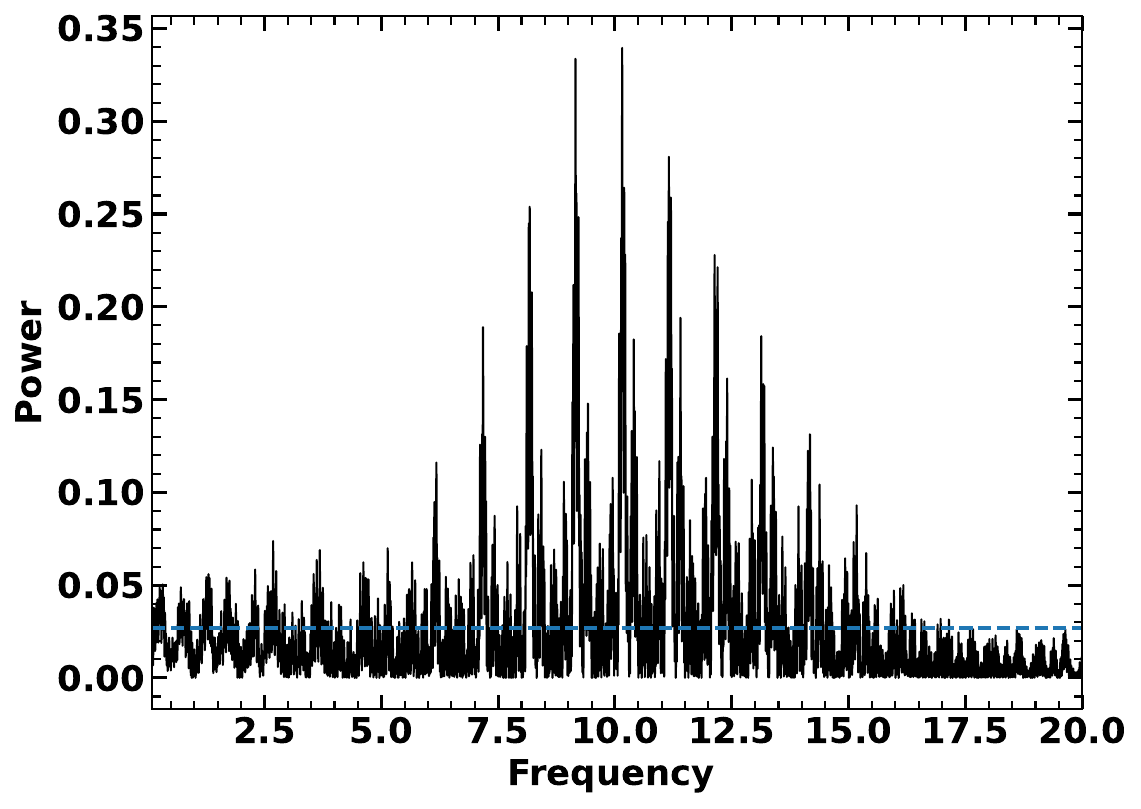}
    \includegraphics[width=4.9cm,height=3.0cm]{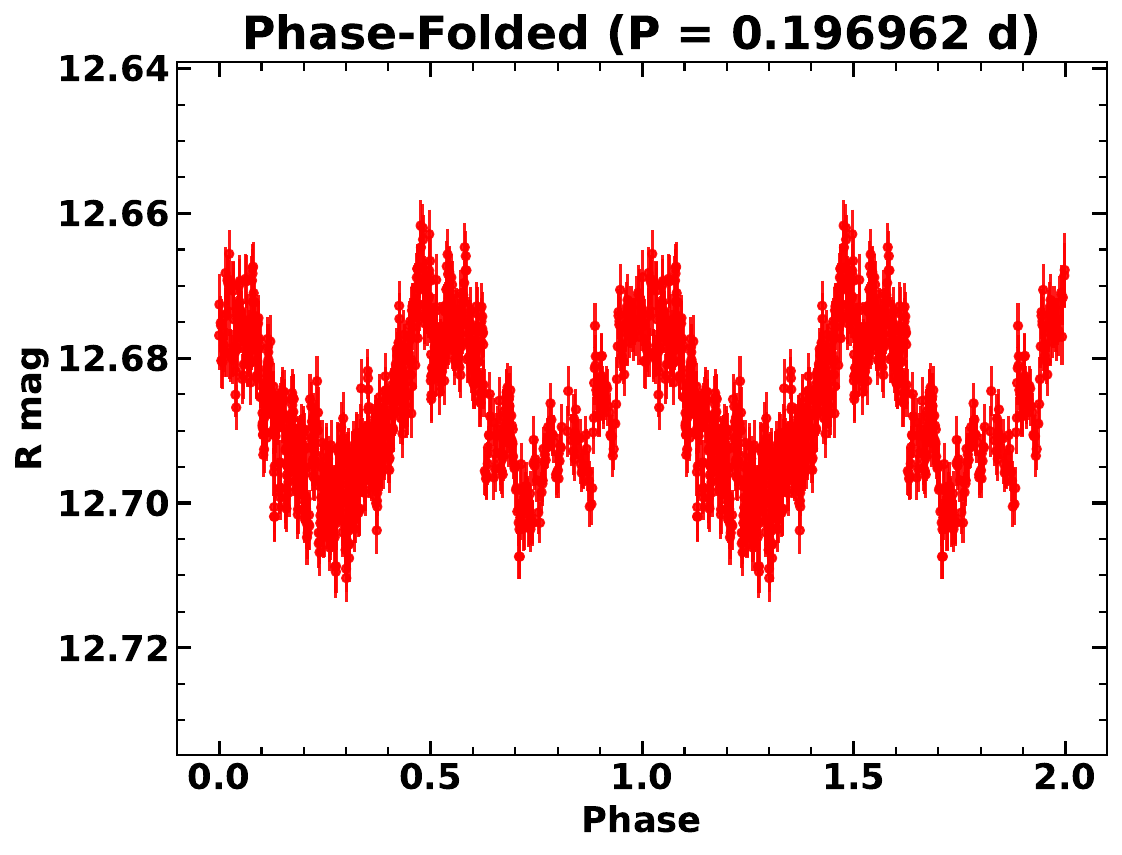}

    \caption{Time-series photometric analysis of the newly identified variable stars V1--V6 in the observed OCs.
\textit{Left panels:} original $R$-band light curves plotted as magnitude versus time (shown as JD$-$JD$_0$ for clarity).
\textit{Middle panels:} Lomb--Scargle periodograms computed from the calibrated light curves, with the dominant frequency($d^{-1}$)
peaks corresponding to the adopted variability periods. Periods and amplitudes derived from these light curves are listed in Table \ref{tab: sed_parameters}. The dashed horizontal line in the periodogram denotes the 1$\%$ false alarm probability (FAP) level computed analytically, indicating a highly significant periodic signal.\textit{Right panels:} phase-folded $R$-band light curves using the best-fit periods indicated in each panel.
The phase-folded curves are shown over two consecutive cycles to emphasise the periodic nature and stability of the variability.
The coherent light-curve morphologies and statistically significant periodogram peaks support the classification of these sources
as pulsating and rotational variables. In all panels, magnitudes are plotted such that brighter values appear toward the top.}
    \label{fig: phase_fold}
\end{figure*}

\begin{figure*}
    \centering
    \includegraphics[width=4.9cm,height=3.0cm]{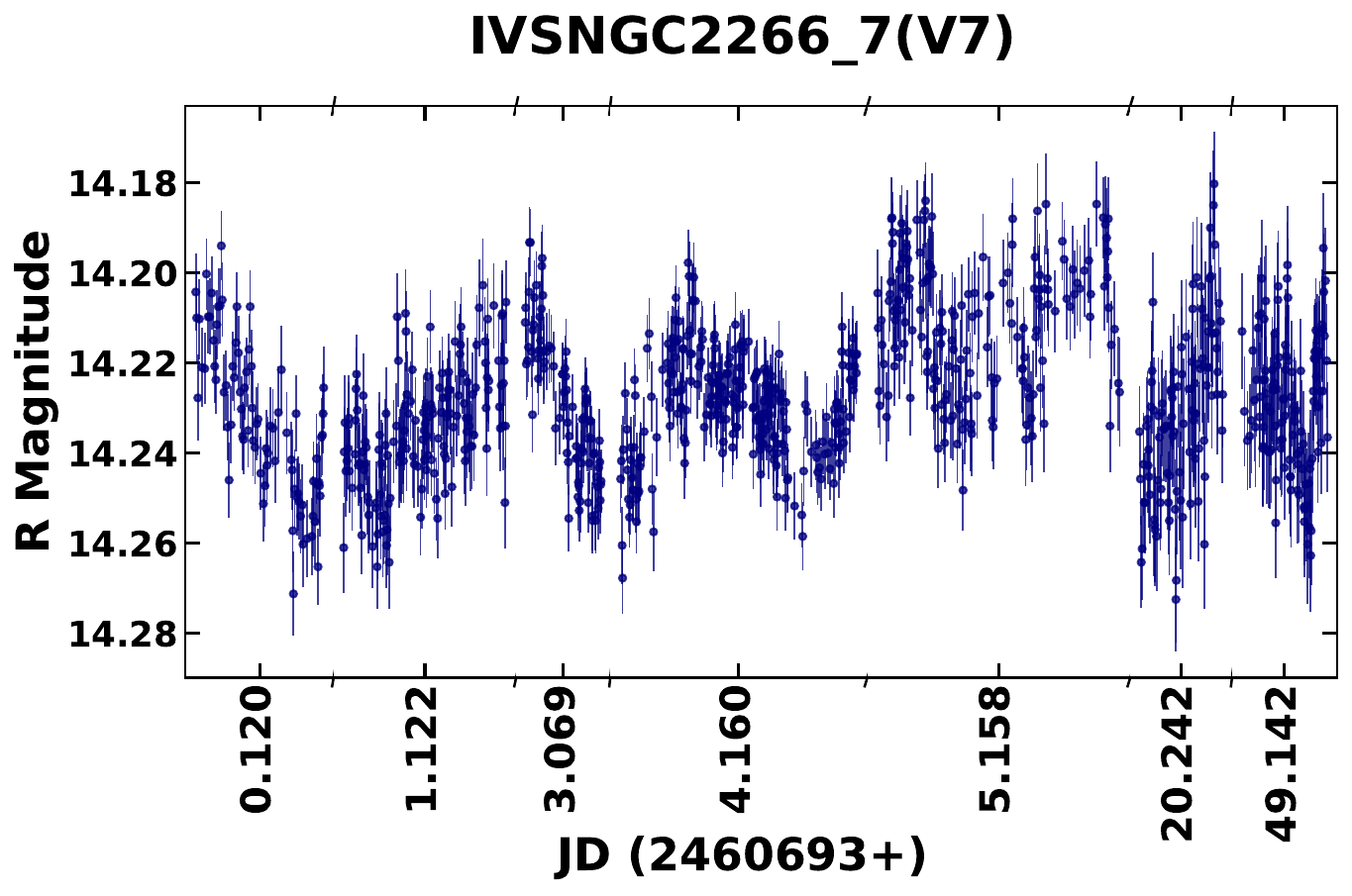}
    \includegraphics[width=4.9cm,height=2.8cm]{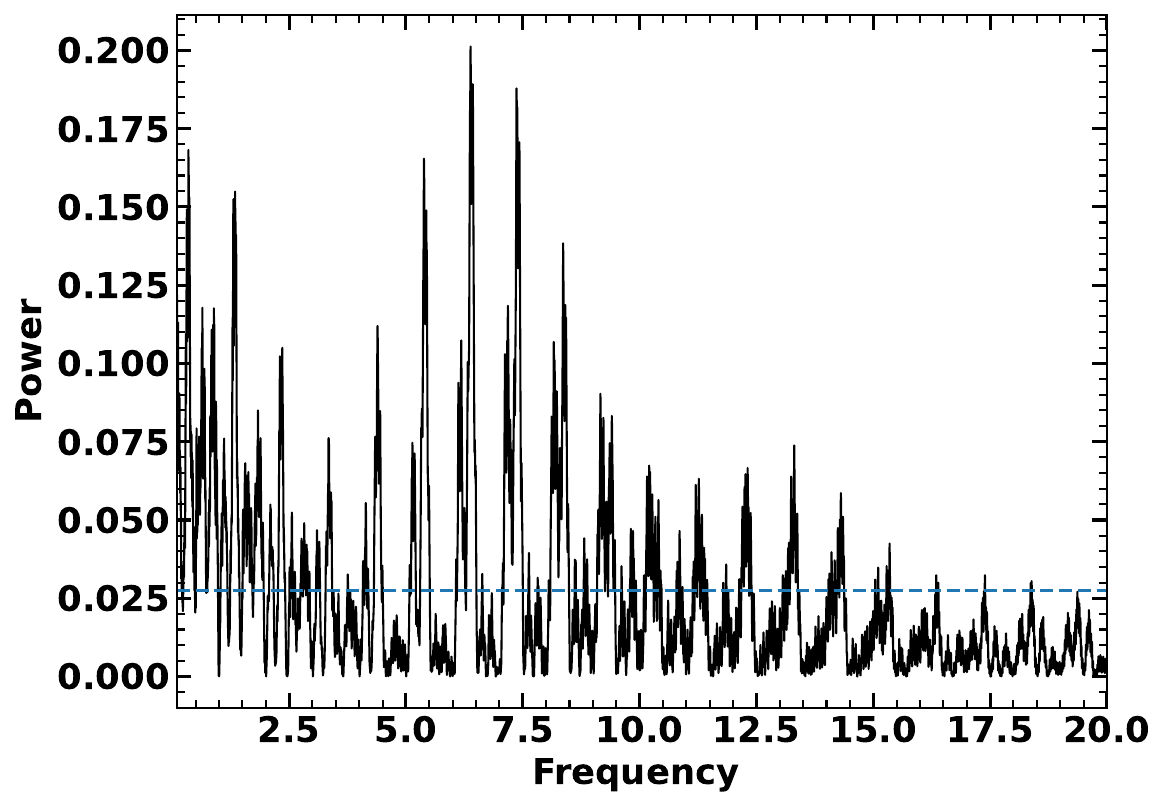}
    \includegraphics[width=4.9cm,height=3.0cm]{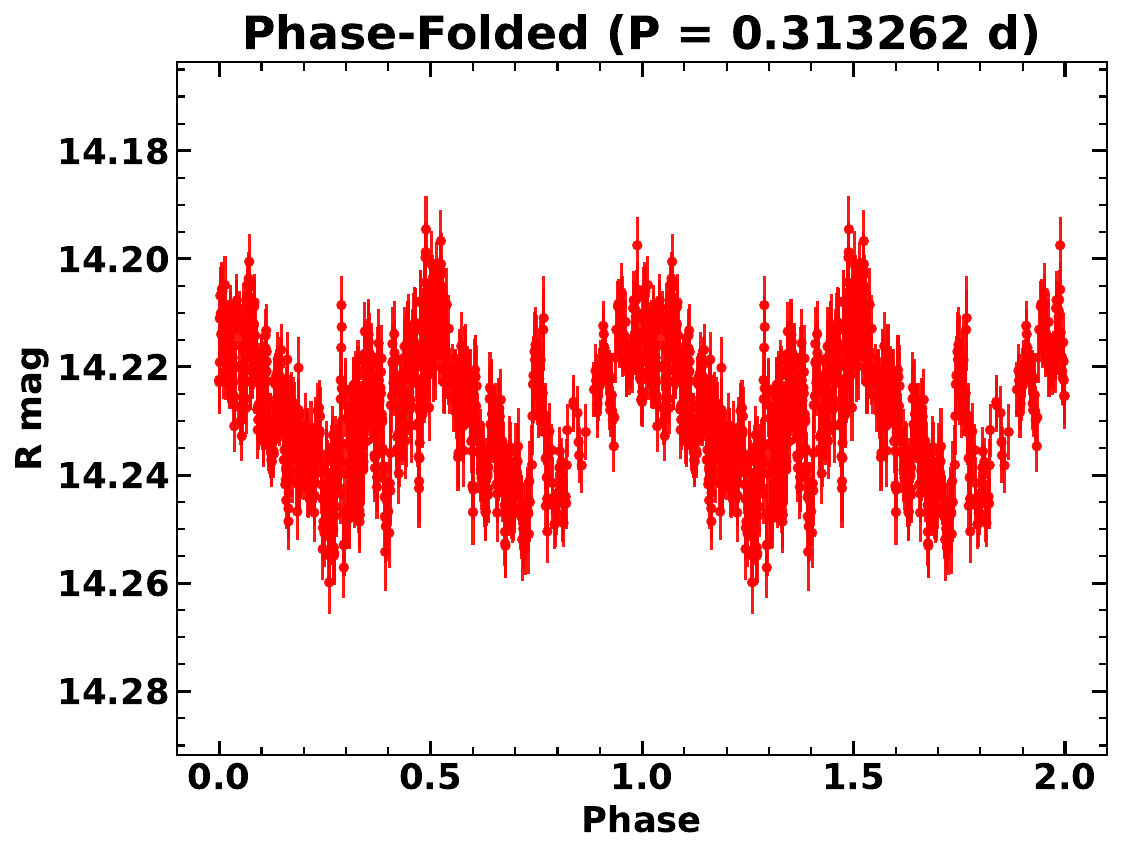}
    
    \includegraphics[width=4.9cm,height=3.0cm]{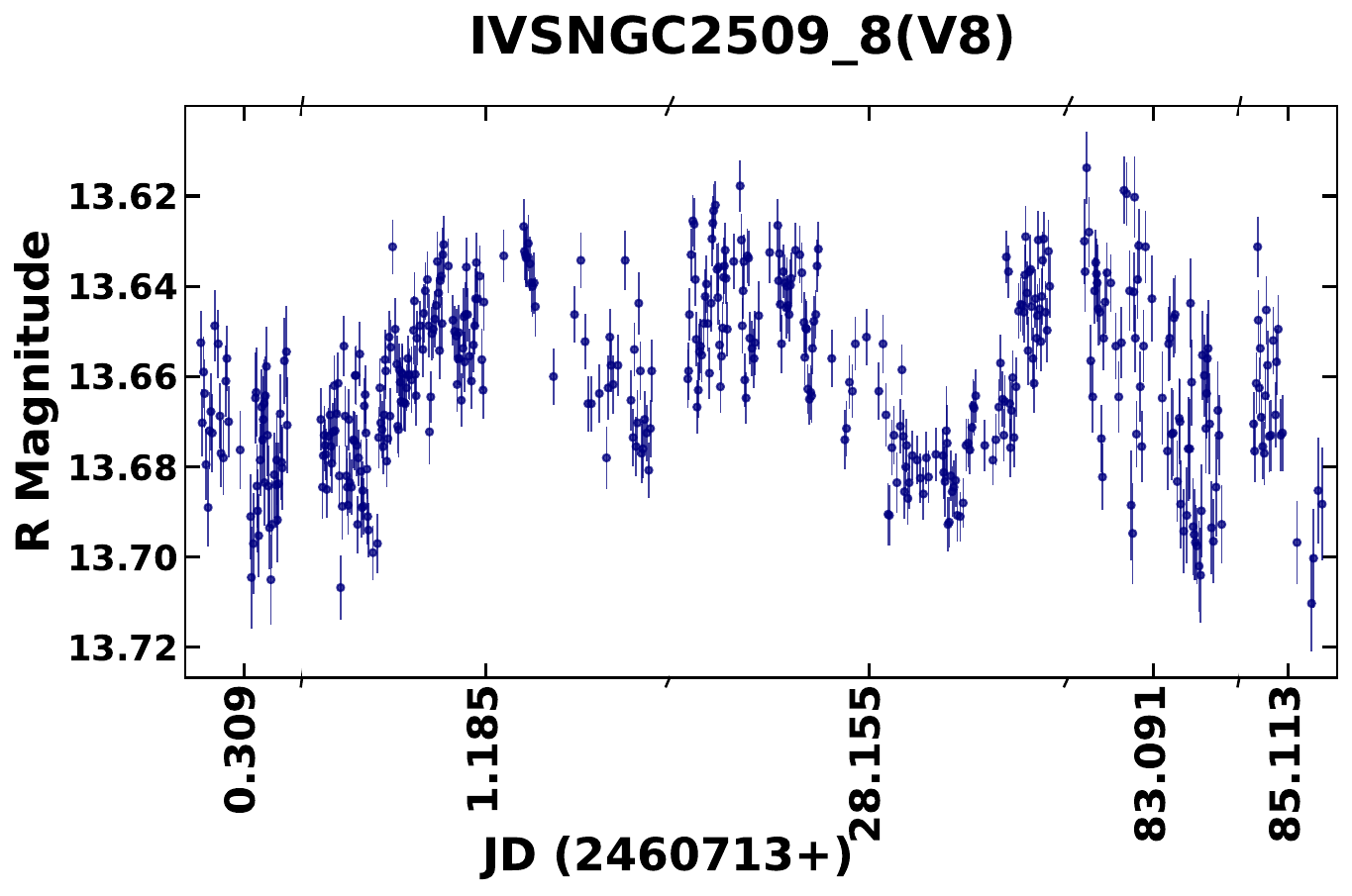}
    \includegraphics[width=4.9cm,height=2.8cm]{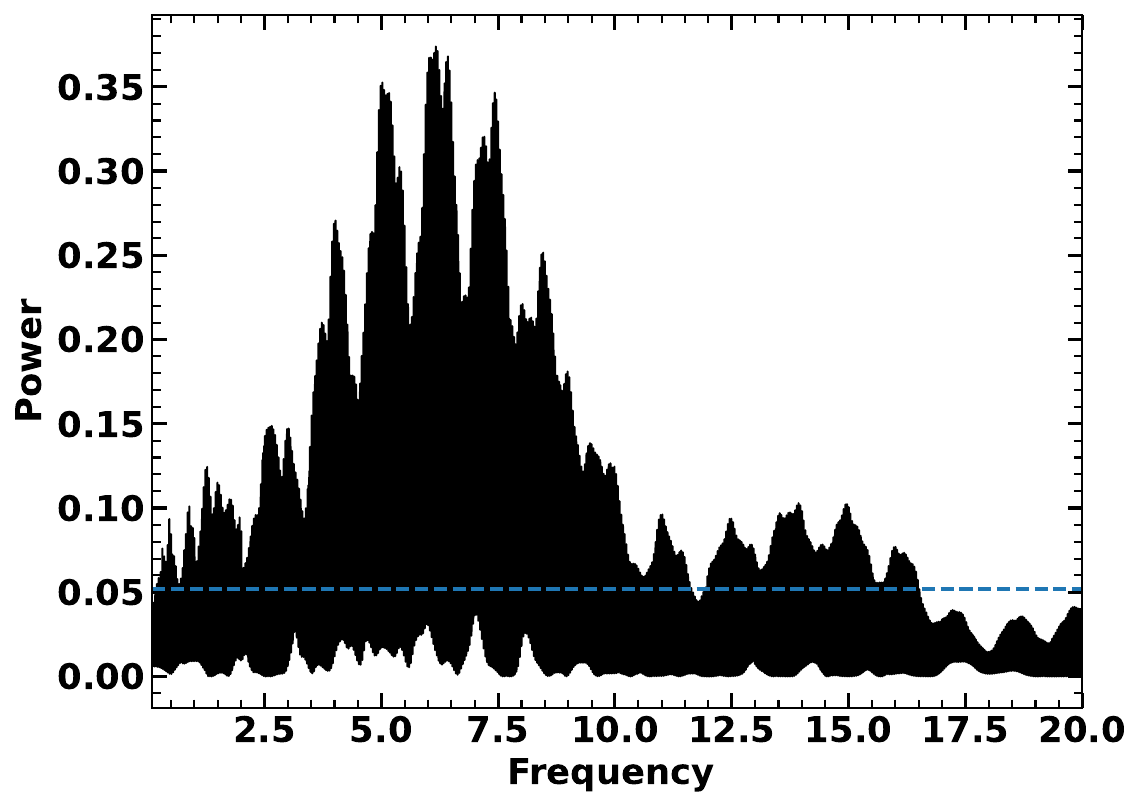}
    \includegraphics[width=4.9cm,height=3.0cm]{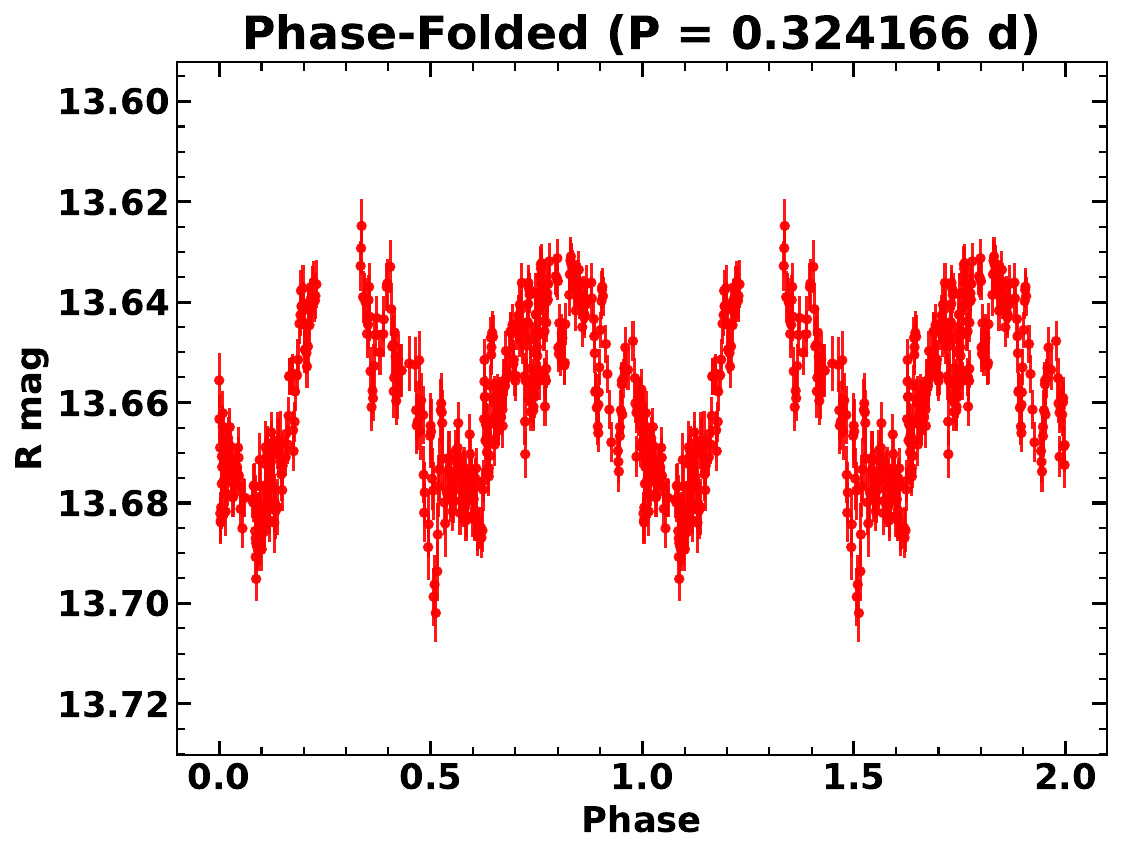}

    \includegraphics[width=4.9cm,height=3.0cm]{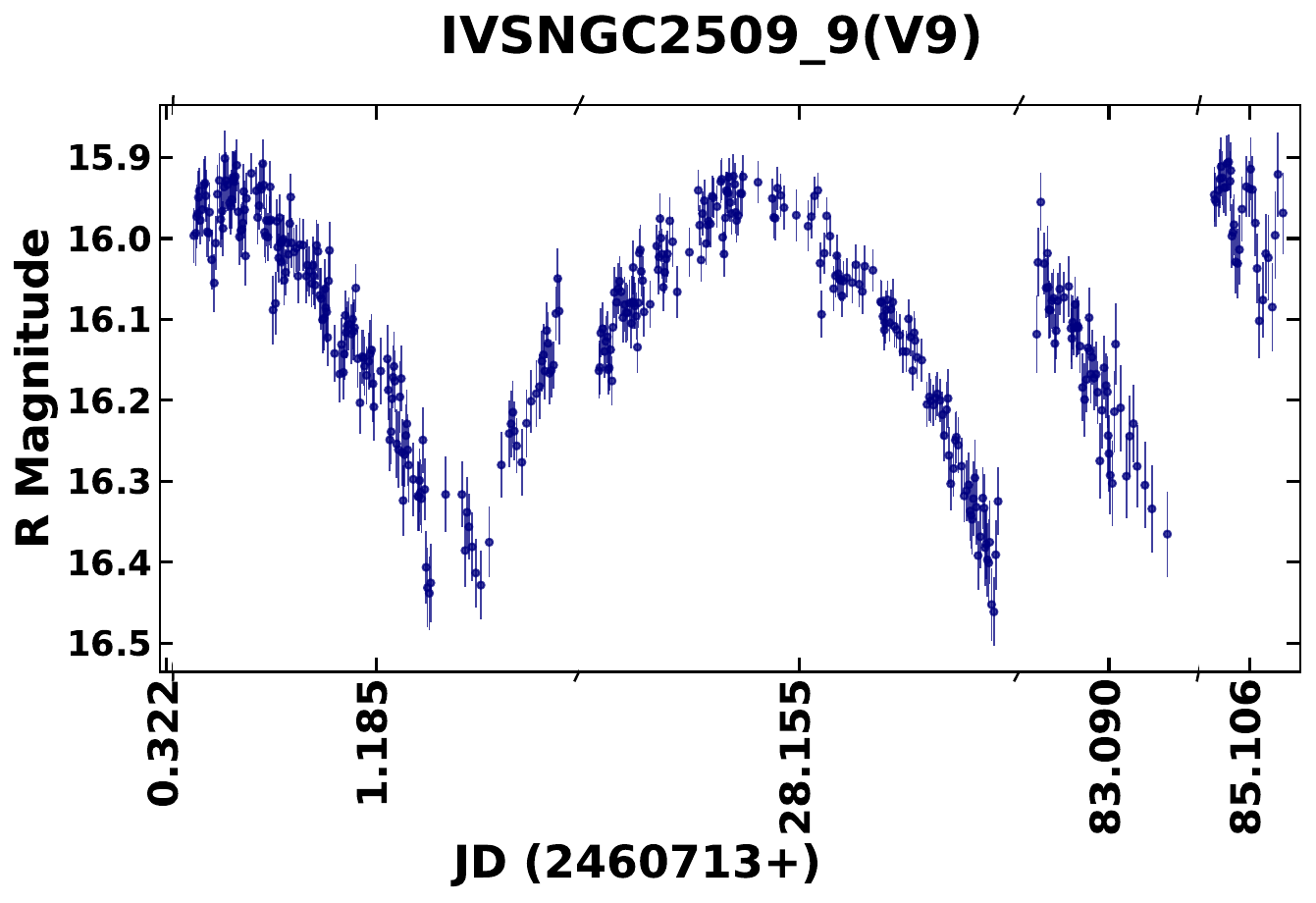}
    \includegraphics[width=4.9cm,height=2.8cm]{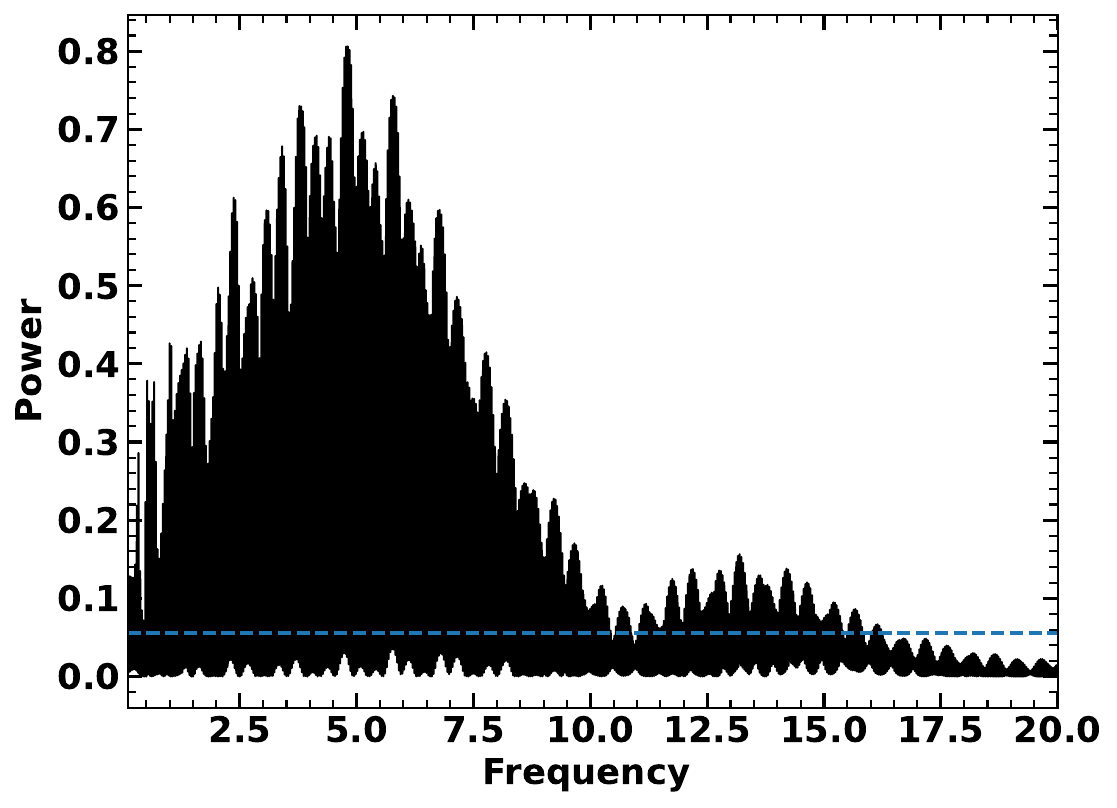}
    \includegraphics[width=4.9cm,height=3.0cm]{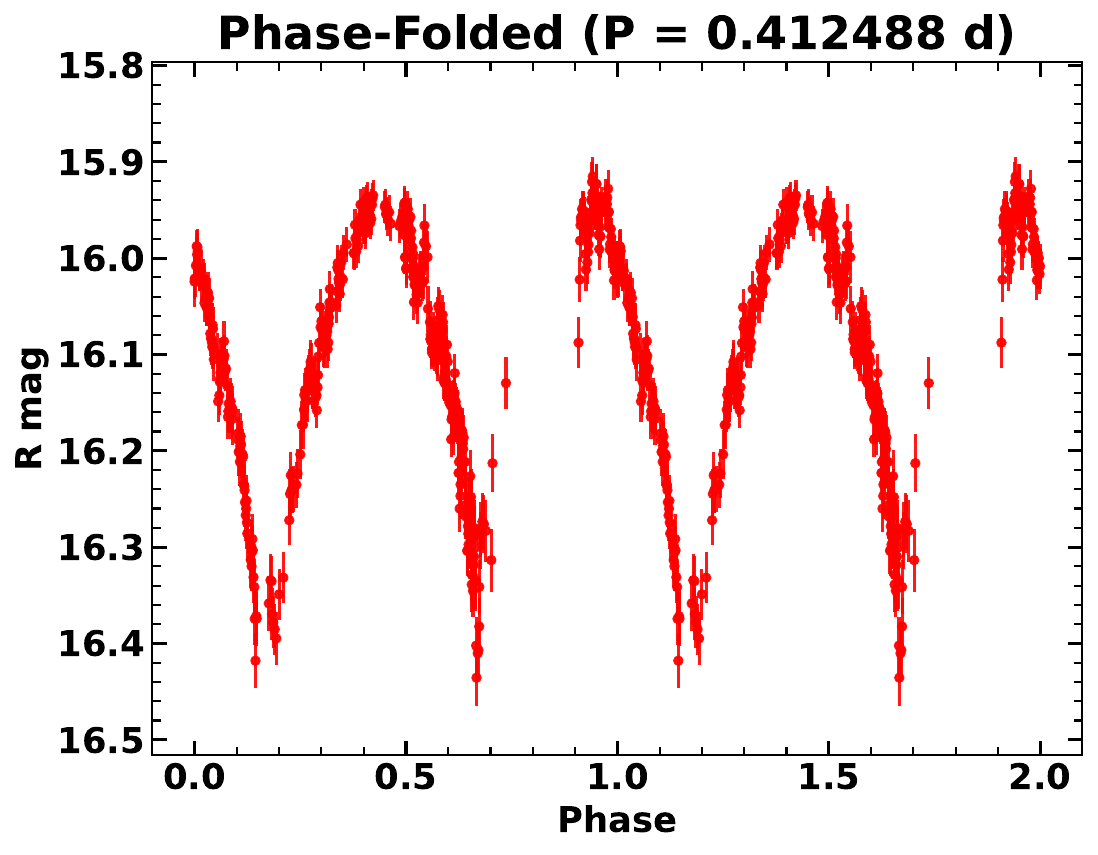}

    \includegraphics[width=4.9cm,height=3.0cm]{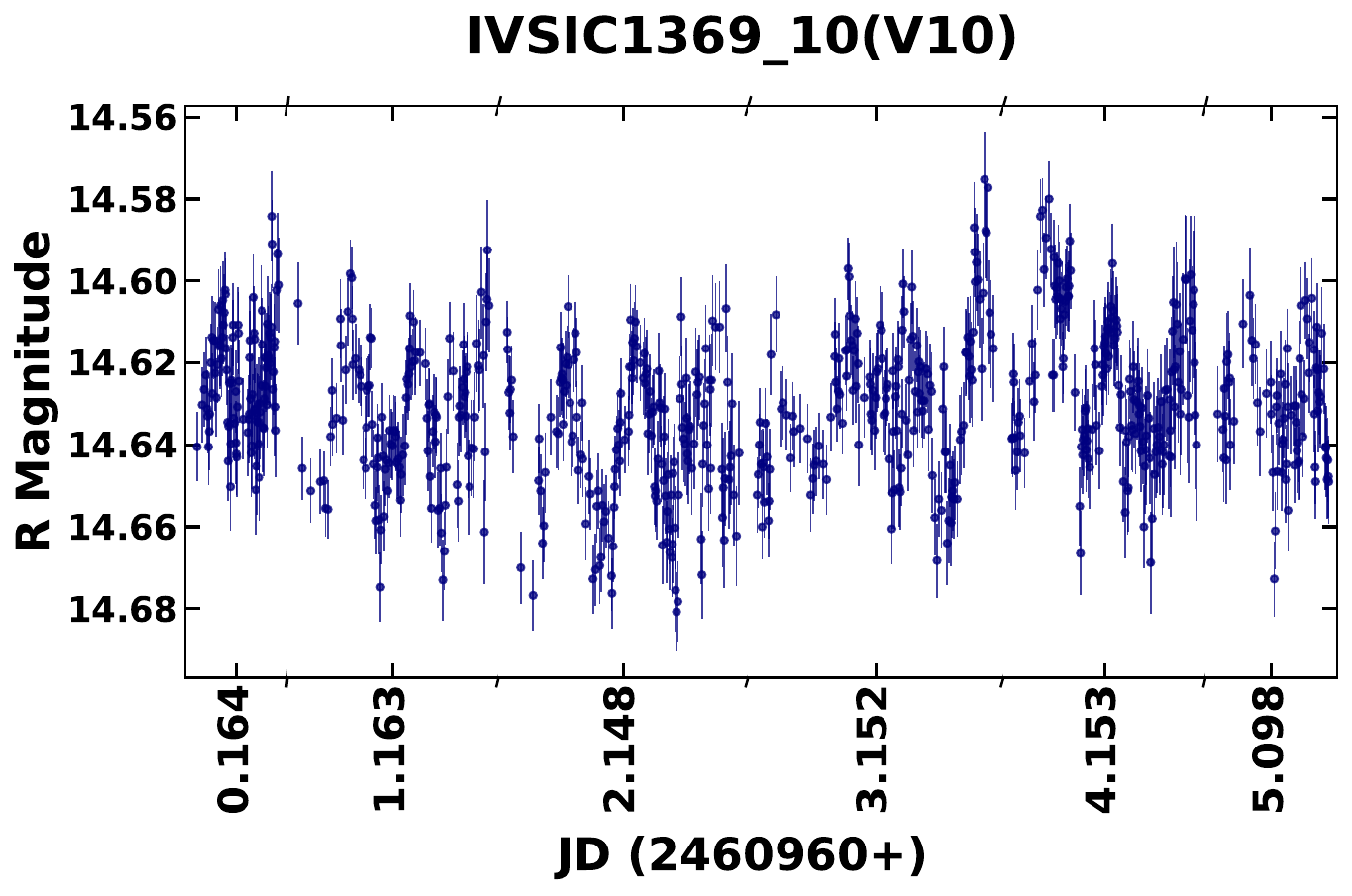}
    \includegraphics[width=4.9cm,height=2.8cm]{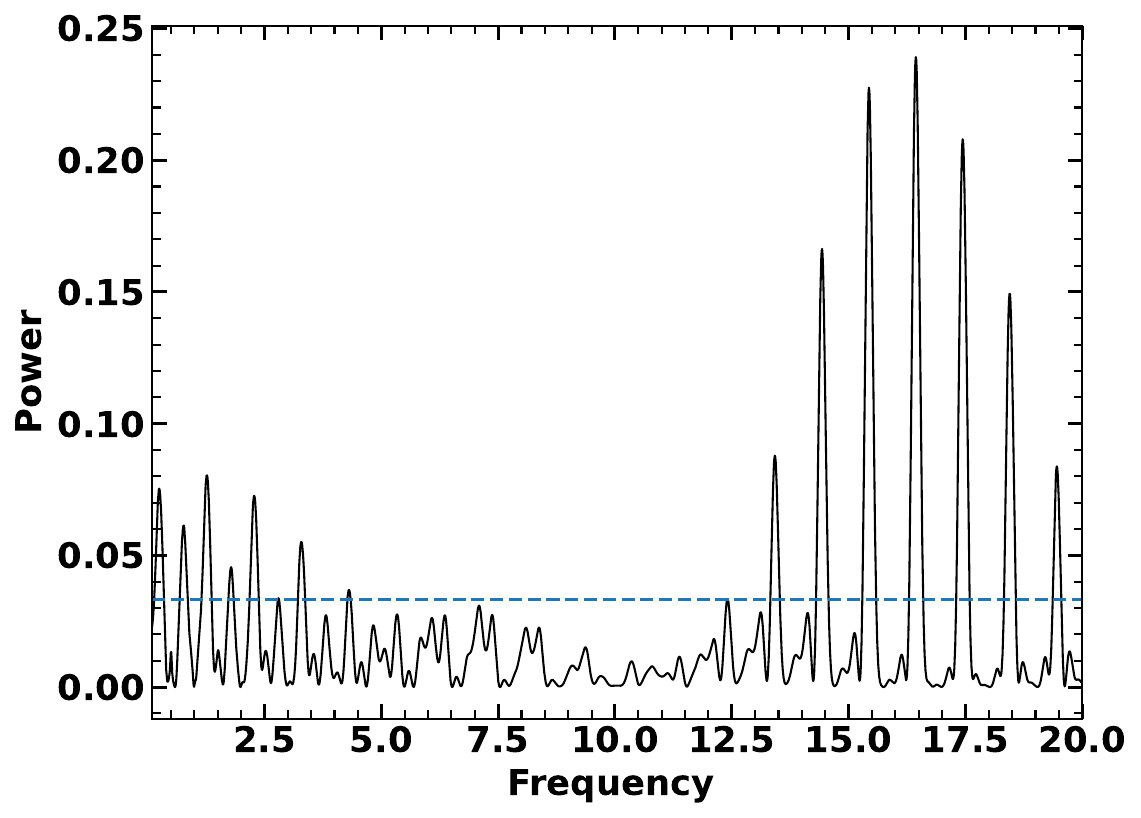}
    \includegraphics[width=4.9cm,height=3.0cm]{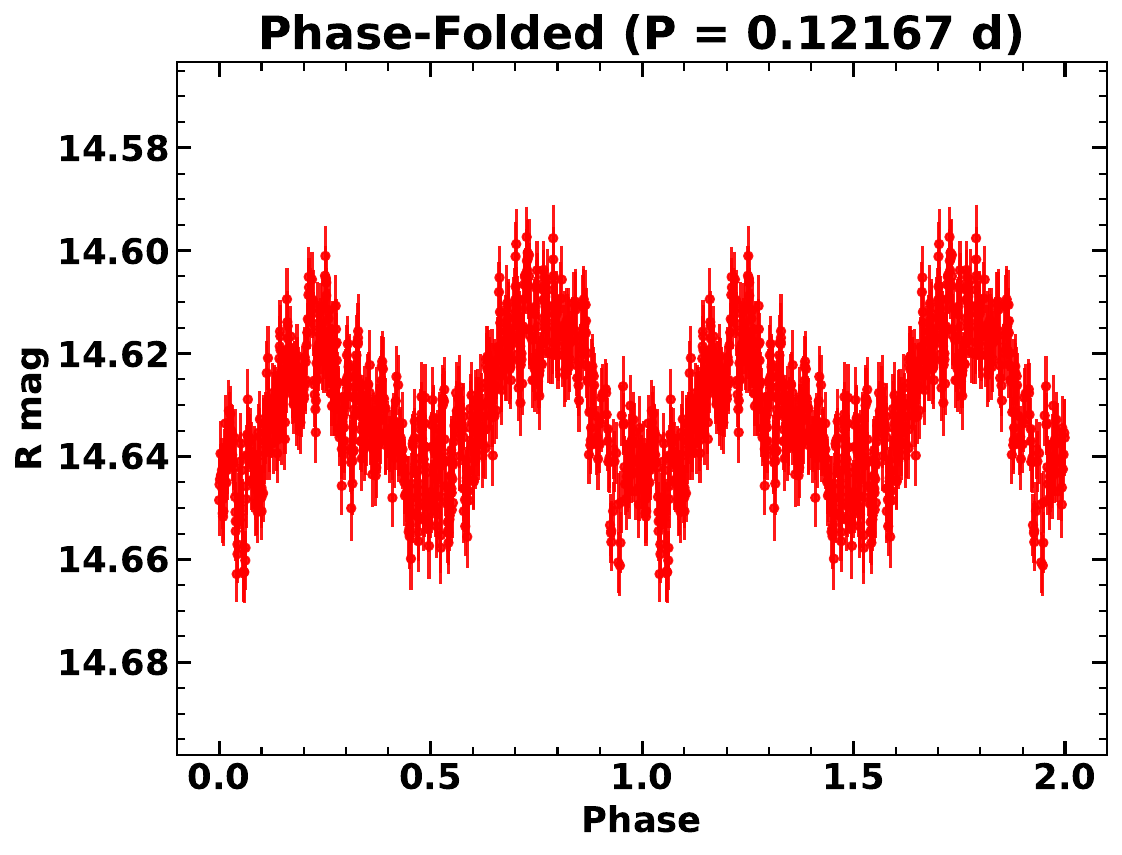}

    \includegraphics[width=4.9cm,height=3.0cm]{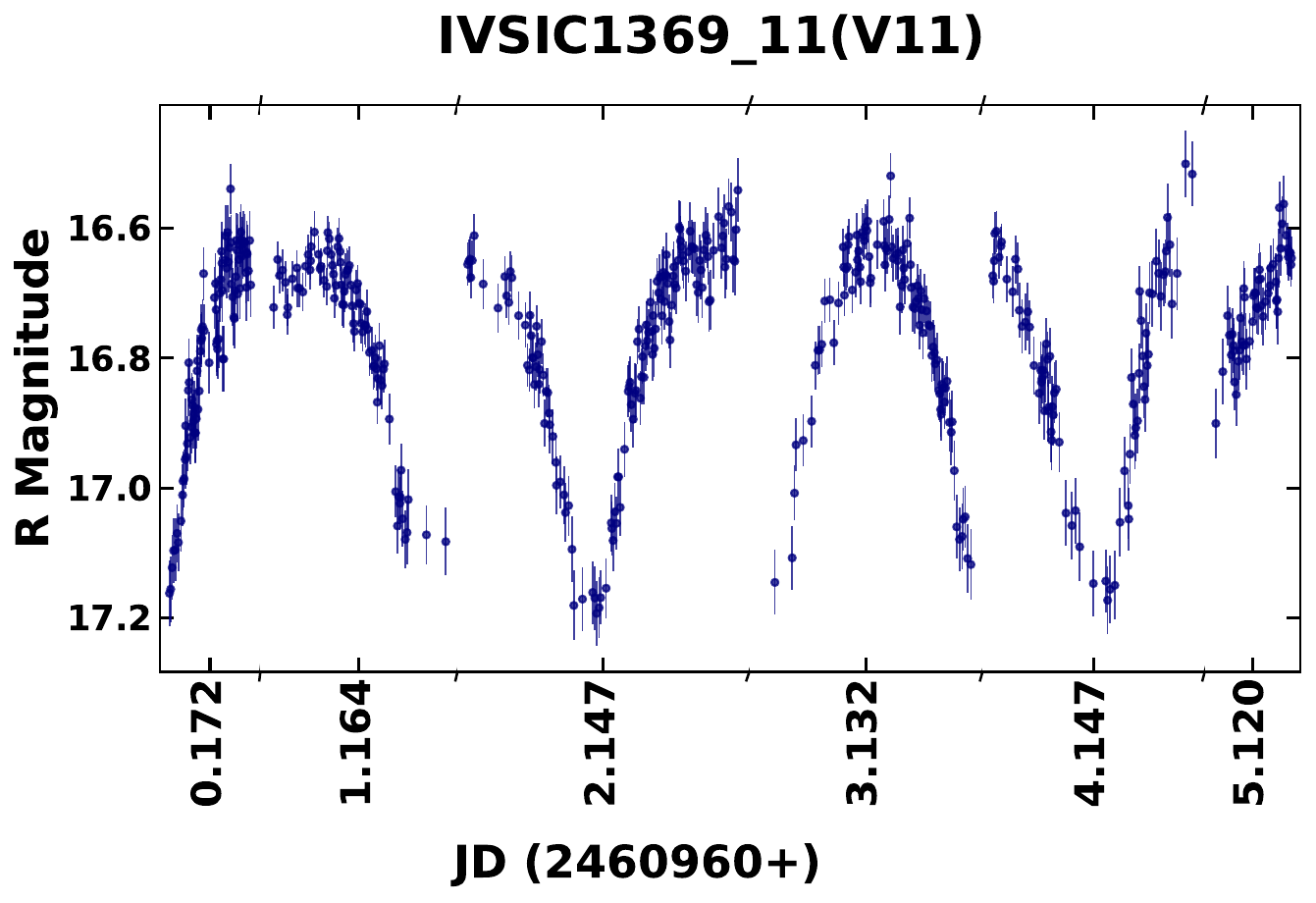}
    \includegraphics[width=4.9cm,height=2.8cm]{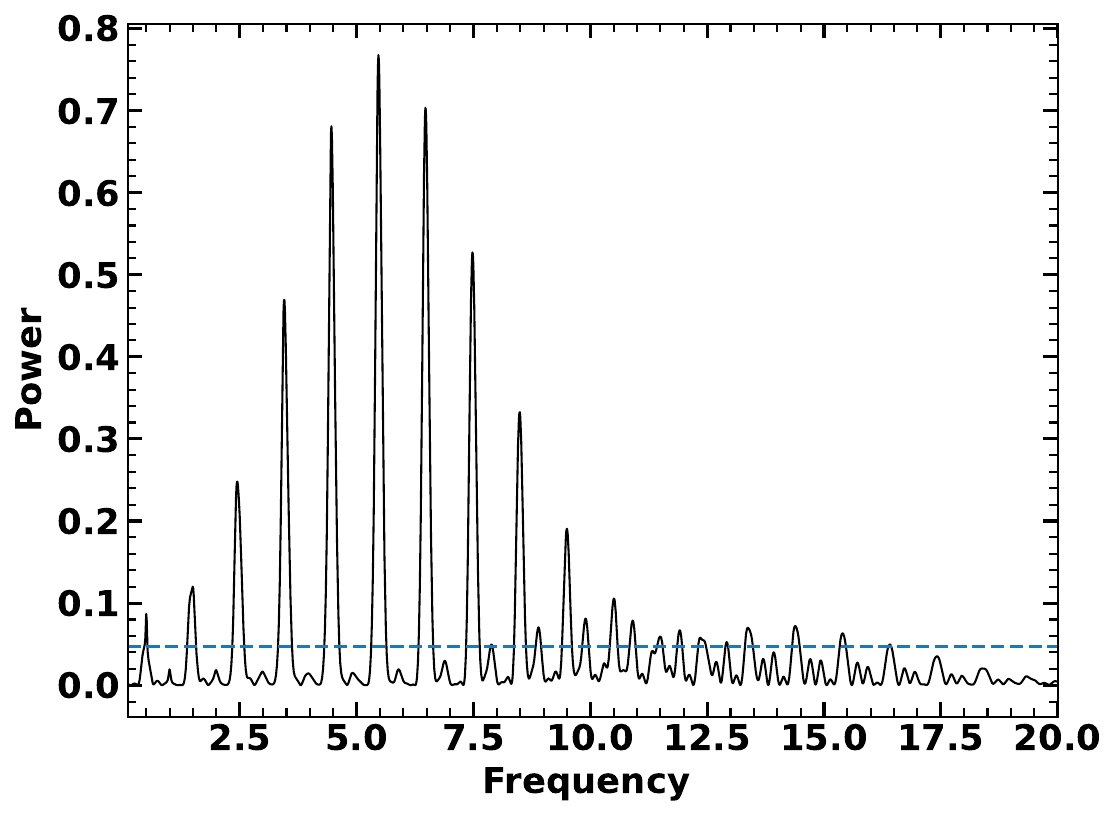}
    \includegraphics[width=4.9cm,height=3.0cm]{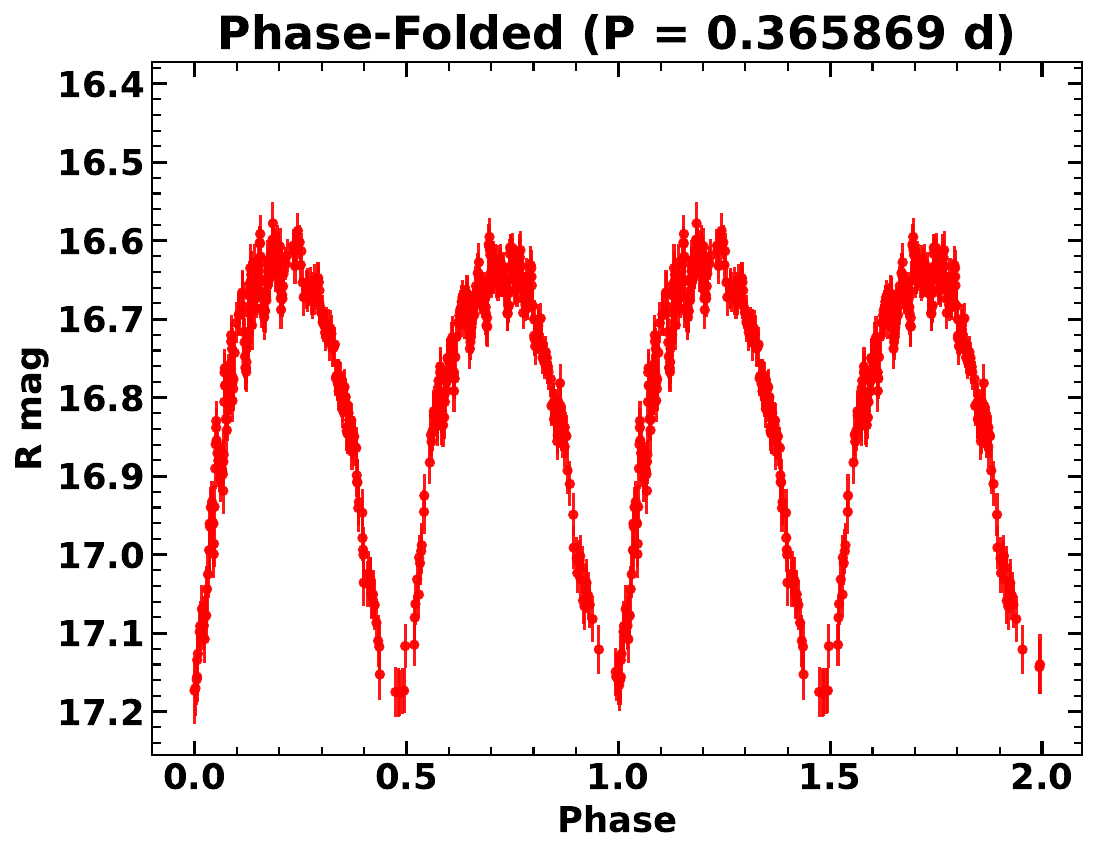}
    \includegraphics[width=4.9cm,height=3.0cm]{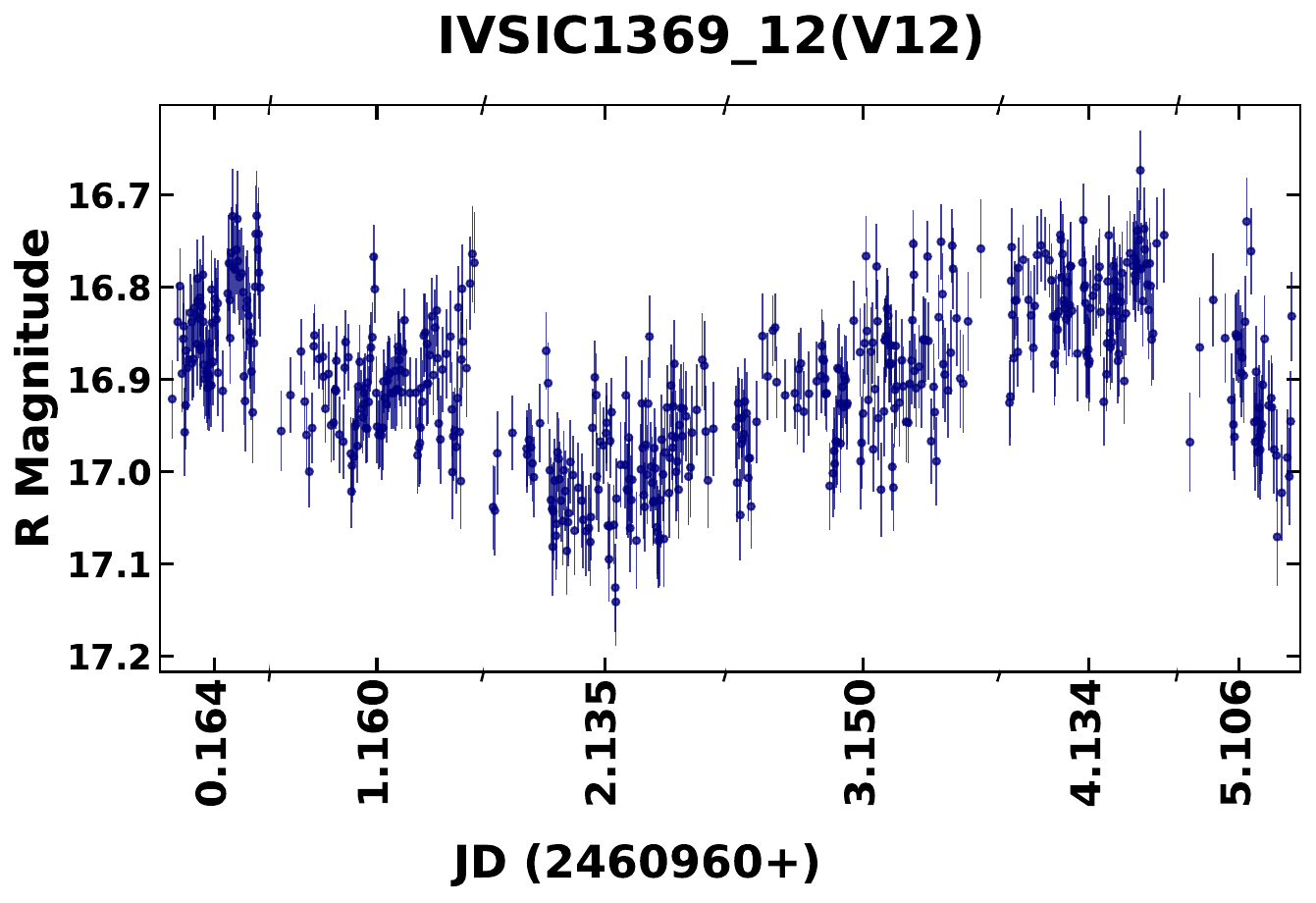}
    \includegraphics[width=4.9cm,height=2.8cm]{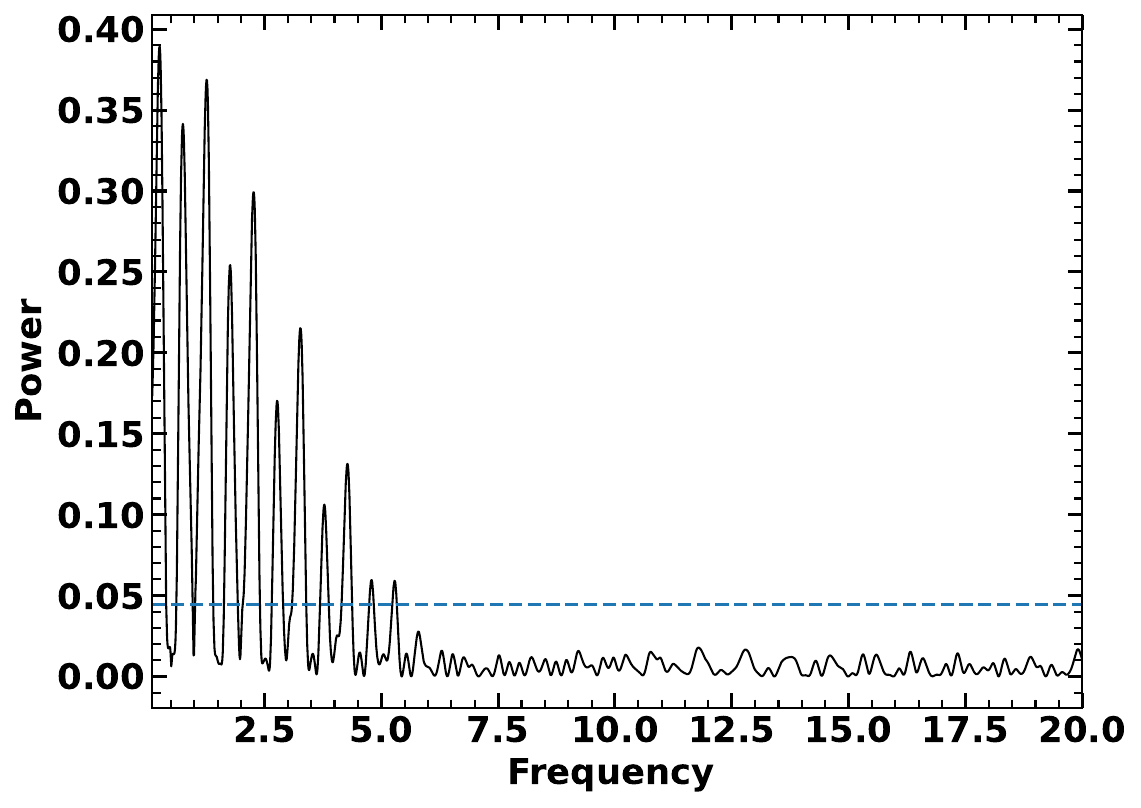}
    \includegraphics[width=4.9cm,height=3.0cm]{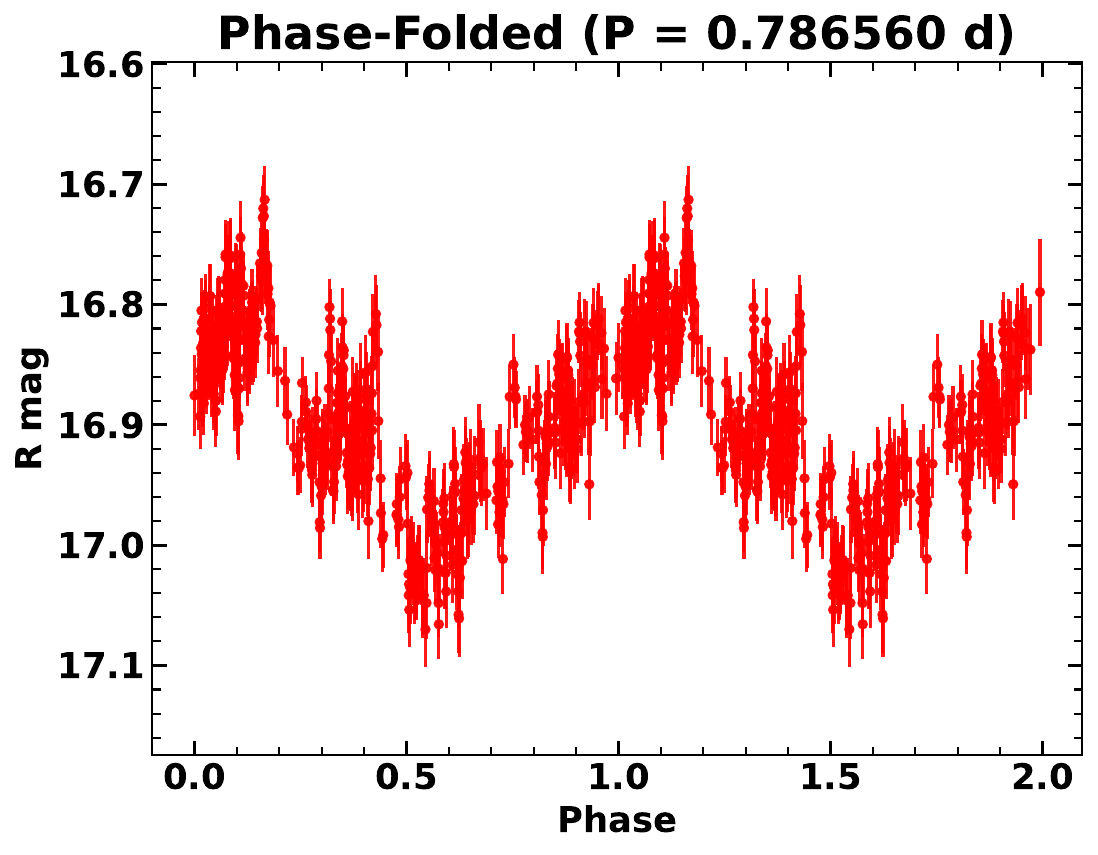}
    \includegraphics[width=4.9cm,height=3.0cm]{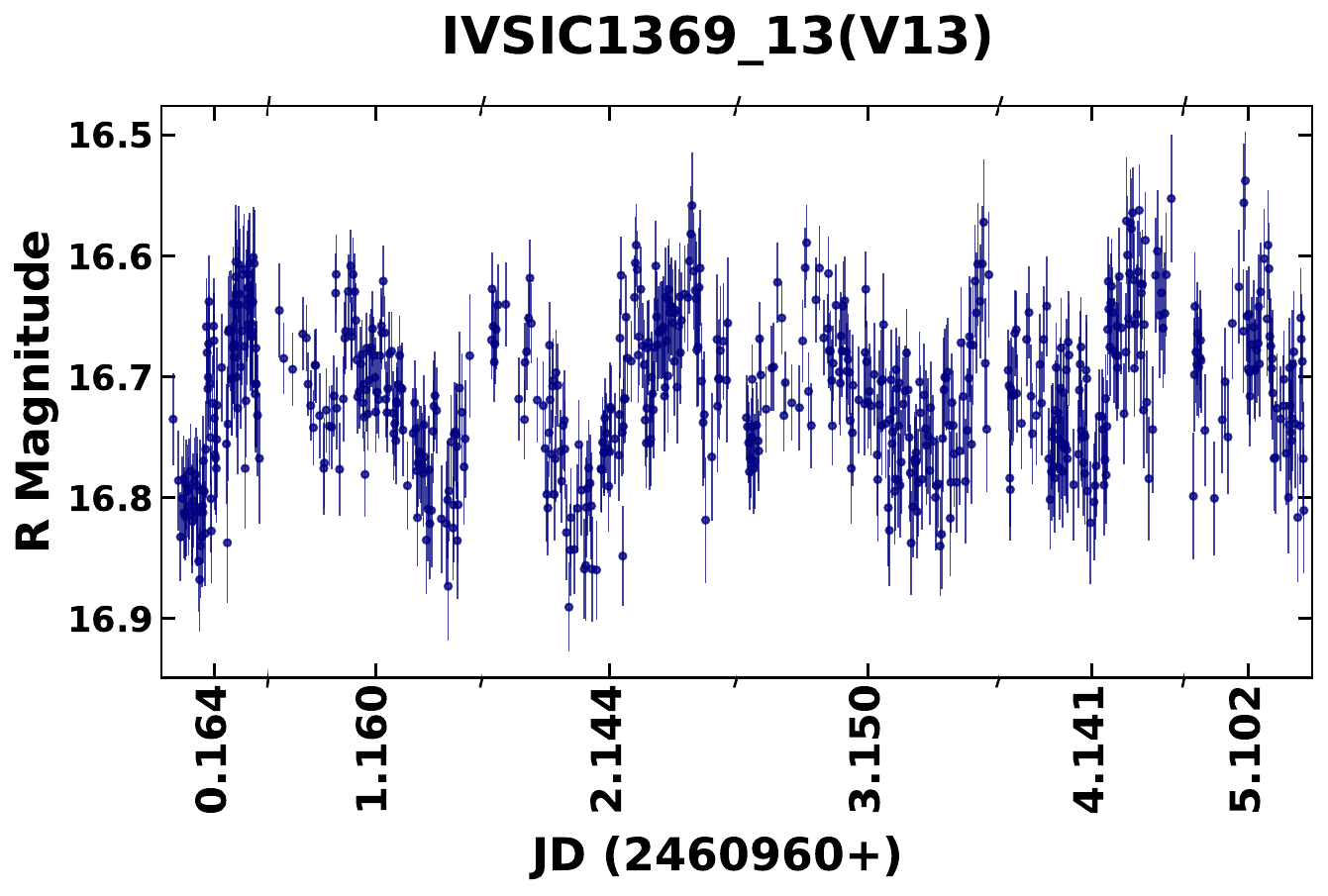}
    \includegraphics[width=4.9cm,height=2.8cm]{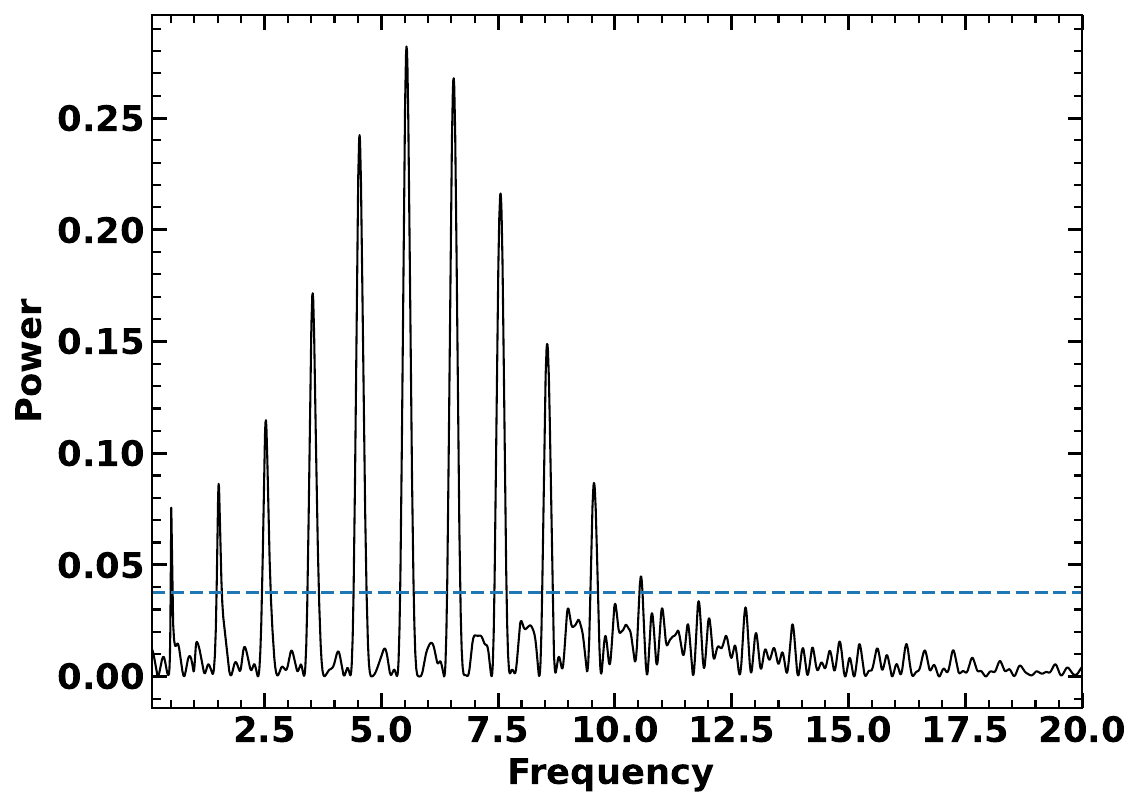}
    \includegraphics[width=4.9cm,height=3.0cm]{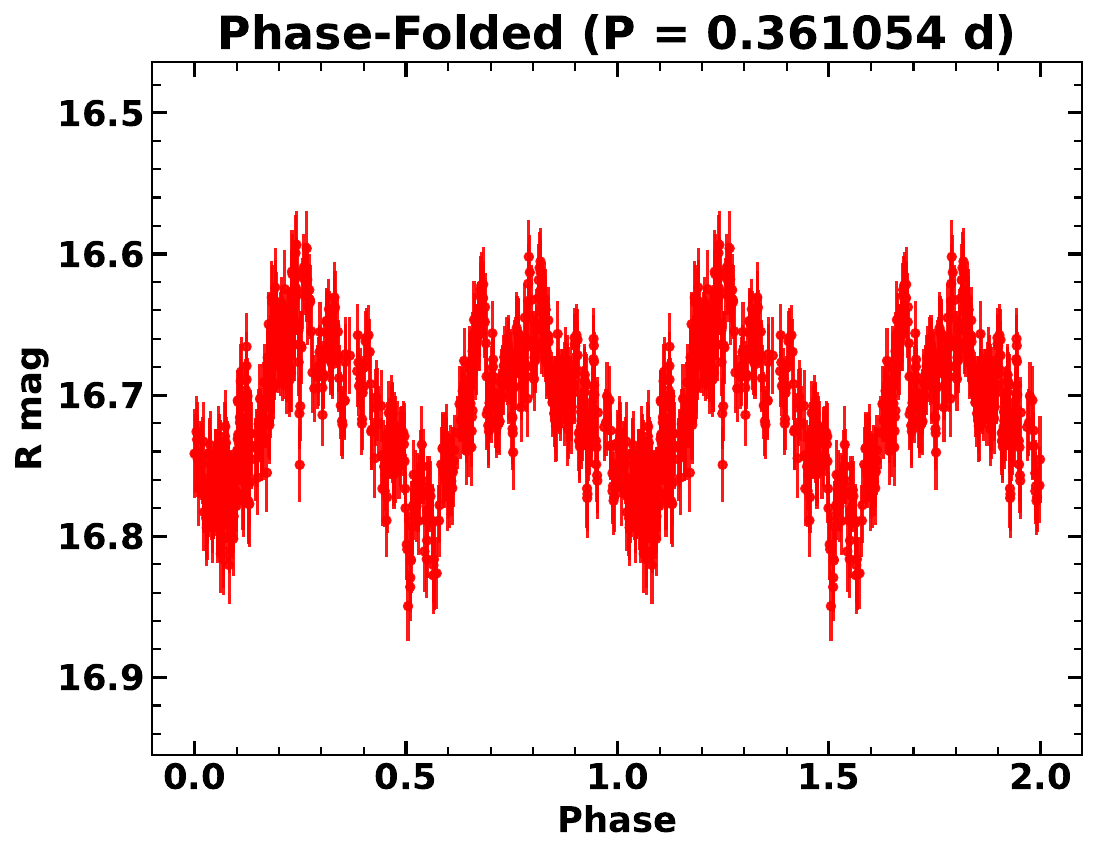}

    \caption{Same as Fig.~\ref{fig: phase_fold}, but for the variable stars V7--V13 identified in the observed OCs. The stars with IDs V7, V8, V12, and V13 are newly identified variable stars in the present study.}
    \label{fig: phase_fold2}
\end{figure*}

The variability analysis includes both cluster members and field stars in the observed regions. Membership information is used to distinguish these two populations; physical interpretation based on cluster distance, reddening, and age is applied only to the member variables, whereas field variables are classified solely from their light-curve properties and SED-based parameters. Time-series monitoring of the four clusters studied here has identified multiple variable stars located primarily along the main-sequence and turn-off regions of their respective colour-magnitude diagrams. 
 Cluster member variables discussed here have Gaia-based membership probabilities p$\ge$ 60\%, with most exceeding p$\ge$ 80\%. Variables with lower membership probabilities are classified as field stars, and their membership is determined solely from their photometric variability and SED-derived physical parameters, without applying cluster-based evolutionary constraints. Periods of the identified variable stars were determined using the Lomb-Scargle periodogram \citep{lomb1976least, scargle1982studies}, which is well-suited for analyzing unevenly spaced time-series data. Fig.~\ref{fig: phase_fold} and~\ref{fig: phase_fold2} present the time-series analysis for representative variables, showing the R-band light curves (left panels), the corresponding Lomb-Scargle periodograms (middle panels), and the phase-folded light curves constructed using the best-fit periods (right panels). The coherent phase-folded light curves and statistically significant periodogram peaks confirm the periodic nature of the detected variability. The variability census is complete for amplitudes greater than approximately 0.05 mag down to R $\sim$ 17.5, while lower-amplitude variability may remain undetected at fainter magnitudes. Therefore, the detected variables represent the high-confidence subset within the survey sensitivity limits, and no statistical conclusions regarding binary fraction, mass segregation, or dynamical evolution are drawn from the present sample.

To investigate the evolutionary status of the variables, we placed them on Hertzsprung-Russell diagrams generated for each cluster. For stars with reliable Gaia DR3 astrophysical parameters, we compared the effective temperatures derived from SED fitting with those from Gaia. The values agree within typical uncertainties (~5–15$\%$), confirming the reliability of the SED-based parameter determination. For stars lacking Gaia parameter estimates or having large uncertainties, SED fitting provides an independent and necessary determination of their physical properties. This approach ensures that all variable stars are consistently placed in the Hertzsprung-Russell diagram. It allows us to distinguish between different pulsation types and binary evolutionary pathways.

\subsection{SED-Based Stellar Parameters and Hertzsprung–Russell Diagram} \label{subsec: sed_hr_diagram}

To constrain the physical properties of the identified variable stars, we performed SED modeling using multi-band photometric data compiled from available all-sky surveys. The analysis was carried out using the Python-based package \texttt{Speedyfit}\footnote{\url{https://speedyfit.readthedocs.io/en/latest/}} \citep{vos2017orbits, vos2018composite, vos2025speedyfit}, which has also been successfully applied in recent eclipsing binaries modeling studies \citep[e.g.,][]{panchal2025exploring}. The package accesses photometric measurements from a wide range of surveys, including GALEX, Gaia, SkyMapper, APASS, SDSS, Str\"omgren--Crawford, 2MASS, and WISE, enabling broad wavelength coverage from the ultraviolet to the infrared.

Speedyfit incorporates several grids of stellar atmosphere models, including Kurucz, Munari, TMAP, and blackbody models, and allows the inclusion of additional catalogues and atmosphere grids. For each variable star, the observed SED was constructed from the available photometric measurements and fitted using a Bayesian framework with Markov Chain Monte Carlo sampling. The fitting simultaneously estimates key stellar parameters, such as effective temperature ($T_{\rm eff}$), radius ($R$), luminosity ($L$), interstellar extinction, stellar mass, and distance, while accounting for observational uncertainties and parameter degeneracies. Priors based on external constraints (e.g., parallax) further restrict the parameter space to physically realistic solutions \citep{vos2025speedyfit}.

\begin{figure*}
    \centering
    \hspace{-1.2cm}
    \includegraphics[width=10cm,height=8cm]{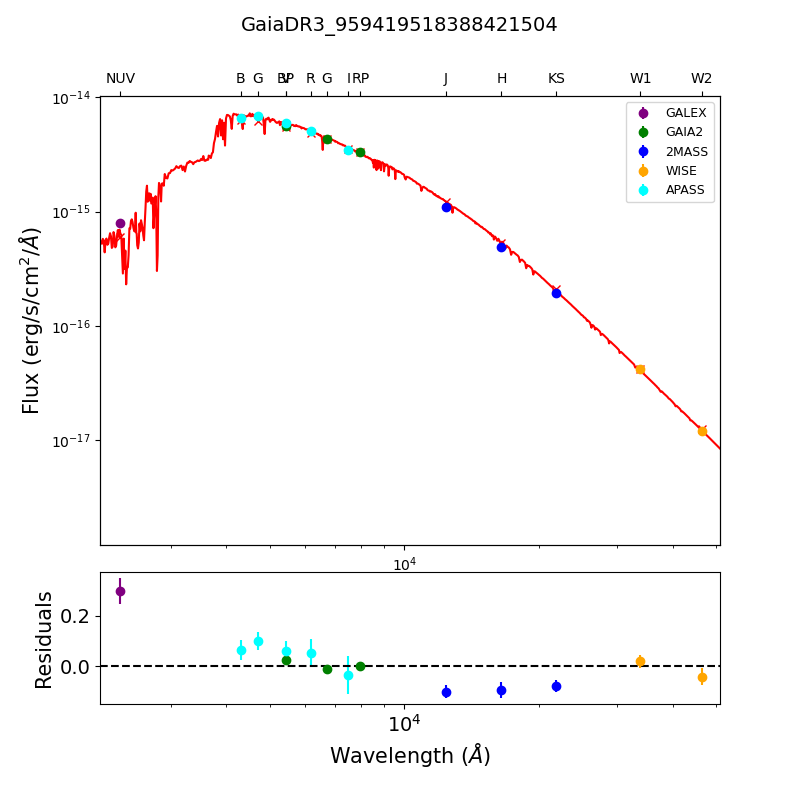}
    \includegraphics[width=8cm,height=7.3cm]{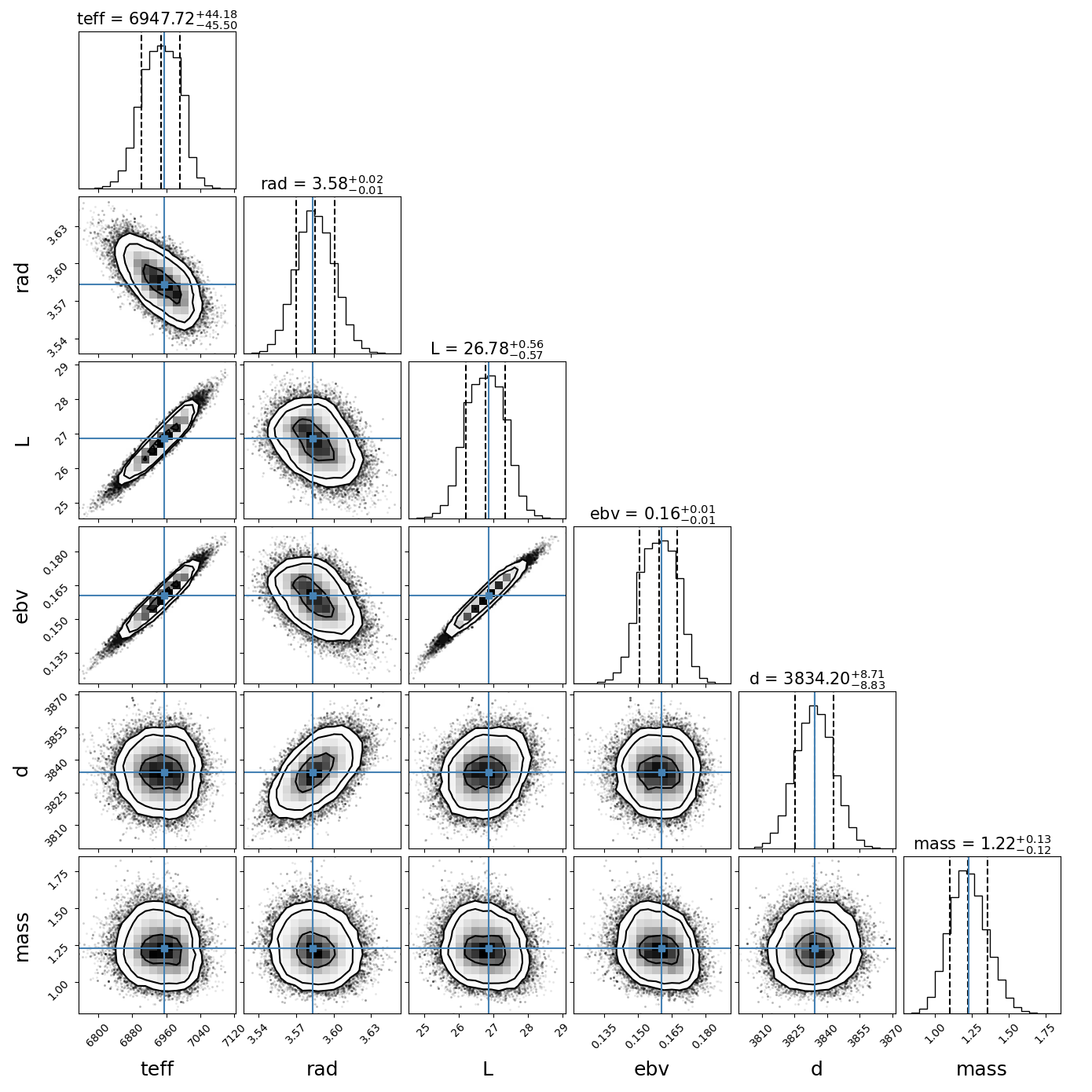}
    
    \caption{Representative SED fitting example for one of the identified variable stars. \textbf{Left:} Observed photometric fluxes from multiple surveys are plotted along with the best–fit model (solid red line). The lower panel displays the corresponding residuals between the observed and model fluxes; the unit of residuals is (mag.). \textbf{Right:} Corner plot illustrating the posterior probability distributions of the derived stellar parameters from the Markov Chain Monte Carlo analysis, including effective temperature ($T_{\rm eff}$), radius ($R$), luminosity ($L$), extinction [$E(B-V)$], distance ($d$), and mass ($M$). The contour levels denote the 1$\sigma$, 2$\sigma$, and 3$\sigma$ confidence regions.}
    \label{fig: one_sed}
\end{figure*}

Fig.~\ref{fig: one_sed} and~\ref{fig: app_all_sed} present representative examples of the SED fitting results, showing the observed photometric fluxes, best-fit model spectra, and corresponding posterior probability distributions. The good agreement between the observed and synthetic SEDs, along with the well-constrained posterior distributions, demonstrates the robustness of the derived parameters.  The complete set of SED-based stellar parameters, along with the mean magnitudes, variability amplitudes, and derived periods of the detected variable stars measured from the calibrated light curves, is summarized in Table~\ref{tab: sed_parameters}.

The derived effective temperatures span a wide range, from $\sim$4700\ K for cooler, likely evolved stars to nearly 10\,000\ K for hotter sources located near the classical instability strip. Correspondingly, the inferred luminosities range from a few solar luminosities up to $\sim$100\ $L_{\odot}$ while stellar radii extend from $\sim$1.3\ $R_{\odot}$ to more than $\sim$4.6$R_{\odot}$. These parameter ranges are fully consistent with expectations for main-sequence and slightly evolved A--F type pulsators, contact eclipsing binaries, and rotational variables commonly found in intermediate- to old-age open clusters.

\begin{figure}
    \centering
    \includegraphics[width=4.1cm,height=4.2cm]{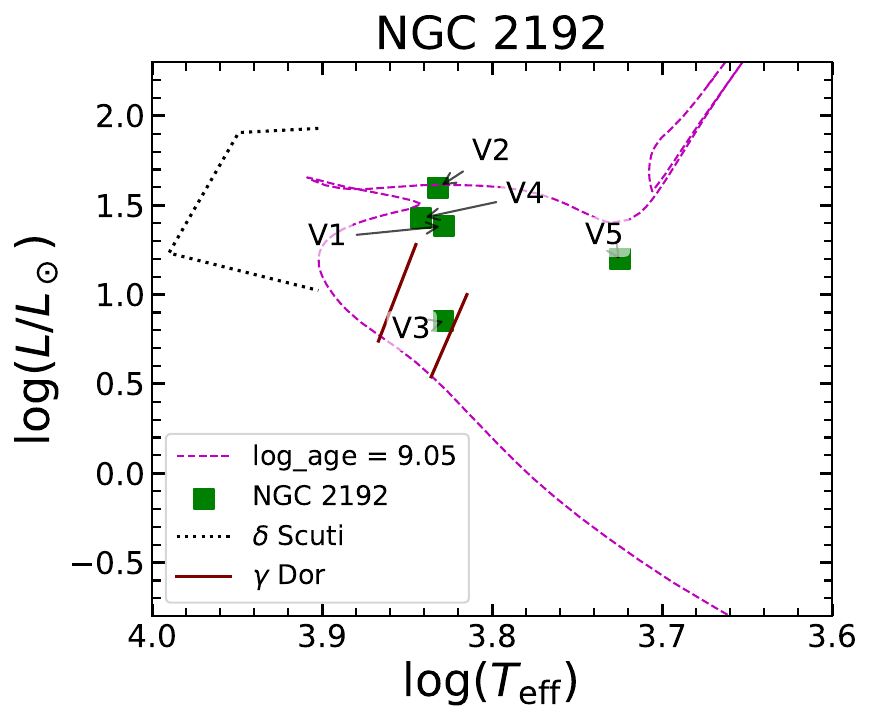}
    \includegraphics[width=4.1cm,height=4.2cm]{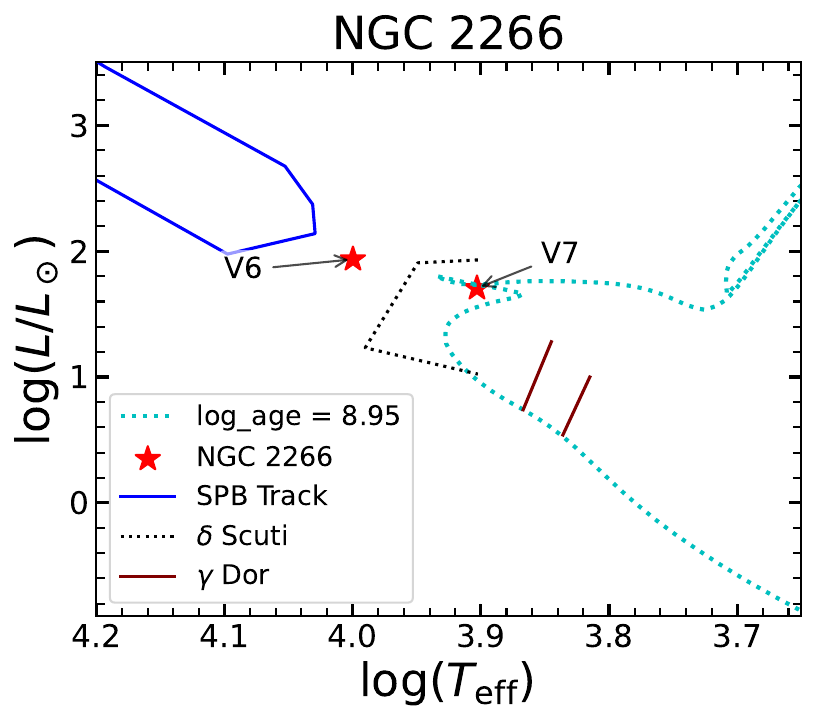}
    \includegraphics[width=4.1cm,height=4.2cm]{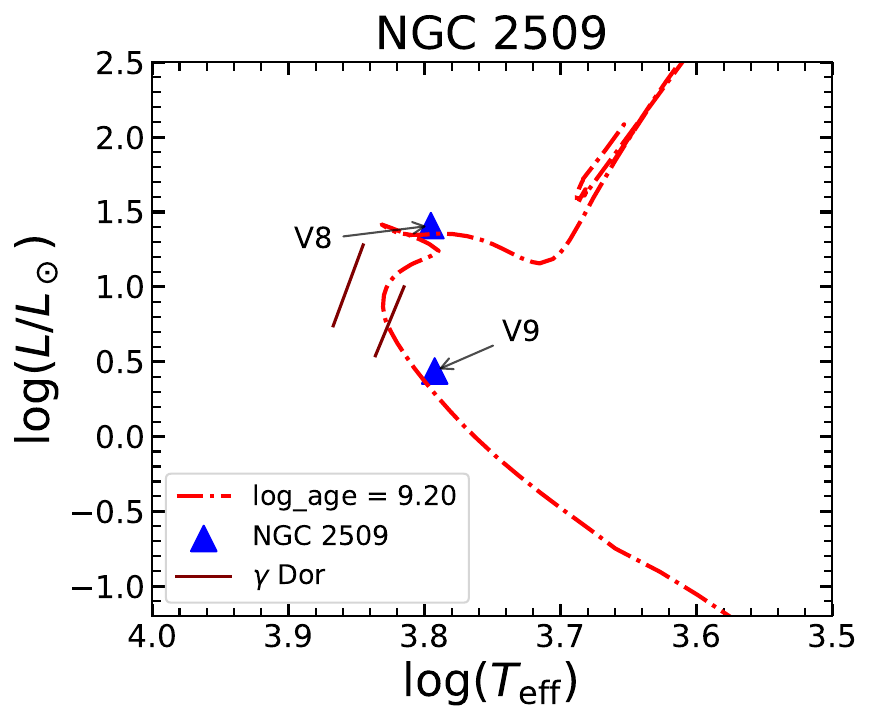}
    \includegraphics[width=4.1cm,height=4.2cm]{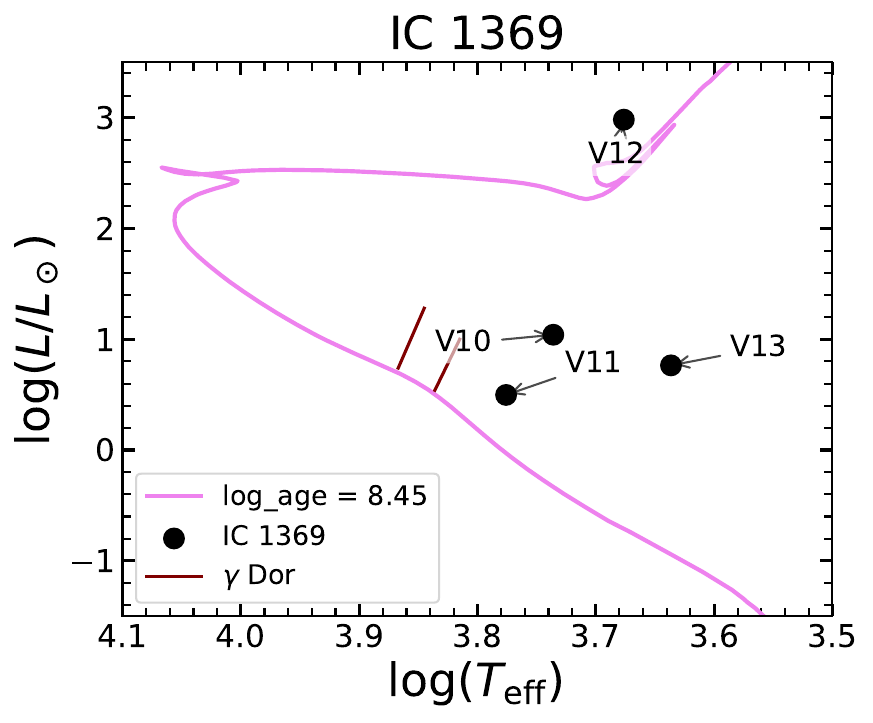}
    \caption{Hertzsprung-Russell diagram showing the locations of the identified variable stars in the cluster field. The filled symbols correspond to the classified variables, with their positions determined from the derived effective temperatures and luminosities. The plotted evolutionary tracks and isochrones correspond to different ages and evolutionary models, including pre-main-sequence, post-main-sequence, and instability-strip boundaries, as indicated in the legend. The instability regions for $\delta$~Scuti, $\gamma$~Dor, and SPB-type variables are also shown for reference. The distribution of the variables across these regions helps constrain their evolutionary stages and potential pulsational nature.}
    \label{fig: H-R Diagram}
\end{figure}

When placed on the Hertzsprung-Russell diagram (Fig.~\ref{fig: H-R Diagram}), the SED-derived stellar parameters enable a clear separation of different variability classes. Short-period pulsators cluster near or within the $\delta$~Scuti and $\gamma$~Doradus instability regions, while eclipsing binaries and rotational variables occupy evolutionary loci associated with binary interaction, stellar deformation, or angular momentum evolution. The agreement between the observed variable positions and the theoretical instability boundaries supports the reliability of the adopted classifications and confirms that the detected variability is consistent with the inferred evolutionary stages of the stars.

Overall, the SED analysis provides a crucial physical interpretation of the time-series photometric results, linking the observed variability characteristics to fundamental stellar parameters. This integrated approach strengthens the connection between photometric variability, stellar evolution, and the dynamical environments of the host open clusters. We note that SED-derived stellar parameters are model-dependent and should be interpreted within the quoted uncertainties, particularly for stars with sparse photometric coverage.

\begin{table*}
\hspace{-2cm}
\centering
\resizebox{\textwidth}{!}{
\begin{tabular}{l c c c c c c c c c c c l}
\hline
ID\_M & Gaia\_ID & RA & DEC.  & $\log g$ & Radius & $\log T_{\rm eff}$ & $\log (L/L_{\odot}) $ & R & Amplitude & Period & Membership & Type \\
 &  &  &  &  & $R_{\odot}$  &  &   & (mag)  & (mag)  & (days) & Probability ($\%$)  &   \\
\hline
IVSNGC2192\_1(V1)  & GaiaDR3\_959420141160113792 & 93.7947 & 39.8585 & 3.327 $\pm$ 0.212 &  3.741 $\pm$ 0.231 & 3.855  $\pm$ 2.241  & 1.178  $\pm$ 0.533 & 14.646 $\pm$ 0.011 &  0.012 $\pm$ 0.001 & 0.2469  $\pm$ 0.0031   &  M (96) & $\delta$ Scuti \\
IVSNGC2192\_2(V2)  & GaiaDR3\_959419385245868928 & 93.8014 & 39.8558 &  3.246 $\pm$ 0.558 & 4.562 $\pm$ 0.112 & 3.861  $\pm$ 2.456  & 1.598  $\pm$ 0.737 & 14.245 $\pm$ 0.010  &  0.013 $\pm$ 0.001  & 0.3105  $\pm$ 0.0043   &  M (90) & Misc \\
IVSNGC2192\_3(V3)  & GaiaDR3\_959419316526393984 & 93.8117 & 39.8414 & 3.862 $\pm$ 0.289 & 2.148 $\pm$ 0.133 & 3.831  $\pm$ 2.459   & 0.852  $\pm$ 0.106 & 16.065 $\pm$ 0.025 &  0.022 $\pm$ 0.002 & 0.3153  $\pm$ 0.0050 &  M (94) & $\gamma$ Dor\\
IVSNGC2192\_4(V4)  & GaiaDR3\_959419518388421504 & 93.8293 & 39.8717 & 3.448 $\pm$ 0.197 &  3.585 $\pm$ 0.077 & 3.848  $\pm$ 2.086 & 1.427  $\pm$ 0.191 & 14.494 $\pm$ 0.010 &  0.011 $\pm$ 0.001 & 0.2607  $\pm$ 0.0044   &  M (96) & $\delta$ Scuti \\
IVSNGC2192\_5(V5)  & GaiaDR3\_959419453965332736 & 93.8360 & 39.8638 &  3.409 $\pm$ 0.281 & 4.707 $\pm$ 0.116 & 3.725  $\pm$ 2.267  & 1.198  $\pm$ 0.190 & 12.784 $\pm$ 0.006 &  0.013 $\pm$ 0.002 & 0.8959  $\pm$ 0.0015   & F (55)  & Rot  \\
IVSNGC2266\_6(V6)  & GaiaDR3\_3385734852722715136 & 100.8232 & 27.0038 & 3.646 $\pm$ 0.236 & 3.105 $\pm$ 0.292 & 3.999   $\pm$ 2.127  & 1.935  $\pm$ 1.238 &  12.693 $\pm$ 0.008 &  0.013 $\pm$ 0.001 & 0.1969  $\pm$ 0.0010  & F (50) & Non-pulsating\\
IVSNGC2266\_7(V7)$^\ast$ & GaiaDR3\_3385734852722879872 & 100.8150 & 27.0028 & 2.004 $\pm$ 0.018 & 3.423 $\pm$ 0.065 & 3.903  $\pm$ 1.966  & 1.705  $\pm$ 0.285  &  14.229 $\pm$ 0.012 &  0.013 $\pm$ 0.001 &  0.313  $\pm$  0.003 & M (91) & $\delta$ Scuti \\
IVSNGC2509\_8(V8)$^\ast$  & GaiaDR3\_5714209865692265984 & 120.1532 & -19.1264 &  3.362 $\pm$ 0.414 &  4.350 $\pm$ 0.087 & 3.795  $\pm$ 2.382  & 1.409  $\pm$ 0.499  &  13.618 $\pm$ 0.010 &  0.020 $\pm$ 0.002   & 0.3241  $\pm$ 0.0001   & F (39) & Misc \\
IVSNGC2509\_9(V9)  & GaiaDR3\_5714219245900570624 & 120.1706 & -19.0498 &  3.964 $\pm$ 0.258 & 1.441 $\pm$ 0.074 & 3.792  $\pm$ 2.344 & 0.438  $\pm$ 0.226  & 15.979 $\pm$ 0.010 &  0.172 $\pm$ 0.010  &  0.4124  $\pm$ 0.0021   & F (55) & EW \\
IVSIC1369\_10(V10) & GaiaDR3\_2164527560019647616 & 318.0646 & 47.7161 & 3.324 $\pm$ 0.283 & 4.013 $\pm$ 0.104 & 3.774  $\pm$ 2.334 & 1.254  $\pm$ 0.305  & 14.655 $\pm$ 0.010 &  0.015 $\pm$ 0.001  &  0.1216   $\pm$ 0.0047  & F (46) & Rot \\
IVSIC1369\_11(V11) & GaiaDR3\_2164531610149292288 & 317.9897 & 47.7742 & 4.003 $\pm$ 0.283 & 1.664 $\pm$ 0.148 & 3.775  $\pm$ 2.455 & 0.497  $\pm$ 0.145  & 16.845 $\pm$ 0.045  &  0.191 $\pm$ 0.010   & 0.3658  $\pm$ 0.0024  &  F (52)  &  EW  \\
IVSIC1369\_12(V12)$^\ast$ & GaiaDR3\_2164528517773330048 & 318.0560 & 47.7541 & 1.939 $\pm$ 0.143 & 46.035 $\pm$ 1.113 & 3.676  $\pm$ 1.814  & 2.982  $\pm$ 1.687  &  16.953 $\pm$ 0.055 &  0.087 $\pm$ 0.010   & 0.7865  $\pm$ 0.0065     & M (73)  & Rot. \\
IVSIC1369\_13(V13)$^\ast$ & GaiaDR3\_2164527972336013824 & 318.1621 & 47.7696 & 2.502 $\pm$ 0.028 & 2.991 $\pm$ 0.241 & 3.636   $\pm$ 1.338  & 0.764  $\pm$ 0.149 & 16.718 $\pm$ 0.044 &  0.051 $\pm$ 0.004  & 0.3610  $\pm$ 0.0038  & F (36) & Rot. \\
\hline
\end{tabular}}
\caption{Derived stellar parameters for the identified variable stars in the cluster field. The table lists the Gaia source identifier, equatorial coordinates (RA, Dec), effective temperature ($T_{\rm eff}$), luminosity ($L$), stellar radius ($R$), and mass ($M$) obtained from the SED fitting and evolutionary tracks. Also included are the variability periods, membership probabilities derived from astrometric data, and the final variability classification for each source. The membership column indicates high-probability members (M) and field candidates (F). ‘Misc’ denotes variables with uncertain or mixed variability characteristics. The uncertainties represent 1$\sigma$ confidence intervals. The stars marked with an asterisk ($^\ast$) denote those newly identified in our study.}
\label{tab: sed_parameters}
\end{table*}

\subsection{Classification of the Variable Stars}

The classification of the detected variable stars was performed based on their light-curve morphology, variability periods, amplitudes, and their positions in the Hertzsprung–Russell diagrams. In this study, we also identify four new variable stars (see Table \ref{tab: sed_parameters}): V7 in NGC~2266, V8 in NGC~2509, and V12 and V13 in IC~1369. Previously, variable stars in these clusters were reported only in the catalogue of \citep{watson2006international}, where two stars (V5 in NGC~2192) were identified as rotational variables without reported periods, and V10 was classified as a $\delta$~Scuti star. Based on our analysis of the pulsation periods and effective temperatures, we reclassify V10 as a rotational variable. In this work, we conduct a detailed time-series analysis to identify reliable periods and, for the first time, classify all detected variable stars.

\subsubsection{Non Pulsating}
The star V2 is located within the gap between the slowly pulsating B-type (SPB) and $\delta$ Scuti instability regions. A population of variable stars occupying this region was first reported by \citet{mowlavi2013stellar} in the open cluster NGC 3766. These stars lie in an area of the Hertzsprung-Russell diagram where standard stellar models do not predict pulsational instability \citep{balona2011kepler, mowlavi2013stellar}. Nevertheless, observations have shown variability in this gap, and such objects are often classified as non-pulsating variables despite exhibiting photometric variability. The observed variability is generally attributed to rapid rotation, which alters the stars' internal structure and excitation conditions \citep{balona2011kepler}.

\citet{mowlavi2013stellar} reported that the periods of these variables typically range from 0.1 to 0.7 days, while \citet{lata2019short} identified similar variables in Stock~8 with periods up to 0.364 days. More recently, \citet{chand2025long} reported newly identified variables in this region with periods spanning 0.09 to 0.74 days.

In the present study, the star V6 is classified as a non-pulsating variable based on its period  (0.196962 d), low variability amplitude (0.013 $\pm$ 0.001 mag); and its position in the Hertzsprung–Russell diagram. This star has previously been reported only in the catalog of \citet{watson2006international} but without a reported variability period or amplitude. The calibrated R-band light curve and phase-folded variation of V6 confirm its periodic variability. The derived amplitude and phase modulation are consistent with rotational variability rather than classical pulsations, supporting its classification as a non-pulsating variable in the instability gap region.

\subsubsection{$\delta$ Scuti Variables}
A small but previously unexplored sample of the detected variables exhibits short-period ($P \sim 0.02$--$0.30$~d) low-amplitude pulsations with nearly sinusoidal light-curve morphologies. These stars are positioned within or very close to the classical instability strip in the Hertzsprung-Russell diagram, typically near the main-sequence turn-off in the observed clusters. Their photometric and evolutionary characteristics identify them as $\delta$~Scuti stars, which are A-F type pulsators undergoing radial and non-radial pressure-mode oscillations driven by the $\kappa$-mechanism. The presence of these pulsators in the studied clusters is consistent with their intermediate ages, as stars of appropriate mass are currently evolving across the instability strip. These $\delta$~Scuti stars therefore provide an effective means to investigate internal stellar structure and helioseismic-like pulsation behaviour in well-constrained cluster environments. Based on their pulsation periods  ($P = 0.246906$--$0.315392$~d), low variability amplitudes ($A = 0.048$--$0.087$~mag and spectral types inferred from their effective temperatures, we identify the stars V1, V4, and V7 as $\delta$~Scuti variables.

\subsubsection{$\gamma$ Doradus Variables}

 Another subset of variables exhibits longer periods ($P \sim 0.3$--$2.0$~d) with smooth, low-frequency brightness variations and is located slightly cooler than the $\delta$~Scuti candidates, consistent with $\gamma$~Doradus or rotationally 
modulated variability. Their variability characteristics are consistent with $\gamma$~Doradus-type pulsation, which is attributed to high-order, non-radial gravity-mode oscillations driven by the convective flux-blocking mechanism at the base of the outer convective envelope. Because $g$-modes are sensitive to deep interior layers, these stars are significant for probing core rotation, mixing efficiency, and angular momentum transport. The detection of $\gamma$~Dor variables in these clusters therefore adds an asteroseismic dimension to the understanding of intermediate-mass stellar evolution in populations of known age and metallicity.
We identify the star V3 as a $\gamma$~Doradus variable based on its position in the H–R diagram, its pulsation period ($P = 0.315392$~d), and low variability amplitude ($A = 0.048 \pm 0.002$~mag, and its spectral type inferred from the effective temperature.

\subsubsection{Rotational Variables}
In rotational variables, photometric variability arises from starspots on the surfaces of low-mass stars, which modulate the observed brightness as the star rotates. Such variability can be effectively identified through time-series analysis of photometric data (e.g., \citet{irwin2009ages}). Rotational variables are commonly found in stellar clusters and serve as essential probes for testing stellar evolutionary models. Stars of G- and K-type spectral classes are particularly indicative of stellar rotation coupled with magnetic activity. The derived effective temperature of the target stars, T$_{\rm eff}$ = 4350 to 5400 K, is consistent with G- and K-type stars, supporting its classification as a rotational variable. In addition, our ground-based observations led to the identification of four additional rotational variables, V5, V10, V12, and V13, with rotation periods ranging from 0.123 to 0.896~days  and variability amplitudes ranging from 0.013 to 0.098~mag, as shown in Fig.~\ref{fig: phase_fold} and \ref{fig: phase_fold2}. The variable star V5 has also been classified as a rotational variable by \citet{watson2006international}, but the period is not reported.
The rotational properties of late-type stars with known ages provide valuable constraints on stellar internal structure and angular momentum evolution. Observations of stellar rotation periods across a wide range of stellar masses and ages are therefore crucial for understanding stellar evolution and the influence of rotation on stellar environments \citep{messina2010race}.

\subsubsection{W~UMa-type Eclipsing Binaries}

In addition to the pulsators, several contact eclipsing binary systems of the W~UMa (EW) type were identified. These stars exhibit continuously varying light curves with two nearly indistinguishable eclipse minima, characteristic of binaries in which both components share a common convective envelope due to Roche-lobe overflow. Such systems are believed to form via angular-momentum loss and tidal evolution in close binaries, and their presence in these clusters links stellar variability to the cluster's dynamical evolution. W~UMa binaries are particularly informative tracers of mass segregation, dynamical relaxation, and binary interaction processes that become increasingly significant as OCs evolve. We classify the stars V9 and V11 as W~UMa-type eclipsing binary systems based on the characteristic morphology of their phase-folded light curves, their short orbital periods ($P = 0.260574$ and $0.310488$~d), and variability amplitudes ($A = 0.082$ and $0.094$~mag). 

This classification is consistent with previous studies that identified similar systems as W~UMa-type contact binaries \citep{heinze2018first}. However, Gaia DR3 2164531610149292288 (V11) has not previously been analysed in detail, and in this work, we perform PHOEBE light-curve modelling for the first time to derive its fundamental parameters, as described in the following section.


\subsection{Light Curve Modeling with PHOEBE}

Among the identified EW-type eclipsing binaries in the present study, V9 and V11 both exhibit contact-binary light-curve morphology. However, the time-series photometry of V9 does not provide sufficiently uniform phase coverage to support reliable light-curve modelling. In particular, the available V9 data lack complete coverage of both eclipse minima, leading to degeneracies in the PHOEBE solutions. We therefore restrict detailed light-curve modelling to V11, which exhibits stable variability and complete phase coverage, enabling robust determination of its physical and geometrical parameters. To investigate the physical and geometrical properties of the eclipsing binary V11, we modelled its R-band light curve using the \texttt{PHOEBE\,1.0} software package \citep{prvsa2005computational}. The orbital period was fixed to the value obtained from our period analysis (0.308568~d). The determination of the mass ratio $q = m_{2}/m_{1}$ constitutes the first step in modeling the photometric light curves. In the absence of radial velocity measurements, the q-search method is employed to estimate the mass ratio. This method has been widely used in previous studies (e.g., \cite{panchal2023optical, li2023five, belwal2025unveiling}). 
The mass ratio of the binary system was estimated using the widely adopted q-search method (e.g., \citep{prvsa2005computational}). In this approach, a series of light-curve solutions is computed for different trial values of the mass ratio $q = m_2/m_1$, with the remaining model parameters optimized for each case. The best-fitting mass ratio is determined by identifying the value that minimizes the $\chi^2$ (or residual) of the light-curve solution. During the q-search procedure, the updated ephemeris was used to convert the light curve from MJD–flux to phase–flux space. The effective temperature of the primary component was fixed. Since contact binaries are expected to possess convective envelopes, the gravity-darkening coefficients (g$_1$ = g$_2$ = 0.32) and bolometric albedos (A$_1$ = A$_2$ = 0.5) were adopted. For a contact configuration, the surface potentials of the two components were assumed to be equal ($\Omega_1 = \Omega_2$). The synchronicity parameters were fixed at 1, and circular orbits were assumed. Limb-darkening coefficients were automatically interpolated by PHOEBE using the square-root law based on the tables of \citet{van1993new}.

During the iterative fitting process, the orbital inclination ($i$), effective temperature of the secondary component ($T_{\rm 2eff}$), surface potential ($\Omega$), and primary luminosity ($l_1$) were allowed to vary. The mass ratio ($q$) was explored over the range 0.1–1.0 in steps of 0.02. For each trial solution, the $\chi^2$ value was minimized using the differential correction algorithm implemented in PHOEBE. The minimization was repeated multiple times with random perturbations of the parameters to avoid convergence to local minima. The solution with the minimum $\chi^2$ was adopted as the best-fit model, yielding a mass ratio of $q = 0.68 \pm 0.02$, a semi-major axis of $a = 3.25 \pm 0.22~R_{\odot}$, and an orbital inclination of $i = 79.57 \pm 0.15^{\circ}$, consistent with deep eclipses in a contact binary configuration. The derived parameters are listed in Table~\ref{tab: Phoebe_parameters}.




The effective temperature of the primary component was fixed at $T_{1eff}$ =  5968$\pm$ 250 ~K based on its SED, while the model yields a secondary temperature of $T_{2eff} = 5750 \pm 115$~K. The small temperature difference is typical of W~UMa-type contact binaries, in which the components share a common envelope and remain in thermal contact. The surface potentials ($\Omega_{1}$ = $\Omega_{2}$), luminosities, and radii ($R_{1} = 1.49 \pm 0.07~R_{\odot}$; $R_{2} = 1.28 \pm 0.06~R_{\odot}$) confirm that both stars fill their Roche lobes and lie well within the contact regime. The corresponding masses are estimated to be $M_{1} = 2.05 \pm 0.10~M_{\odot}$ and $M_{2} = 1.39 \pm 0.11~M_{\odot}$.

\begin{figure}
    \centering
    \hspace{-1.2cm}\includegraphics[width=9cm,height=5.5cm]{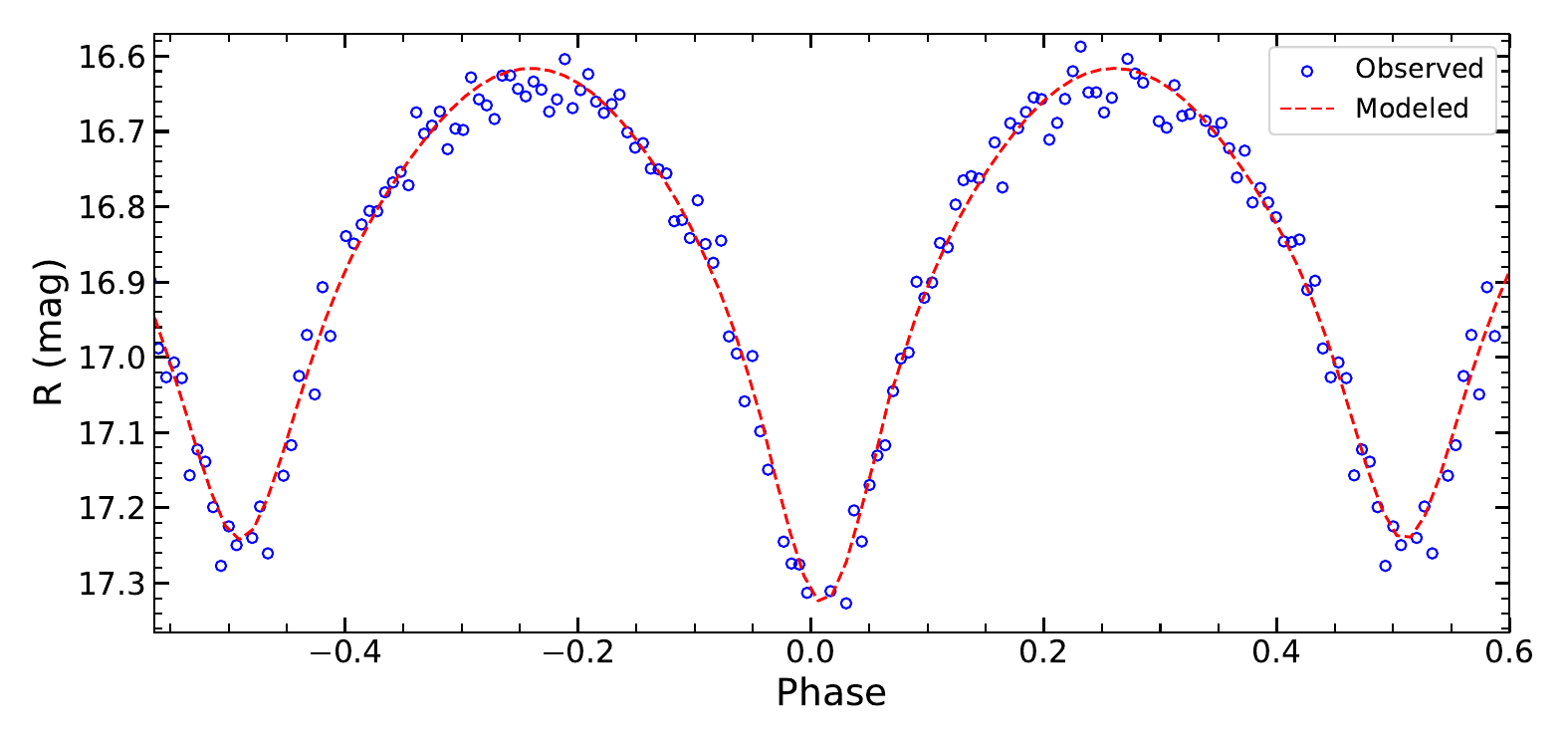}
    \caption{Observed (blue points) and PHOEBE-modeled (red curve) R-band light curve of the eclipsing binary IC~1369 plotted as a function of orbital phase. The solid curve shows the best-fitting model from \texttt{PHOEBE\,1.0}, demonstrating excellent agreement between the observed data and the synthetic light curve.}
    \label{fig: phobemodel}
\end{figure}

Fig.~\ref{fig: phobemodel} shows the phased light curve along with the synthetic model. The close agreement between the observed photometry and the PHOEBE fit demonstrates that the adopted configuration reliably reproduces the system’s variability. 

 The detailed modeling of V11 provides important constraints on the structure and evolution of low-mass contact binaries, which exchange mass and energy through a common convective envelope. Such systems are valuable laboratories for studying angular momentum loss, thermal relaxation oscillations, and possible merger pathways in close binaries. In addition, well-determined parameters help calibrate empirical relations for W~UMa binaries, such as period–luminosity and mass–radius correlations, and contribute to the growing sample of well-characterized contact binaries in open clusters, thereby enabling future comparative studies of binary evolution.

\begin{table}
\centering
\vspace{-0.0cm}
\caption{Absolute parameters derived from the light-curve modeling of the W UMa-type eclipsing binary in IC 1369 using the PHOEBE 1.0 software. The Gaia DR3 identifier of the system is Gaia DR3 2164531610149292288. The parameters include the orbital period (in days), mass ratio ($q = M_2/M_1$), semi-major axis ($a$ in $R_\odot$), orbital inclination ($i$ in degrees), effective temperatures of the primary ($T_{1\mathrm{eff}}$) and secondary ($T_{2\mathrm{eff}}$) components in Kelvin, luminosities of the components ($l_1$, $l_2$ in $L_\odot$), surface potentials ($\Omega_1$, $\Omega_2$), and radii ($R_1$, $R_2$ in $R_\odot$) and masses ($M_1$, $M_2$ in $M_\odot$) of the components.}
\begin{tabular}{ |c |c |}
\hline
Parameters & \bf{IC 1369} \\
\hline
\hline
Period (day/fixed) & 0.308568 \\
q &  0.68 $\pm$ 0.02 \\
a (R$_\odot$) & 3.25 $\pm$ 0.22  \\
i $\deg$ & 79.57 $\pm$ 0.15  \\
T$_{1eff}$ K (fixed) & 5968$\pm$ 250  \\
T$_{2eff}$ K & 5750 $\pm$ 115   \\
l$_{1}$ (L$_\odot$)  & 3.21 $\pm$ 0.14    \\
l$_{2}$ (L$_\odot$)  & 2.92 $\pm$ 0.14   \\
$\Omega_{1}$ & 2.95 $\pm$ 0.05   \\
$\Omega_{2}$ & $\Omega_{1}$  \\  
R$_{1}$ (R$_\odot$) & 1.49 $\pm$ 0.07         \\
R$_{2}$ (R$_\odot$) & 1.28 $\pm$ 0.06         \\
M$_{1}$ (M$_\odot$) & 2.05 $\pm$ 0.10         \\
M$_{2}$ (M$_\odot$) & 1.39 $\pm$ 0.11         \\

\hline
  \end{tabular}
\label{tab: Phoebe_parameters}
\end{table}

\subsection{Astrophysical Implications}

The detected variables include $\delta$~Scuti, $\gamma$~Doradus, rotational, and W~UMa-type systems, providing insight into stellar variability within coeval cluster populations. The $\delta$~Scuti and $\gamma$~Doradus pulsators probe complementary regimes of stellar interiors through pressure and gravity modes, respectively, and their presence near the main-sequence turnoff is consistent with expectations for intermediate-age clusters. Their locations on the H–R diagram, together with the measured periods and amplitudes, provide useful constraints on stellar evolution under well-defined cluster ages and metallicities. Rotational variables trace stellar spin and magnetic activity in late-type main-sequence stars, offering insight into angular-momentum evolution in clusters of known age. The W~UMa-type binary identified in IC~1369 provides an initial reference for future studies of contact-binary populations in clusters of comparable age.

\section{Conclusions} \label{sec: conclusions}

This work presents the first ground-based time-series photometric study of the Galactic open clusters NGC~2192, NGC~2266, NGC~2509, and IC~1369 using the 0.6~m VASISTHA telescope at IERCOO, combined with \textit{Gaia}~DR3 astrometric data. Although variability in some of these sources has been reported in earlier studies, the present work provides, for the first time, a detailed photometric analysis and physical characterization of the variable stars based on dense ground-based time-series observations. Our main findings are summarized as follows:

\begin{itemize}

\item Using \textit{Gaia}~DR3 astrometry, we identified clean samples of probable cluster members with minimal field contamination. The analysis yields 196, 285, 273, and 293 member stars in NGC~2192, NGC~2266, NGC~2509, and IC~1369, respectively. The resulting colour–magnitude diagrams indicate cluster ages of $\sim$0.3–1.6~Gyr, extinction values of $(A_V)\sim0.23–2.68$~mag, and distances of $\sim$2.5–3.5~kpc, consistent with recent literature estimates.

\item Radial density profiles fitted with King models yield compact core radii of $r_c = 1.20$–$2.22$ arcmin and tidal radii of $r_t = 11.6$–$16.6$ arcmin. The resulting core-to-tidal radius ratios ($r_c/r_t \approx 0.09$–$0.13$) indicate centrally concentrated stellar distributions, suggesting that all four clusters are gravitationally bound systems that have likely experienced significant dynamical evolution.

\item From $\sim$34.7~h of calibrated $R$-band time-series photometry, we identify four previously unreported variable stars, including $\delta$~Scuti, $\gamma$~Doradus, and rotational variables. SED fitting for selected sources yields effective temperatures of $\sim$4300–10,000~K, luminosities of a few to $\sim$100~$L_\odot$, and radii of $\sim$1.3–46~$R_\odot$, thereby placing them on the Hertzsprung–Russell diagram and supporting their evolutionary classifications.

\item The detected variables predominantly exhibit short periods ($\sim$0.12–0.35~d) and low-to-moderate amplitudes ($\sim$0.01–0.10~mag), consistent with $\delta$~Scuti pulsations typically observed near the main-sequence turn-off of intermediate-age clusters. In contrast, longer periods (up to $\sim$0.9~d) and larger amplitudes are found for rotational variables and eclipsing binaries, reflecting variability produced by stellar rotation or orbital modulation rather than pulsation. These period–amplitude characteristics are consistent with those commonly observed in OCs of similar ages.

\item A contact eclipsing binary detected in IC~1369 was modelled using \texttt{PHOEBE}. The system is confirmed as a W~UMa-type binary with a mass ratio $q=0.68\pm0.02$, orbital inclination $i\simeq79.6^\circ$, and component radii of $R_1\simeq1.49,R_\odot$ and $R_2\simeq1.28,R_\odot$. This represents the first physically modelled compact binary system associated with this cluster.

\end{itemize}

\section*{Data Availability}

 The photometric time-series data used to construct the light curves of the four newly identified variable stars in this study are provided in machine-readable format as supplementary material associated with this article, including the observation times, calibrated $R$-band magnitudes, and corresponding uncertainties.

\section*{Acknowledgements}

We thank the anonymous referee for their constructive comments and valuable suggestions, which significantly improved the clarity and scientific quality of this manuscript. Ing-Guey Jiang acknowledges support from the National Science and Technology Council (NSTC), Taiwan, under grants (NSTC 113-2112-M-007-030 and NSTC 114-2112-M-007-029). The observations presented in this work were carried out using the 0.6\,m VASISTHA telescope at the Ionospheric and Earthquake Research Centre and Optical Observatory (IERCOO), Indian Centre for Space Physics (ICSP), Kolkata. We acknowledge the Higher Education Department, Government of West Bengal, for its support in establishing the observational facilities at ICSP. KB and MB acknowledge ICSP for financial support during this research. This research made use of data from the European Space Agency’s \textit{Gaia} mission, processed by the \textit{Gaia} Data Processing and Analysis Consortium (DPAC). We thank the developers of the open-source software and analysis tools used in this work, including DAOPHOT II, DAOmatch, DAOMaster, VARTOOLS, PHOEBE, and relevant Python scientific libraries.

Python modules: NUMPY (1.19.0;\citep{harris2020array}), SCIPY (v1.3.1 \citep{Eri01}), and MATPLOTLIB (v3.1.1 \citep{hunter2007matplotlib}) 

\bibliography{main}{}
\bibliographystyle{aasjournal}



\section{Appendix}

This appendix presents supplementary figures that support the analyses discussed in the main text. Fig.~\ref{fig:membership_appendix} shows the spatial, proper-motion, and parallax distributions of stars in the fields of the open clusters NGC~2192, NGC~2266, and NGC~2509, illustrating the effectiveness and consistency of the membership selection procedure adopted in this work. Fig~\ref{fig: app_all_sed} presents SED fitting results for the remaining variable stars not shown in the main text, included here for completeness and to demonstrate the robustness of the SED-based stellar parameter estimation.

\setcounter{figure}{0}
\renewcommand{\thefigure}{A\arabic{figure}}

\label{appendix:A1}
\begin{figure}[htbp]
   \centering
   \includegraphics[width=15cm,height=4.5cm]{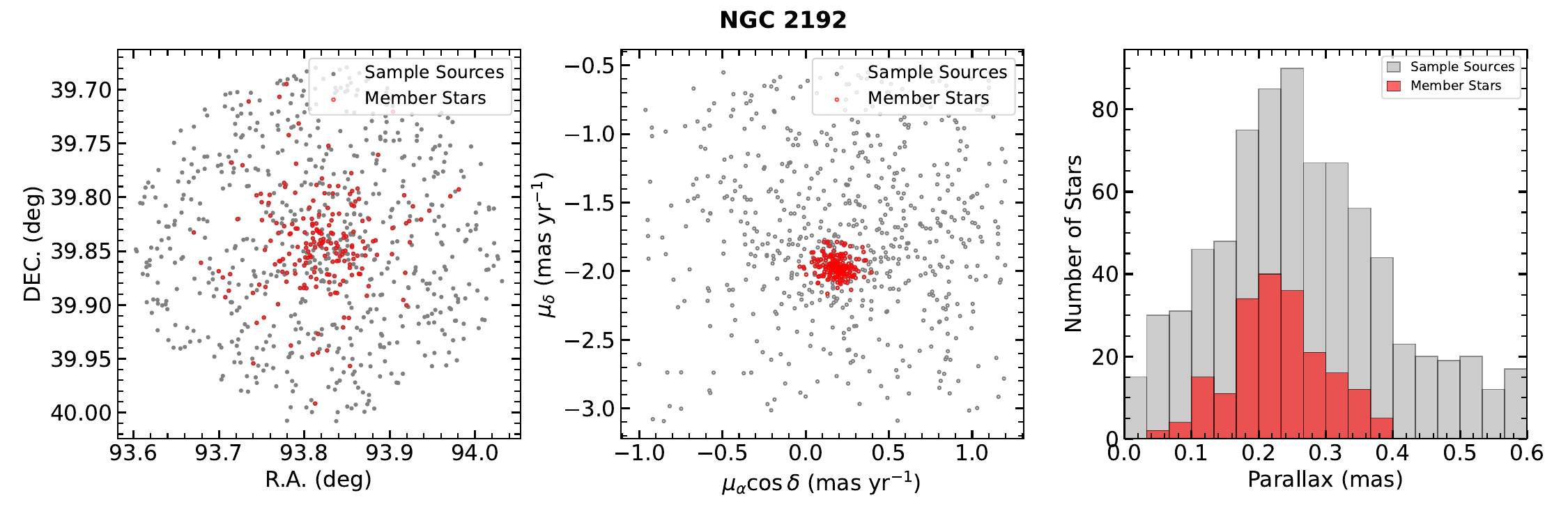}
   \includegraphics[width=15cm,height=4.5cm]{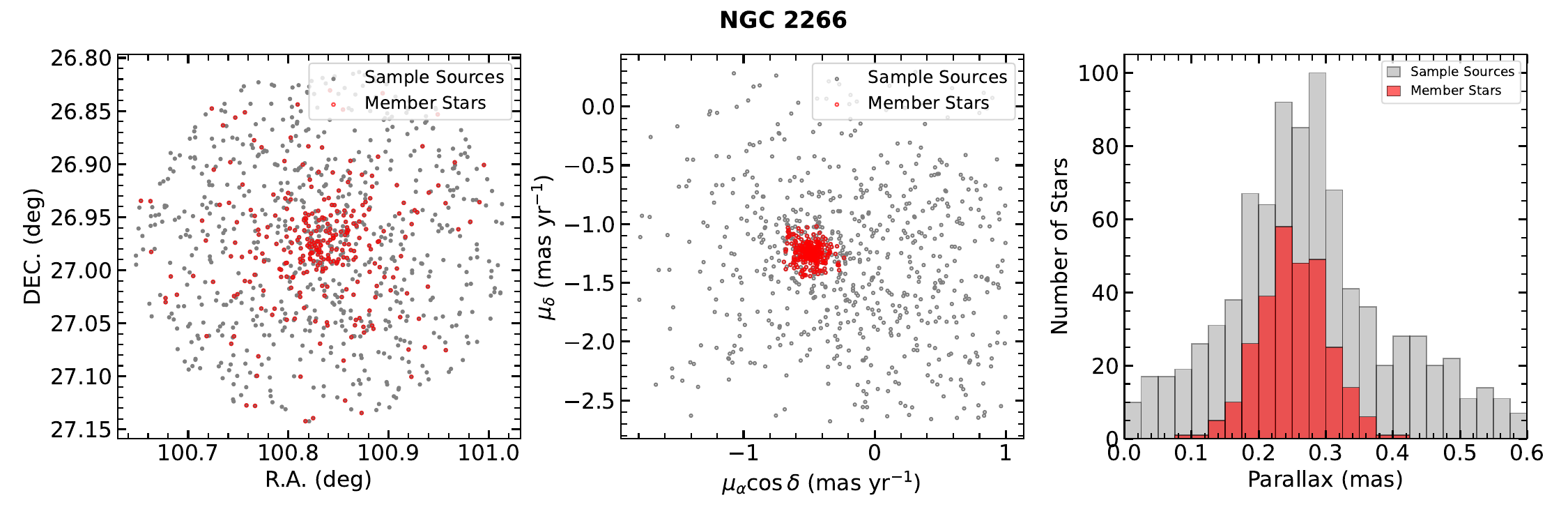}
   \includegraphics[width=15cm,height=4.5cm]{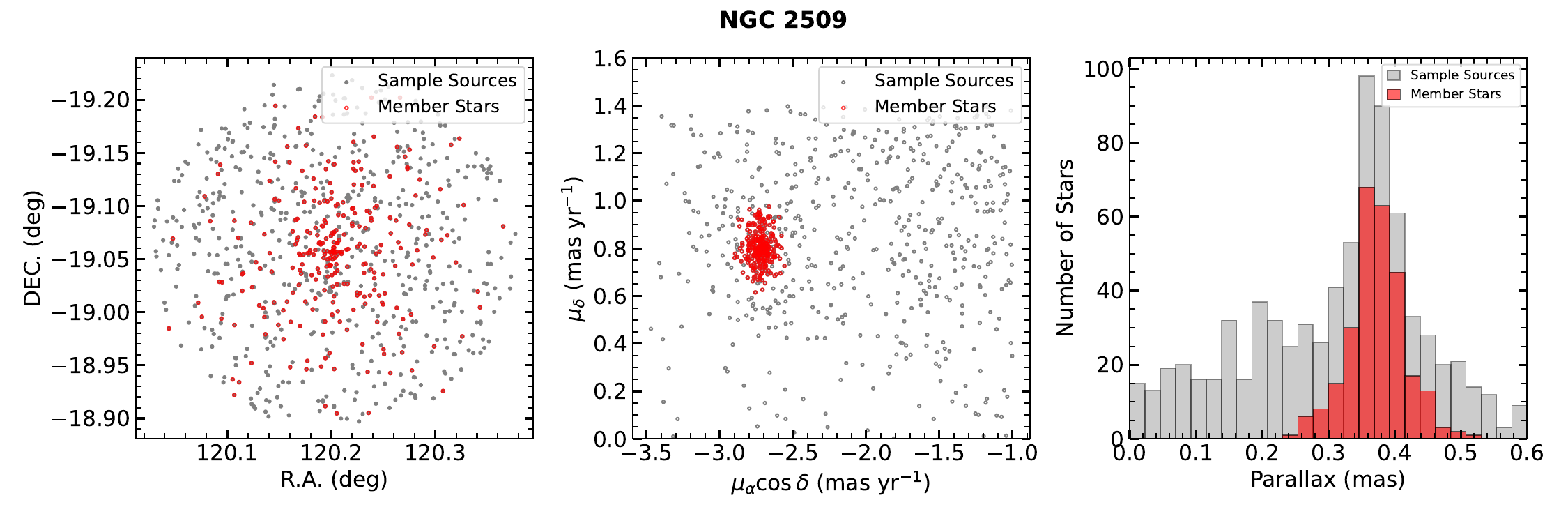}
   \caption{Spatial, proper-motion, and parallax distributions of stars in the field of NGC~2192, NGC~2266, and NGC~2509, illustrating the membership selection procedure based on \textit{Gaia}~DR3 astrometry. In each row, the left panel shows the spatial distribution of all sources within the adopted cluster radius (grey points), with probable cluster members highlighted in red. The middle panel shows the corresponding proper-motion distribution, which demonstrates the tight kinematic clustering of member stars. The right panel shows the parallax histogram, where the red distribution indicates the selected cluster members and the grey distribution represents the full sample. These plots demonstrate the effectiveness of the Gaussian Mixture Model membership analysis applied uniformly to all clusters studied in this work.}
   \label{fig:membership_appendix}
\end{figure}



\label{appendix:A2}

\begin{figure}[htbp]
    \centering
    \includegraphics[width=5.5cm,height=5.0cm]{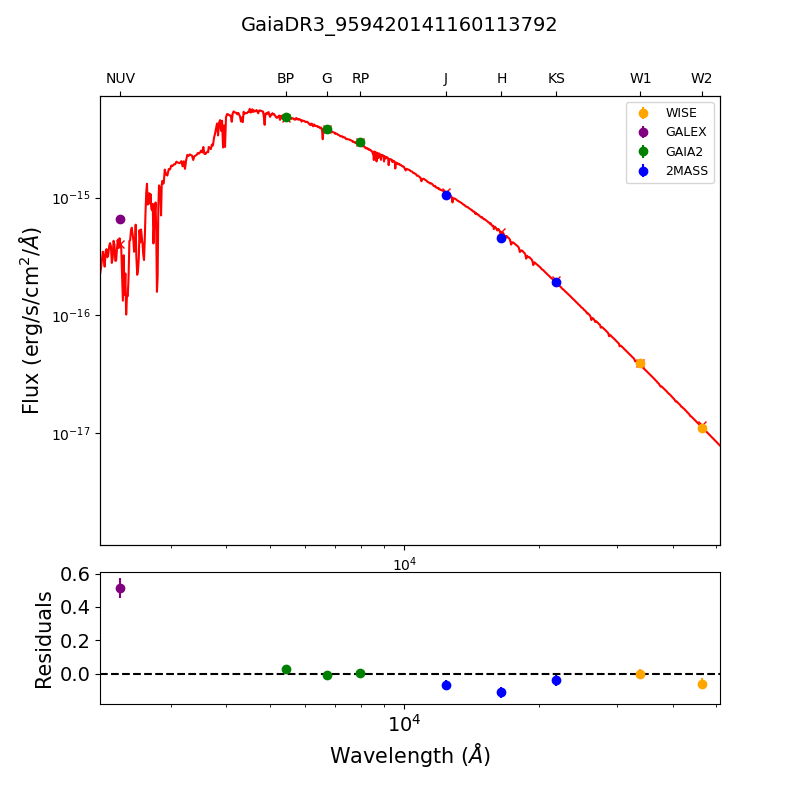}
    \includegraphics[width=5.5cm,height=5.0cm]{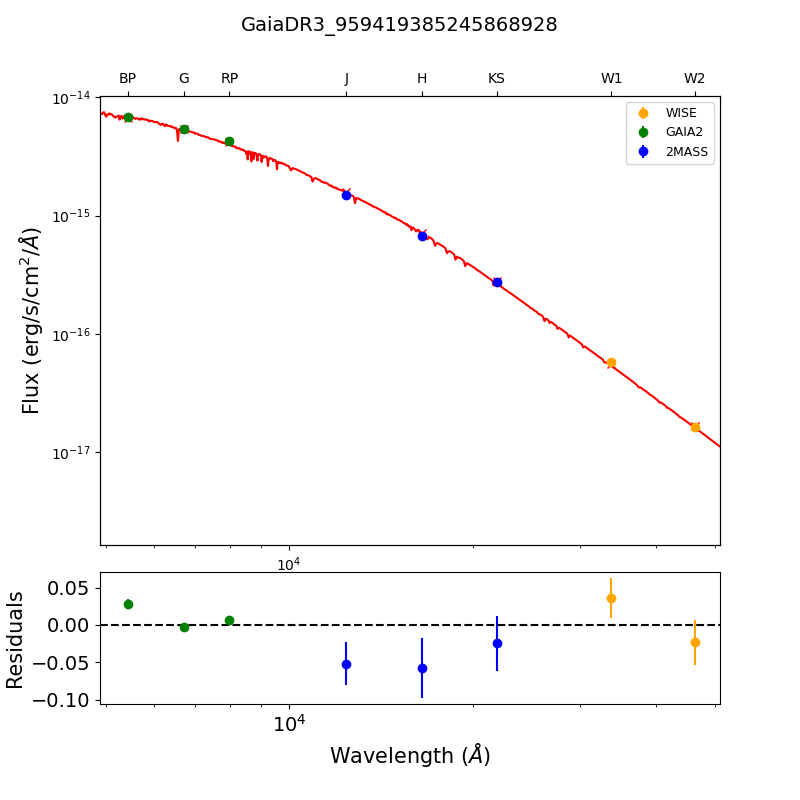}
    \includegraphics[width=5.5cm,height=5.0cm]{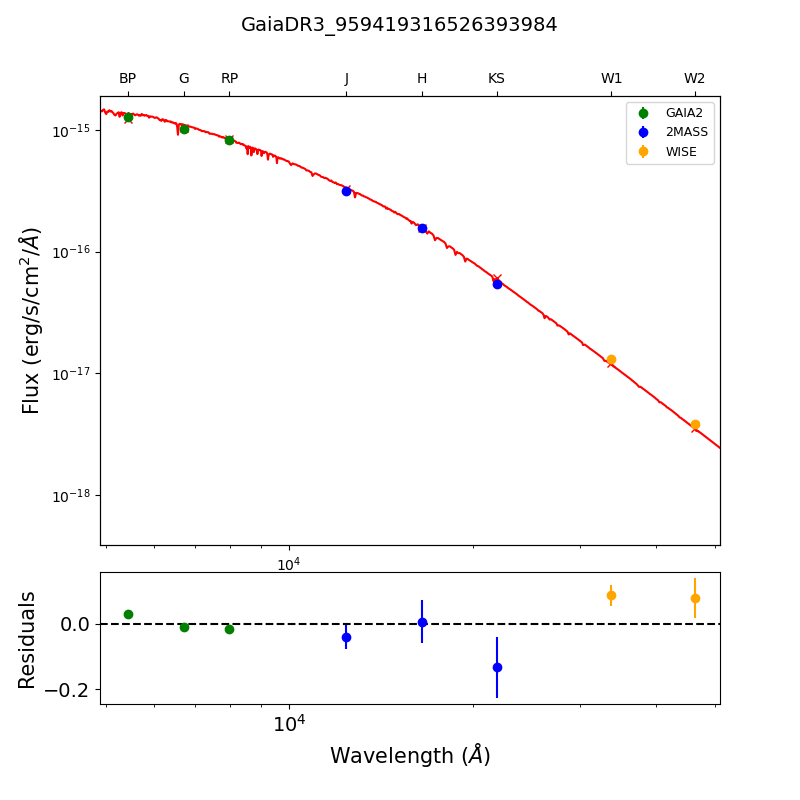}
    \includegraphics[width=5.5cm,height=5.0cm]{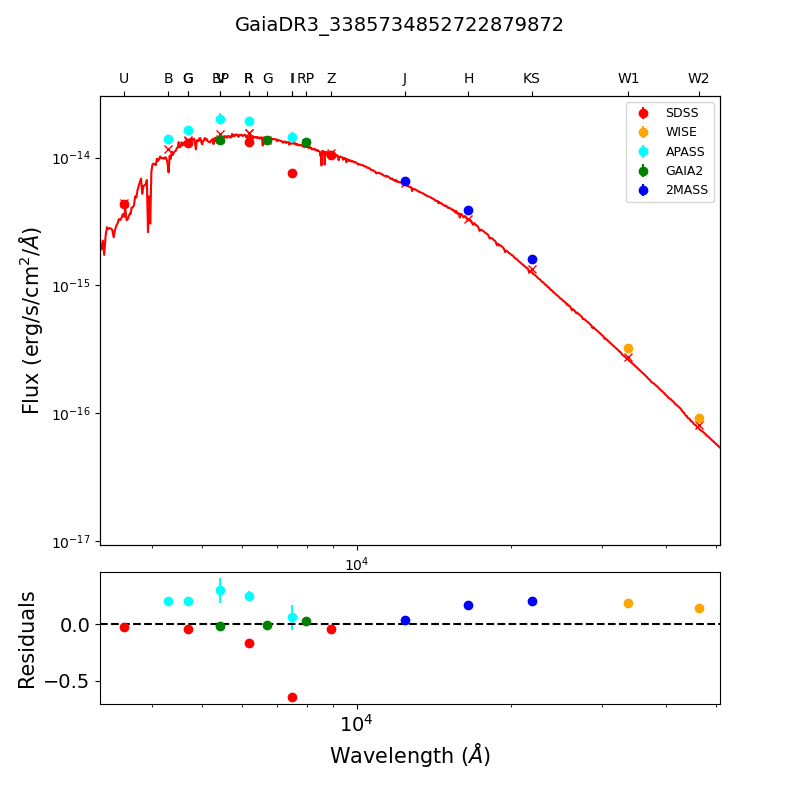}
    \includegraphics[width=5.5cm,height=5.0cm]{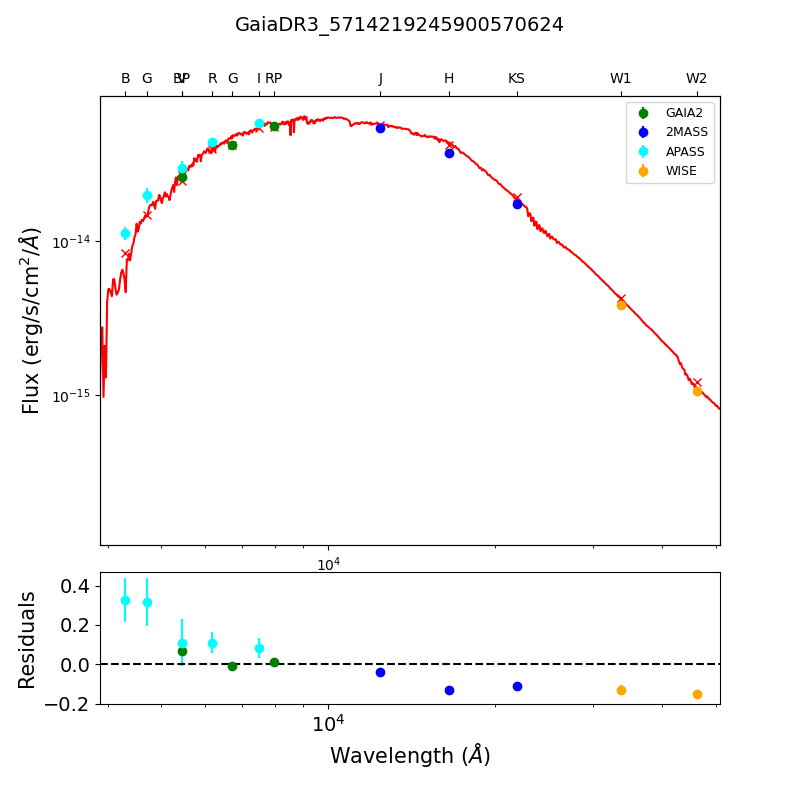}
    \caption{Spectral energy distribution (SED) fits for the remaining variable stars identified in the observed OCs, shown here for completeness. The observed multi-band photometric fluxes are plotted together with the best-fitting Kurucz model atmospheres, while the lower panels display the corresponding residuals. These SEDs were used to estimate the stellar parameters reported in Table~\ref{tab: sed_parameters}.}
    \label{fig: app_all_sed}
\end{figure}


\end{document}